\documentclass[longauth]{aa}  
\usepackage{graphicx,multirow}
\usepackage{txfonts}
\usepackage{xcolor}
\usepackage{hyperref}
\hypersetup{colorlinks = false,linkcolor = .,linkbordercolor = green, urlbordercolor = green, }
\usepackage{gensymb}
\usepackage{bm}
\usepackage{booktabs}
\usepackage{placeins}
\usepackage{csquotes}
\usepackage{float}
\defcitealias{Wright25}{W25}
\defcitealias{KiDS+DES}{DES+KiDS}

\begin{document} 

\title{KiDS-Legacy: Consistency of cosmic shear measurements and joint cosmological constraints with external probes}
\titlerunning{KiDS-Legacy: Internal consistency and joint cosmological constraints}
\author{Benjamin St\"olzner\inst{1}\thanks{stoelzner@astro.rub.de}\and
        Angus H. Wright\inst{1}\and
        Marika Asgari\inst{2}\and
        Catherine Heymans\inst{1,3}\and
        Hendrik Hildebrandt\inst{1}\and
        Henk Hoekstra\inst{4}\and
        Benjamin Joachimi\inst{5}\and
        Konrad Kuijken\inst{4}\and
        Shun-Sheng Li\inst{1,4}\and
        Constance Mahony\inst{22,23,1}\and
        Robert Reischke\inst{1,6}\and
        Mijin Yoon\inst{4}\and
        Maciej Bilicki\inst{19}\and
        Pierre Burger\inst{7,8,6}\and
        Nora Elisa Chisari\inst{9,4}\and
        Andrej Dvornik\inst{1}\and
        Christos Georgiou\inst{21}\and
        Benjamin Giblin\inst{3}\and
        Joachim Harnois-D\'eraps\inst{2}\and
        Priyanka Jalan\inst{19}\and
        Anjitha John William\inst{19}\and
        Shahab Joudaki\inst{10,11}\and
        Giorgio Francesco Lesci\inst{13,14}\and
        Laila Linke\inst{12}\and
        Arthur Loureiro\inst{24,25}\and
        Matteo Maturi\inst{26,27}\and
        Lauro Moscardini\inst{13,14,15}\and
        Nicola R. Napolitano\inst{28,29,30}\and
        Lucas Porth\inst{6}\and
        Mario Radovich\inst{16}\and
        Tilman Tr\"oster\inst{20}\and
        Edwin Valentijn\inst{31}\and
        Maximilian von Wietersheim-Kramsta\inst{17,18}\and
        Anna Wittje\inst{1}\and
        Ziang Yan\inst{1}\and
        Yun-Hao Zhang\inst{3,4}
        }

\institute{
Ruhr University Bochum, Faculty of Physics and Astronomy, Astronomical Institute (AIRUB), German Centre for Cosmological Lensing, 44780 Bochum, Germany 
\and School of Mathematics, Statistics and Physics, Newcastle University, Herschel Building, NE1 7RU, Newcastle-upon-Tyne, United Kingdom
\and Institute for Astronomy, University of Edinburgh, Royal Observatory, Blackford Hill, Edinburgh, EH9 3HJ, United Kingdom
\and Leiden Observatory, Leiden University, P.O.Box 9513, 2300RA Leiden, The Netherlands
\and Department of Physics and Astronomy, University College London, Gower Street, London WC1E 6BT, United Kingdom
\and Argelander-Institut für Astronomie, Universität Bonn, Auf dem Hügel 71, D-53121 Bonn, Germany
\and Waterloo Centre for Astrophysics, University of Waterloo, Waterloo, ON N2L 3G1, Canada
\and Department of Physics and Astronomy, University of Waterloo, Waterloo, ON N2L 3G1, Canada
\and Institute for Theoretical Physics, Utrecht University, Princetonplein 5, 3584CC Utrecht, The Netherlands
\and Centro de Investigaciones Energéticas, Medioambientales y Tecnológicas (CIEMAT), Av. Complutense 40, E-28040 Madrid, Spain
\and Institute of Cosmology \& Gravitation, Dennis Sciama Building, University of Portsmouth, Portsmouth, PO1 3FX, United Kingdom
\and Universität Innsbruck, Institut für Astro- und Teilchenphysik, Technikerstr. 25/8, 6020 Innsbruck, Austria
\and Dipartimento di Fisica e Astronomia "Augusto Righi" - Alma Mater Studiorum Università di Bologna, via Piero Gobetti 93/2, I-40129 Bologna, Italy
\and Istituto Nazionale di Astrofisica (INAF) - Osservatorio di Astrofisica e Scienza dello Spazio (OAS), via Piero Gobetti 93/3, I-40129 Bologna, Italy
\and Istituto Nazionale di Fisica Nucleare (INFN) - Sezione di Bologna, viale Berti Pichat 6/2, I-40127 Bologna, Italy
\and INAF - Osservatorio Astronomico di Padova, via dell'Osservatorio 5, 35122 Padova, Italy
\and Institute for Computational Cosmology, Ogden Centre for Fundament Physics - West, Department of Physics, Durham University, South Road, Durham DH1 3LE, United Kingdom
\and Centre for Extragalactic Astronomy, Ogden Centre for Fundament Physics - West, Department of Physics, Durham University, South Road, Durham DH1 3LE, United Kingdom
\and Center for Theoretical Physics, Polish Academy of Sciences, al. Lotników 32/46, 02-668 Warsaw, Poland
\and Institute for Particle Physics and Astrophysics, ETH Zürich, Wolfgang-Pauli-Strasse 27, 8093 Zürich, Switzerland
\and Institut de Física d’Altes Energies (IFAE), The Barcelona Institute of Science and Technology, Campus UAB, 08193 Bellaterra (Barcelona), Spain
\and Department of Physics, University of Oxford, Denys Wilkinson Building, Keble Road, Oxford OX1 3RH, United Kingdom
\and Donostia International Physics Center, Manuel Lardizabal Ibilbidea, 4, 20018 Donostia, Gipuzkoa, Spain
\and The Oskar Klein Centre, Department of Physics, Stockholm University, AlbaNova University Centre, SE-106 91 Stockholm, Sweden
\and Imperial Centre for Inference and Cosmology (ICIC), Blackett Laboratory, Imperial College London, Prince Consort Road, London SW7 2AZ, United Kingdom
\and Zentrum für Astronomie, Universit\"at Heidelberg, Philosophenweg 12, D-69120 Heidelberg, Germany
\and Institute for Theoretical Physics, Philosophenweg 16, D-69120 Heidelberg, Germany
\and Department of Physics “E. Pancini” University of Naples Federico II C.U. di Monte Sant’Angelo Via Cintia, 21 ed. 6, 80126 Naples, Italy
\and INAF – Osservatorio Astronomico di Capodimonte, Salita Moiariello 16, I-80131 Napoli, Italy
\and INFN, Sez. di Napoli, via Cintia, 80126, Napoli, Italy
\and Kapteyn Institute, University of Groningen, PO Box 800, NL 9700 AV Groningen
}
\date{Received 31 March 2025 / Accepted 31 July 2025}
 
\abstract
{
We present a cosmic shear consistency analysis of the final data release from the Kilo-Degree Survey (KiDS-Legacy). By adopting three tiers of consistency metrics, we compared cosmological constraints between subsets of the KiDS-Legacy dataset split by redshift, angular scale, galaxy colour, and spatial region. We also reviewed a range of two-point cosmic shear statistics. As all the data passed our set of consistency metric tests, we demonstrate that KiDS-Legacy is the most internally consistent KiDS catalogue to date. In a joint cosmological analysis of KiDS-Legacy and DES Y3 cosmic shear, combined with data from the Pantheon+ Type Ia supernovae compilation and baryon acoustic oscillations from DESI Y1, we report constraints that are consistent with {\it Planck} measurements of the cosmic microwave background, with
$S_8\equiv \sigma_8\sqrt{\Omega_{\rm m}/0.3} = 0.814^{+0.011}_{-0.012}$ and $\sigma_8 = 0.802^{+0.022}_{-0.018}$.
}


\maketitle
\tableofcontents
\section{Introduction}
In the current era of precision cosmology, weak lensing surveys have probed the standard $\Lambda$-cold dark matter ($\Lambda$CDM) cosmological model to an unprecedented level of precision. The weak gravitational lensing effect causes distortions in the shape of galaxy images, known as cosmic shear. This allows us to map the distribution of gravitating matter along the line of sight, which is sensitive to the shape and amplitude of the matter power spectrum. Since its first detection \citep{Kaiser00,Wittman00,vanWaerbeke00,Bacon00}, cosmic shear has become a primary cosmological probe for imaging galaxy surveys. Recent analyses from the three current stage-III weak lensing surveys, namely the ESO Kilo-Degree Survey \citep[KiDS;][]{Kuijken15}, the Dark Energy Survey \citep[DES;][]{DES6}, and the Subaru Hyper Suprime Cam Subaru Strategic Program \citep[HSC;][]{HSC1}, have showcased the potential of cosmic shear measurements as a probe of the cosmological model, making it a main science driver for upcoming galaxy surveys conducted with the Vera C. Rubin Observatory \citep{LSST}, the {\it Euclid} satellite \citep{Euclid24}, the \textit{Nancy Grace Roman} Space Telescope \citep{Spergel15}, and the Chinese Space Station Telescope \citep{Gong19}.
        
Weak lensing studies mainly constrain the structure growth parameter $S_8=\sigma_8\sqrt{\Omega_{\rm m}/0.3}$, which combines the matter density parameter, $\Omega_{\rm m}$, and the standard deviation of matter density perturbations in spheres of $8\, h^{-1}{\rm Mpc}$ radius, denoted as $\sigma_8$. Current cosmic shear measurements have yielded $S_8$ values that are lower \citep{Asgari21,Heymans21,Abdalla22,Amon22,Secco22,Dalal23,LiHSC23,KiDS+DES} than values derived from observations of the cosmic microwave background \citep[CMB;][]{Planck2018} at a $\sim2\sigma$ level. However, there is no consensus about whether this feature, which is commonly referred to as the \enquote*{$S_8$ tension}, is a result of systematics in the data analysis or theoretical modelling, statistical fluctuations, or effects beyond the standard flat $\Lambda$CDM model.

Given the unclear nature of the apparent $S_8$ tension, recent works have studied probes of the late Universe, reporting a consistency between independent cosmic shear surveys \citep{Amon22b,Amon23,Longley23,KiDS+DES}. In addition to tests of the consistency between independent datasets, additional tests of the internal consistency of a given dataset are of particular importance to rule out systematic effects within one dataset as the source of the inconsistency between different datasets \citep[][]{Efstathiou18,Koehlinger19,Raveri20,Li21}.

This work is part of a series of KiDS-Legacy papers. The production process of all data products in the fifth KiDS data release (DR5), including shape measurements and the KiDS-Legacy sample selection, is presented in \citet{Wright23_DR5}. The calibration of the photometric redshift distribution is described in \citet{Wright23_redshifts} and multi-band image simulations enabling a joint shear and redshift calibration are presented in \citet{Li23b}. The modelling of the covariance for the three main summary statistics is summarised in \citet{Reischke23} and the angular clustering of KiDS-Legacy galaxies is analysed in \citet{Yan25}. The fiducial cosmic shear analysis was conducted in \citet[][hereafter W25]{Wright25}. In this work, we performed several internal consistency tests of the KiDS-Legacy data, focussing on their impact on cosmological constraints inferred in the cosmic shear analysis. In particular, we split the KiDS-Legacy dataset into various subsets based on redshift, spatial region, angular scale, and colour. We then performed a split cosmological analysis by modelling the observed data in each subset with a separate set of cosmological parameters. By evaluating several consistency metrics, we quantified the level of agreement between the data subsets. As established in the previous KiDS-1000 analysis \citep{Asgari21}, constraints on cosmology were inferred from three different two-point statistics (COSEBIs, band powers, and two-point correlation functions). Here, we quantified the internal consistency between summary statistics.

While current cosmic shear surveys cannot constrain both $\Omega_{\rm m}$ and $\sigma_8$ separately, the addition of external data allows for a breaking of the degeneracy. For this purpose, we employed data from recent measurements of baryon acoustic oscillations (BAOs), redshift space distortions (RSDs), and Type Ia supernovae (SN Ia), which place tight constraints on the matter density. We performed a consistency test and a joint cosmological analysis of KiDS-Legacy data combined with data from the recent DESI Y1 BAO analysis \citep{DESI24}, the earlier eBOSS \citep{Alam21} BAO and RSD analysis, and the Pantheon+ SN Ia compilation \citep{Scolnic22,Brout22}. Additionally, we conducted a joint analysis with DES Y3 cosmic shear data \citep{Amon22,Secco22} and quantified the consistency between KiDS-Legacy and {\it Planck} CMB constraints \citep{Planck2018}.

This paper is structured as follows. In Sect. \ref{sec:data}, we provide a brief summary of the KiDS-Legacy data and external data employed in this work. In Sect. \ref{sec:methodology}, we review the theoretical model for weak lensing observables and discuss the metrics quantifying the internal consistency of the data. In Sect. \ref{sec:results}, we present the results of the internal consistency tests. We provide the results of our joint cosmological constraints with external data in Sect. \ref{sec:results_external} and present our conclusions in Sect. \ref{sec:conclusions}. Appendix \ref{ap:catlevel} summarises the data properties of the KiDS-Legacy catalogue divided into sub-samples. In Appendix \ref{ap:neff}, we provide details on our estimations of the effective number of constrained parameters in our analysis. Appendix \ref{ap:sensitivity} presents a sensitivity analysis of our consistency metrics.
\section{Data}
\label{sec:data}
\subsection{KiDS-Legacy}
The Kilo-Degree Survey \citep{Kuijken15,Dejong15,Dejong17,Kuijken19,Wright23_DR5} is a public survey conducted by the European Southern Observatory with the VLT Survey Telescope (VST). KiDS and the complementary VISTA Kilo-Degree Infrared Galaxy Survey \citep[VIKING;][]{Edge13} combine optical and near-infrared imaging in nine photometric bands. In this work, we analyse weak lensing data from the fifth and final data release (DR5). Here, we provide a brief summary of the survey and refer to \citet{Wright23_DR5} for details. 

The KiDS-DR5 dataset consists of imaging data covering an area of 1347 ${\rm deg}^2$ on-sky, divided into two distinct stripes across the celestial equator in the North Galactic Cap and across the South Galactic Pole, respectively. All sources were observed in four optical bands ($u,g,r$, and $i$) with the VST as well as in five near-infrared bands ($Z,Y,J,H$, and $K_{\rm s}$) from VIKING. In comparison to the previous data release, KiDS-DR5 features a second pass of $i$-band observations and an increase of $34\%$ in survey area. The lensing sample was obtained via a masking process, selecting sources with high-quality data in all photometric bands and applying a sequence of cuts on magnitude, colour, and lensing-related quantities. This sample, dubbed KiDS-Legacy, contains approximately 43 million sources on $967\,{\rm deg}^2$ of sky, corresponding to an effective number density of $n_{\rm eff}=8.79\,{\rm arcmin}^{-2}$. 

Photometric redshift estimates of KiDS-Legacy sources were computed via the Bayesian Photo-z code \citep[{\sc bpz};][]{Benitez00}. The deeper $i$-band depth and a significantly larger spectroscopic calibration dataset allow for a higher photometric redshift limit of $z_{\rm B}=2$ compared to previous KiDS analyses, enabling the addition of another redshift bin. The sources in the KiDS-Legacy lensing sample were divided into six approximately equi-populated bins via their photometric redshift estimates. The redshift distributions of the resulting bins were inferred via a direct calibration with deep spectroscopic surveys using self-organising maps \citep[SOMs;][]{Lima08,Masters15,Wright20}. While this calibration method was utilised in earlier KiDS analyses \citep{Wright20b,Hildebrandt21}, the KiDS-Legacy redshift calibration features several improvements such as the use of one SOM per tomographic bin instead of one overall SOM and a selection of sources via SOM-derived gold weights, while also accounting for prior volume effects. Furthermore, the SOM redshift distributions were calibrated with the multi-colour SKiLLS simulations \citep{Li23b}. For a detailed discussion of the improved calibration method, we refer to the KiDS-Legacy redshift calibration manuscript \citep{Wright23_redshifts}.

Shape measurements in KiDS-Legacy were performed with an updated version of the {\sc lensfit} algorithm \citep{Miller13,Fenech17} and calibrated with the SKiLLS image simulations as established in \citet{Li23b}. The shape measurements were validated with a series of systematics tests as outlined in \citetalias{Wright25}.
\subsection{External data}
\label{sec:external_data}
We employed several external datasets featuring measurements of BAOs, RSDs, and SN Ia to quantify the consistency of the KiDS results and infer joint constraints with the KiDS-Legacy cosmic shear data. In this section, we briefly describe the external datasets. In this work, we use publicly available likelihoods, which are implemented in the {\sc CosmoSIS} standard library \footnote{\url{https://github.com/joezuntz/cosmosis-standard-library/tree/v4.0/likelihood}}.

We made use of BAO measurements from the first data release of the Dark Energy Spectroscopic Instrument \citep[DESI;][]{Aghamousa16,Abareshi22} survey. In particular, DESI targets four different classes of extragalactic objects: a bright galaxy sample \citep[][]{Hahn23}, luminous red galaxies \citep[LRGs;][]{Zhou23}, emission line galaxies \citep[ELGs;][]{Raichoor23}, and quasi-stellar objects \citep[QSOs;][]{Chaussidon23}. Cosmological results from DESI BAO measurements were presented in \citet{DESI24}. The DESI likelihood provides BAO measurements from the four DESI sub-samples, covering a wide range of redshifts. In particular, we employ measurements from the bright galaxy sample ($0.1<z<0.4$), two LRG samples ($0.4<z<0.6$ and $0.6<z<0.8$, respectively), an ELG sample ($1.1<z<1.6$), a combined LRG and ELG sample ($0.8<z<1.1$), a QSO sample ($0.8<z<2.1$), and a Lyman-$\alpha$ forest sample ($1.77<z<4.16$).

As an alternative to recent BAO measurements from DESI, we employ data from the Sloan Digital Sky Survey's (SDSS) Baryon Oscillation Spectroscopic Survey (BOSS) and Extended Baryon Oscillation Spectroscopic Survey (eBOSS). This dataset provides BAO measurements as well as measurements of RSDs. The likelihoods provide constraints from the SDSS DR7 main galaxy sample \citep{Ross15,Howlett15}, BOSS DR12 \citep{Alam17}, eBOSS DR16 ELGs \citep{Tamone20,Raichoor21,deMattia21}, eBOSS DR16 LRGs \citep{Bautista21,GilMarin20}, eBOSS DR16 QSOs \citep{Neveux20,Hou21}, and the eBOSS DR16 Lyman-$\alpha$ forest \citep{Bourboux20}.

Additionally, we employed measurements of SN Ia from the Pantheon+ compilation \citep{Scolnic22}. This dataset consists of 1701 light curves of 1550 spectroscopically confirmed SN Ia with redshifts $z\in (0.001,2.26)$. Cosmological constraints from this dataset were presented in \citet{Brout22}.

We employed cosmic shear measurements from DES \citep{DES05,DES6,DES15}. In particular, we make use of the \enquote*{KiDS-excised} DES data vector presented in \citet[][hereafter DES+KiDS]{KiDS+DES}, which is based on the analysis of the DES Y3 cosmic shear measurements \citep{Amon22, Secco22} and excludes 8\% of DES data in the overlap region between KiDS and DES. Following the methodology of this study, we neglected the cross-covariance between the two surveys, which was shown to be sufficiently small. Furthermore, we adopted the \enquote*{$\Lambda$CDM-optimised} angular scale cuts of \citet{Amon22} and \citet{Secco22}.

Finally, we adopted CMB measurements from \citet{Planck2018}. Here, we make use of the compressed {\it Planck} likelihood of \citet{Prince19}, which approximates the $\ell<30$ temperature likelihood by two Gaussian data points and employs the {\it Planck} {\tt plik-lite} TTTEEE likelihood for $\ell>30$. Following the methodology outlined in this work, we impose a Gaussian prior on the optical depth to re-ionisation, $\tau$, which is derived from base $\Lambda$CDM parameter constraints from {\it Planck}.
\section{Methodology}
\label{sec:methodology}
\subsection{Cosmic shear model}
\label{sec:WL_model}
Our cosmic shear analysis pipeline is based on \citetalias{Wright25} and is implemented in the public \textsc{CosmoPipe}\footnote{\url{https://github.com/AngusWright/CosmoPipe/tree/KiDSLegacy_CosmicShear}} infrastructure. In this section, we summarise the theoretical modelling of the cosmic shear signal. We employed three summary statistics as established in the previous KiDS-1000 analysis \citep{Asgari21}. In addition to real space shear two-point correlation functions (2PCFs), which are commonly used in cosmic shear studies, we made use of two additional summary statistics which are derived from 2PCF measurements. First, we computed complete orthogonal sets of E/B-integrals \citep[COSEBIs; ][]{Schneider10,Asgari12}, which provide a clean separation of E and B modes. This is of particular advantage since, for current surveys, we expect only the E modes to carry the cosmic shear signal to first order, allowing for the B modes to be used as a null test for residual systematics. Second, we employed band power spectra inferred from correlation functions \citep{Schneider02,Becker16,vanUitert18}. This statistic enables an approximate separation of E and B modes and follows the underlying angular power spectra. In this work, we used COSEBIs as our fiducial statistic when reporting analysis results, following the choice made in \citetalias{Wright25}.
                
In general, we model the signal for each summary statistic $S$ via a linear transformation of the cosmic shear power spectrum, $C^{(ij)}_{\varepsilon\varepsilon}$:
        \begin{equation}
        \label{eq:summarystatistic}
                S^{(ij)}=\int_0^\infty {\rm d} \ell\, C^{(ij)}_{\varepsilon\varepsilon}(\ell)W_S(\ell)\;, 
        \end{equation}
        where $W_S(\ell)$ is a weight function depending on the angular scale and the summary statistic itself. The cosmic shear power spectrum is given by the sum of the gravitational lensing power spectrum (GG), the intrinsic alignment of galaxies (II), and the corresponding cross term (GI):
        \begin{equation}
        \label{eq:IA}
                C^{(ij)}_{\varepsilon\varepsilon}(\ell) = C^{(ij)}_{\rm GG}(\ell) + C^{(ij)}_{\rm II}(\ell) + C^{(ij)}_{\rm GI}(\ell) + C^{(ij)}_{\rm IG}(\ell)\;.
        \end{equation}
        Under the assumption of the extended Limber approximation \citep{Kaiser92,Loverde08,Kilbinger17}, the gravitational lensing power spectrum can be written as
        \begin{equation}
C^{(ij)}_{\rm GG}(\ell) = \int_0^{\chi_{\rm H}} {\rm d}\chi\, \frac{W^{(i)}_{{\rm G}}(\chi)W^{(j)}_{\rm G}(\chi)}{f_{\rm K}^2(\chi)}P_{\rm m, nl}\left(\frac{\ell+1/2}{f_{\rm K}(\chi)},z(\chi)\right),\;
        \label{eq:shear_powerspectrum}
        \end{equation}
where $P_{\rm m, nl}$ denotes the non-linear matter power spectrum and $f_{\rm K}$, $\chi$, and $\chi_{\rm H}$ are the comoving angular diameter distance, the comoving radial distance, and the comoving horizon distance, respectively. The weak lensing kernel $W^{(i)}_{\rm G}(\chi)$ is given, for example, in eq. 2 in \citetalias{Wright25}. Furthermore, the underlying galaxy sample is divided into tomographic bins using estimates of the photometric redshift, allowing for an increase in constraining power \citep{Hu99}. Therefore, Eq. \eqref{eq:shear_powerspectrum} refers to the cross cosmic shear signal between combinations of tomographic bins $i$ and $j$, where the probability distribution of comoving distances of galaxies per bin enters the window function, $W^{(i)}_{\rm G}(\chi)$.

In \citetalias{Wright25}, we explored a range of models of the intrinsic alignment (IA) of galaxies, finding no significant impact of the IA model on the $S_8$ constraints. In the present work, we therefore employed the fiducial mass-dependent IA model, dubbed NLA-$M$. This model extends the non-linear linear alignment (NLA) model \citep[][]{Bridle07}, incorporating an alignment of red, early-type galaxies, while assuming zero alignment of blue, late-type galaxies. The fraction of early-type galaxies is inferred by selecting galaxies with spectral type $T_{\rm B}<1.9$. The spectral type is inferred with {\sc bpz}, which uses a set of six model templates of the spectral energy distribution ordered approximately based on the star formation activity and determines the best-fitting spectral energy distribution by interpolating between templates. The cut on $T_{\rm B}$ selects galaxies with contributions of an elliptical galaxy spectrum. We modelled the IA power spectrum of red galaxies as a power law dependent on the average halo mass within a tomographic bin \citepalias[see eq. 11 in][]{Wright25}. In this model, we employed two nuisance parameters, $A_{\rm IA}$ and $\beta$, which parameterise the IA amplitude and the slope of the IA mass scaling, respectively. We adopted the joint posterior on $A_{\rm IA}$ and $\beta$ from \citet{Fortuna24} as a prior in our analysis, which we approximated by a bivariate Gaussian distribution. Additionally, we employed a multivariate Gaussian prior on the halo mass per tomographic bin. For a detailed description of the IA model, we refer to Section 2.3.4 and appendix B in \citetalias{Wright25}.
\subsection{Consistency metrics}
\label{sec:metrics}
\begingroup
\renewcommand{\arraystretch}{1.05}
\begin{table*}
\caption{Model parameters and their priors.}
\label{tab:parameters}
\centering
\begin{tabular}{cllll}
\hline\hline
Type & Parameter & Prior & Duplicated & Description\\
\hline
\multirow{5}{*}{\rotatebox[origin=c]{90}{Cosmological}} 
&$\omega_{\rm cdm}$             & $\mathcal{U}(0.051,0.255)$                                              & Yes & Reduced cold dark matter density\\
&$\omega_{\rm b}$               & $\mathcal{U}(0.019,0.026)$                                              & Yes & Reduced baryon density\\
&$h$                                        & $\mathcal{U}(0.64,0.82)$                                                    & Yes & Reduced Hubble parameter\\
&$n_{\rm s}$                    & $\mathcal{U}(0.84,1.1)$                                                         & Yes & Spectral index of the primordial power spectrum\\
&$S_8$                          & $\mathcal{U}(0.5,1.0)$                                                          & Yes & Structure growth parameter \\
\hline
\multirow{5}{*}{\rotatebox[origin=c]{90}{Nuisance}} 
&$\log_{10}T_{\rm AGN}$     & $\mathcal{U}(7.3,8.3)$                                                      & Yes & Baryon feedback parameter\\
&$A_{\rm IA}$                           & $\mathcal{N}(5.74,\mathbf{C}_{A_{\rm IA}, \beta})$ & No + red-blue split only   & Amplitude of intrinsic galaxy alignments for red galaxies\\
&$\beta$                                        & $\mathcal{N}(0.44,\mathbf{C}_{A_{\rm IA}, \beta})$ & No  & Slope of the mass scaling of intrinsic galaxy alignments\\
&$\log_{10}M_{\rm i}$           & $\mathcal{N}(\bm{\mu},\mathbf{C}_M)$                             & No  & Mean halo mass of early-type galaxies per tomographic bin $i$\\
&$\delta_{{\rm z},i}$                   & $\mathcal{N}(\bm{\mu},\mathbf{C}_z)$                                   & No + red-blue split only & Shift of the mean of the redshift distribution per tomographic bin $i$\\
\hline
\end{tabular}
\tablefoot{The first two columns provide parameter names and types of the sampling parameters. The third column lists the adopted prior with uniform priors denoted by $\mathcal{U}$ and Gaussian priors denoted by $\mathcal{N}$. The fourth column indicates whether or not the parameter is duplicated in the split cosmological analysis. The fifth column provides a brief parameter description. The priors on $A_{\rm IA}$ and $\beta$ are centred at the mean of the posterior reported by \citet{Fortuna24} including a correlated prior with correlation coefficient $r=-0.59$. The prior on $M_{\rm i}$ is inferred from halo mass measurements as discussed in appendix B in \citetalias{Wright25}. The $\delta_{\rm z}$ priors are centred at the mean of the shift of the redshift distribution and correlated through a covariance matrix estimated following the methodology of \citet{Wright23_redshifts}.}
\end{table*}
\endgroup
To assess the internal consistency of the KiDS-Legacy dataset, we followed the methodology of \citet{Koehlinger19} and subdivided the data in many ways before analysing the subsets jointly. In particular, we applied splits at the data vector level by redshift and angular scale and at the catalogue level by spatial region and colour. Additionally, we constructed a joint data vector of different summary statistics. For each split, we analysed the cosmic shear data with two modelling setups:
\begin{enumerate}
        \item Fiducial cosmological model: one set of parameters models the full dataset;
        \item Split cosmological model: two sets of parameters model two mutually exclusive (but generally correlated) subsets of the data.
\end{enumerate}
For splits at the data vector-level, the first setup is equivalent to the fiducial cosmic shear analysis setup of \citetalias{Wright25}. For splits at the catalogue level and splits by summary statistic, we constructed a different data vector, whose information content may differ from the fiducial data vector; for example, the red-blue split excludes shape correlations between red and blue galaxies. The second modelling setup features two independent sets of cosmological parameters, which allow us to assess whether or not different subsets of the data prefer different cosmologies. A list of cosmological and nuisance parameters is given in Table \ref{tab:parameters}. The cosmic shear analysis features a number of nuisance parameters which are marginalised over, assuming Gaussian or top-hat priors. Since the posterior distribution of the nuisance parameters is entirely driven by the prior \citepalias[see][]{Wright25} we did not duplicate these parameters in the split cosmological analysis; instead, we kept them shared between both data subsets. An exception is the colour-based split, as discussed in Sect. \ref{sec:redblue}.
 
In practice, we conducted a likelihood analysis for each data split employing both the fiducial model and the split cosmological model. We then evaluated various consistency metrics, testing whether the split cosmological model is preferred over the fiducial model. There are a variety of statistical tools available that allow for a model comparison and an estimation of the significance of the preference of a specific model. These can be grouped into techniques that compress the full likelihood or posterior into a single summary statistic, parameter-space methods that focus on differences in single or multiple model parameters, as well as methods that quantify differences in data vector space. In this work, we performed three tiers of consistency tests, following the nomenclature of \citet{Koehlinger19}. We note that only the tier 1 test requires a cosmological analysis with the fiducial model since it performs a model comparison. The remaining tiers focus on the analysis with the split cosmological model to quantify the consistency between data subsets.
\subsubsection{Tier 1: Evidence-based metric}
\label{sec:tier1_methods}
The first tier of consistency tests includes tests compressing the full likelihood or posterior into a single summary statistic. A widely used example of such a metric is the Bayes ratio, which in logarithmic form is given by the difference between Bayesian evidence,
\begin{equation}
        \log_{10} R = \log_{10}\mathcal{Z}_{\rm fiducial} - \log_{10}\mathcal{Z}_{\rm split}\;,
\end{equation}
where $\mathcal{Z}_{\rm fiducial/split}$ denotes the Bayesian evidence of the two models, which can generally be computed by integrating the product of the likelihood, $\mathcal{L}$, and the prior, $\pi$,
\begin{equation}
\mathcal{Z} = \int {\rm d}{\bm \theta}\,\mathcal{L}({\bm \theta})\pi({\bm \theta})\;,
\end{equation} 
over the parameters represented by ${\bm \theta}$. We note that for two independent datasets, A and B, the evidence for the split model simplifies to $\mathcal{Z}_{\rm split}=\mathcal{Z}_{\rm A}\mathcal{Z}_{\rm B}$. However, this equality does not hold for correlated datasets, namely, for most data splits considered in this work. Therefore, the evidence needs to be computed from the joint posterior distribution of the two subsets, modelled with two sets of parameters, and taking the cross-correlation between subsets into account.

In general, values of $\log_{10} R>0$ correspond to preference for the fiducial model while $\log_{10} R <0$ indicates preference for the split model. The Bayes ratio is usually interpreted using Jeffreys' scale \citep{Jeffreys39}, which provides limits for the degree of preference for a specific model. This scale associates values of $|\log_{10} R| > [\frac12,1,2]$ with \enquote*{substantial}, \enquote*{strong}, and \enquote*{decisive} preference for the specific model, respectively, but this lacks a clear motivation and there is no consensus on when to report tension between models. Additionally, the Bayes ratio suffers from a prior dependence which is suboptimal when using wide, uninformative priors. This makes the Bayes ratio particularly suboptimal for the analysis presented in this work, since it involves a duplication of the parameter space and the corresponding prior volume. 
                        
To circumvent these issues, \citet{Handley19b} proposed the so-called suspiciousness parameter, $S$, expressed as
\begin{equation}
    \ln S = \ln R - \ln I\;.
\end{equation} 
Here, the information ratio $\ln I = \mathcal{D}_{\rm split} - \mathcal{D}_{\rm fiducial}$ is defined through the Kullback-Leibler divergence $\mathcal{D}$ \citep{Kullback51}, which quantifies the information gain between the prior and the posterior. The suspiciousness is designed to remove the effect of the prior volume from the Bayes ratio. As discussed in \citet{Handley19b}, it is insensitive to the choice of prior as long as the prior does not affect the shape of the posterior distribution. Under the assumption of Gaussian posteriors, a tension probability can be identified via the quantity $d-2\ln S$. Here, $d$ denotes the difference between the effective number of constrained parameters by the two models
\begin{equation}
\label{eq:d_eff}
        d = N_{\Theta}^{\rm split} - N_{\Theta}^{\rm fiducial},\;
\end{equation}                           
with the effective number of free parameters constrained by the posterior distribution, $N_{\Theta}$. We note that the suspiciousness can be rephrased in terms of the expectation value of the log-likelihood \citep[see Appendix G.3 in][]{Heymans21} and therefore does not strictly require the computation of the evidence.

In a prior-informed analysis with correlated sampling parameters, such as the cosmic shear analysis in this work, this quantity is smaller than the number of free parameters and is non-trivial to determine. There exist several estimators, such as the Bayesian model complexity \cite[][]{Spiegelhalter02} and the Bayesian model dimensionality \citep[BMD;][]{handley19}. However, \citet{Joachimi21} reported that commonly used dimensionality measures in general are biased estimators of the effective number of parameters. As an alternative, they proposed an estimation via $\chi^2$ minimisation of a set of mock data vectors, which was generally found to reproduce an unbiased estimate of the true value of $N_{\Theta}$. Therefore, we adopted this strategy as the fiducial method in the analysis. For comparison, we additionally computed the BMD via
\begin{equation}
\label{eq:BMD}
        N_{\Theta}/2 = \langle\ln\mathcal{L}^2\rangle_P - \langle\ln\mathcal{L}\rangle_P^2\;,
\end{equation}
where $\langle\rangle_P$ denotes the average over the posterior. This quantity can be directly obtained as a byproduct of common posterior sampling algorithms, such as Markov chain Monte Carlo or nested sampling; therefore, it is available at no additional computational cost. For Gaussian posteriors, the tension probability inferred from the suspiciousness statistic is then determined by
\begin{equation}
        p_{\rm t} =\int\limits_{d-2\ln S}^{\infty} {\rm d}x\,\chi_{d}^2(x)\;,
\end{equation}
where $\chi_d^2(x)$ denotes the probability density function of a $d$-dimensional $\chi^2$-distribution. The corresponding number of sigma can then be inferred from the tension probability via
\begin{equation}
\label{eq:nsigma}
    N_{\sigma,S} = \sqrt{2}\;{\rm erf}^{-1}\left(1-p_{\rm t}\right)\;.
\end{equation}
\subsubsection{Tier 2: Multi-dimensional parameter metric}
\label{sec:tier2_methods}
The second tier consists of an analysis of the posterior distribution of parameter duplicates in the split cosmological model. Given that the posterior distribution for several sampling parameters is prior-dominated, we restricted the tier 2 test to a subset of parameters, $\bm{\theta}$, that are constrained by the data while marginalising over the remaining parameters $\bm{\theta}^{\rm marg}$. We calculated the difference via 
\begin{equation}
        \Delta{\bm\theta} = {\bm\theta}_1 - {\bm\theta}_2
\end{equation}
for each data point in the chain, where $\bm{\theta}_{\rm 1/2}$ denotes the two parameter instances in the subspace of parameters of interest. We then analysed the posterior $P(\Delta\bm{\theta})$ in the subspace of parameters of interest. The posterior of parameter differences is given by
\begin{equation}
        P(\Delta{\bm\theta}) = \int {\rm d}{\bm\theta}_1\, P^{\rm marg}({\bm\theta}_1, {\bm\theta}_1-\Delta{\bm\theta}),\;
\end{equation}
where $P^{\rm marg}({\bm\theta}_1,{\bm\theta}_2)$ denotes the posterior distribution marginalised over the remaining (unconstrained) parameters. In the absence of internal tension in the data, we expect $P(\Delta{\bm\theta})$ to be centred on the origin, while internal inconsistencies may shift the posterior away from the origin. To quantify the deviation of the posterior from zero, we followed the approach of \citet{Koehlinger19} and modelled the posterior with a kernel density estimator. We evaluated the kernel density estimator at the origin and determined the volume of the posterior where the probability of a shift is higher than the probability of no shift. Mathematically, this is equivalent to 
\begin{equation}
        V = \int\limits_{P(\Delta{\bm \theta})<P(0)} \!\!\!\!\!\!{\rm d} \Delta{\bm \theta}P(\Delta{\bm\theta}).\;
\end{equation}
We note that in practice, we compute the fraction of (weighted) samples with a posterior value lower than $P(0)$. The significance of the shift $m\sigma$ is computed by identifying $V$ with the probability mass of a one-dimensional Gaussian distribution outside of the interval $[-m\sigma, m\sigma]$. Thus, the tension in levels of sigma is given by 
\begin{equation}
        m=\sqrt{2}\;{\rm erf}^{-1}(1-V)\;.
\end{equation}
We note that this approach is mathematically equivalent to the Monte Carlo exact parameter shift method adopted in \citet{Raveri20} and \citet{KiDS+DES}.
\subsubsection{Tier 3: Posterior predictive distribution metric}
\label{sec:tier3_methods}
The third tier of consistency tests compares the observed data with predictions generated from the posterior distribution of model parameters. This is usually probed via the posterior predictive distribution (PPD), which describes the distribution of data realisations given the observed data and assuming a particular model. Given a set of observed data, ${\bm d}$, and a model, $\mathcal{M}$, the distribution of data realisations, $\hat{\bm{d}}$, is given by 
\begin{equation}
        {\rm P}(\hat{\bm{d}}|\bm{d},\mathcal{M}) = \int{\rm d} {\bm \theta}\,P(\hat{\bm{d}}|{\bm \theta},\mathcal{M})P({\bm \theta}|\bm{d},\mathcal{M})\;.
\end{equation} 
By testing whether the observed data is compatible with being drawn from the PPD, the data can be probed for internal inconsistencies. This is typically achieved by drawing data realisations from the PPD and evaluating a test statistic for both the synthetic and the observed data. This allows us to calculate $p$-values representing the probability of getting a higher test statistic for synthetic data realisations than for the observed data, which serves as a measure of consistency in the data \citep{Doux21}. 
                        
As an alternative to a test of the PPD, Köhlinger et al. (2019)
introduced the so-called translated posterior distribution (TPD) as a special case of the PPD, which can be obtained by translating posterior samples back into model predictions. Therefore, it describes the distribution of model predictions given the uncertainty of model parameters. Since the TPD can directly be generated as a byproduct of the sampling process, we adopt the TPD in our consistency test in data space and employ the $\chi^2$ values as test statistic for the internal consistency between subsets of the data. Prior to the consistency analysis, we conducted sensitivity tests on internally inconsistent mock data, which confirmed that our TPD-based consistency metric yields estimates of the significance of the internally inconsistency that is compatible with the estimates inferred with the tier 1 and tier 2 metrics (see Appendix \ref{ap:sensitivity}).

Considering the split cosmological model, we infer the TPD for each set of cosmological parameters, ${\bm \theta}_{\rm A}$ and ${\bm \theta}_{\rm B}$ consisting of theory predictions for the full data vector, ${\bm t}({\bm \theta}_{\rm A})$ and ${\bm t}({\bm \theta}_{\rm B})$. We then quantify to what extent the observed data in one subset ${\bm d}_{\rm A}$ is compatible with the TPD inferred from the other subset and vice versa. To do so, we draw a realisation of ${\bm d}_{\rm A}$ conditioned on the TPD of subset B for each sample, denoted by ${\bm d}^{\rm sim}_{\rm A}$. Since the conditional distribution of one set of variables conditioned on the other is Gaussian if both sets are jointly Gaussian, the simulated data points are given by a multivariate Gaussian distribution $\mathcal{N}(\mu^{\rm sim}_{\rm A},\mathbf{C}^{\rm sim}_{\rm A})$ with \citep[see e.g. ][]{Bishop}\ 
\begin{align}
\mu^{\rm sim}_{\rm A}&={\bm t}_{\rm A}({\bm \theta}_{\rm B})+\mathbf{C}_{\rm AB}\mathbf{C}^{-1}_{\rm BB}\left[{\bm d}_{\rm B}-{\bm t}_{\rm B}({\bm \theta}_{\rm B})\right],\\
\mathbf{C}^{\rm sim}_{\rm A} &= \mathbf{C}_{\rm AA}-\mathbf{C}_{\rm AB}\mathbf{C}^{-1}_{\rm BB}\mathbf{C}_{\rm BA}\;,
\end{align}
where $\mathbf{C}_{\rm AB}$ denotes the cross-covariance between the two data subsets. For each simulated data vector we then compute the $\chi^2$-statistic given by
\begin{equation}
        \chi^2\left[{\bm d},{\bm t}({\bm \theta})\right]=\left[{\bm d}-\bm{t}({\bm \theta})\right]^T\mathbf{C}^{-1}\left[\bm{d}-\bm{t}({\bm \theta})\right]\;.
\end{equation}
We quantified the consistency between data regions in terms of the $p$-value, $p(A|B)$, which is given by the fraction of posterior samples with 
\begin{equation}
\chi^2\left[{\bm d}^{\rm sim}_{\rm A},{\bm t}_{\rm A}({\bm \theta}_{\rm B})\right]>\chi^2\left[{\bm d}_{\rm A},{\bm t}_{\rm A}({\bm \theta}_{\rm B})\right].\;
\end{equation}

The $p$-value quantifies the probability of the data in data subset A being a realisation of the TPD of subset B. Thus, low $p$-values indicate an internal inconsistency of the data. We note that \cite{Doux21} show that this method of quantifying consistencies can result in $p$-values that are biased low if the two posteriors prefer vastly dissimilar regions in parameter space. This can be circumvented by calibrating the $p$-value with simulated data. In this work, we interpret the TPD-based $p$-value as a conservative metric for the internal consistency since it can exaggerate a potential tension in the dataset and reserve a further calibration of the $p$-value for cases for which it fails to pass our adopted threshold for internal consistency.
\section{Results of internal consistency tests}
\label{sec:results}
In this section, we present the results of our internal consistency analysis. We adopted the fiducial {\sc CosmoPipe} pipeline and sample the parameter space via the {\sc Nautilus}\footnote{\url{https://github.com/johannesulf/nautilus/tree/v0.7.2}} sampler \citep{Lange23}, interfaced with the cosmological parameter estimation code {\sc Cosmosis} \citep{Zuntz15}. A list of model parameters is provided in Table \ref{tab:parameters}. As is standard practice in stage-III cosmic shear analyses, we adopted a multivariate Gaussian likelihood. In the split cosmological analysis, we duplicated parameters with uniform priors, while parameters with informative Gaussian priors were not duplicated. An exception is the colour-based split of the catalogue, for which we calibrated separate redshift nuisance parameters and allowed for different intrinsic galaxy alignments between the subsets. We modelled the theoretical prediction for the observed signal in each subset with the two independent sets of cosmological parameters. When comparing theory and data, both subsets were linked through the data covariance matrix, which was computed analytically using the {\sc OneCovariance}\footnote{\url{https://github.com/rreischke/OneCovariance/tree/v1.0.0}} code \citep{Reischke23}. We note that our covariance model adopts the NLA model of intrinsic alignments, as opposed to the fiducial NLA-$M$ model. As was shown by \citet{Reischke23}, however, the contribution of intrinsic alignments has a negligible impact on the covariance of KiDS-Legacy data. In \citetalias{Wright25}, we considered $p$-values of $p>0.01$, corresponding to a $2.36\sigma$ offset, to be consistent with the null hypothesis for systematics tests. Therefore, we adopted the same threshold for the internal consistency tests conducted in this work. 

Before the unblinding of the KiDS-Legacy catalogue, we conducted the full consistency analysis on one blind for all three summary statistics. The blinding process, adopted from \citet{Kuijken15}, involves the generation of two additional catalogues with systematic differences in the measured galaxy shapes, which result in up to $\pm2\sigma$ shifts in the inferred $S_8$. As the consistency tests are not sensitive to the overall $S_8$ value, we did not expect the blinding to have an impact on the internal consistency of the KiDS-Legacy dataset. With the exception of the split by angular scales with 2PCFs reported in Sect. \ref{sec:angular}, we found no differences between consistency tests for the three statistics. We therefore limited our analysis to COSEBIs, except for the split of the data vector by scale, for which we employed all three summary statistics. When evaluating the consistency in parameter space, we focussed on $\Omega_{\rm m}$ and $S_8$, which are the two parameters that are mostly constrained by cosmic shear data. Prior to the cosmological analysis, we determined the number of constrained parameters $N_{\Theta}$ that is required for the suspiciousness test, as discussed in Sect. \ref{sec:tier1_methods}. The results of this analysis are presented in Appendix \ref{ap:neff}. In Appendix \ref{ap:sensitivity}, we present a sensitivity analysis with mock realisations of the fiducial data vector, showing that in the absence of internal tension each metric yields a level of consistency that is compatible with typical noise fluctuations in the data.
\subsection{Splits at the data vector level}
\label{sec:results_datavector}
In the case of data vector-level splits, we employed the fiducial KiDS-Legacy cosmic shear data vector, covariance matrix, and the prior on the shift in the mean of the redshift distribution per tomographic bin of \citetalias{Wright25} when conducting the likelihood analysis. 
\begin{figure*}
\includegraphics[width=\textwidth]{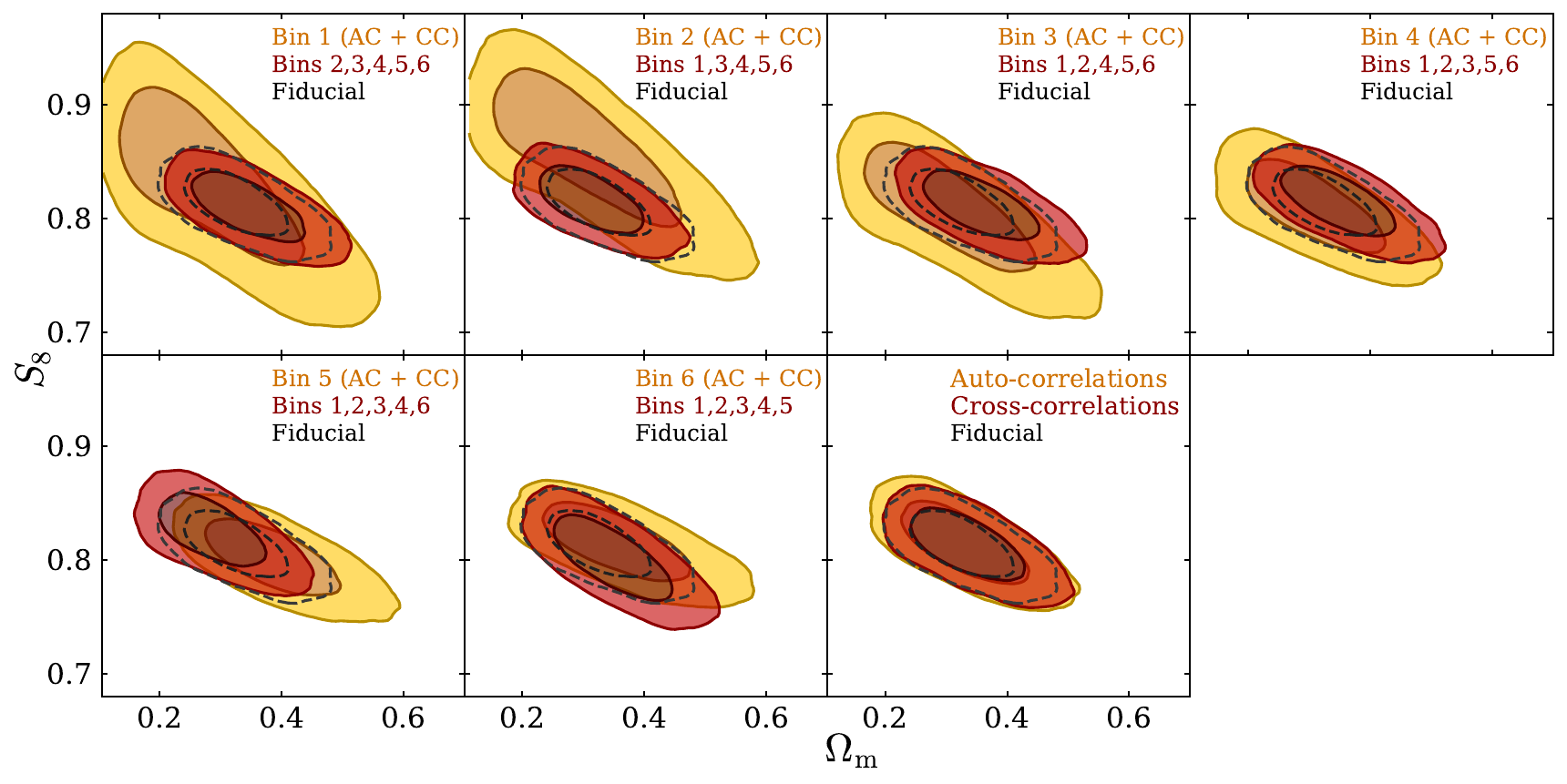}
\caption{Posterior distribution of the two instances of cosmological parameters in a split by redshift bin for COSEBIs. The yellow contours show the posterior of parameters modelling one specific redshift bin and its cross-correlation with the other bins, while the red contours show the posterior distribution of the parameters modelling the auto- and cross-correlation signal of the remaining redshift bins. The dashed contours show the fiducial constraints for reference. The final panel presents the posterior distribution in a split between auto-correlations of all redshift bins and their cross-correlations. When running the chains, both regimes are linked through the cross-covariance between redshift bins. The inner and outer contours of the marginalised posteriors correspond to the 68\% and 95\% credible intervals, respectively.}
\label{fig:datavectorlevel_cosebis}
\end{figure*}
\begin{table*}
\caption{Consistency metrics for data vector level splits of KiDS-Legacy data.}
\label{tab:datavectorlevel}
\centering
\begin{tabular}{llcc|ccc|cc}
\hline\hline
&&\multicolumn{2}{c}{Tier 1}&\multicolumn{3}{c}{Tier 2}&\multicolumn{2}{c}{Tier 3}\\
Statistic &Split&  $\log_{10}R$ & $N_{\sigma,S}$ & $N_{\sigma}(\Delta\Omega_{\rm m})$ & $N_{\sigma}(\Delta S_{\rm 8})$ & $N_{\sigma}(\Delta(\Omega_{\rm m}, S_{\rm 8}))$ & $p(A|B)$ & $ p(B|A)$\\
\hline
\multirow{8}{*}{COSEBIs}
& Bin 1    & 0.91 &  0.50 & 0.81 & 0.61 & 0.35 & 0.40 & 0.49\\
& Bin 2    & 0.31 &  1.16 & 0.34 & 1.09 & 1.34 & 0.14 & 0.74\\
& Bin 3    & 1.07 &  0.65 & 0.66 & 0.02 & 0.61 & 0.72 & 0.39\\
& Bin 4    & 0.89 &  1.39 & 0.65 & 0.22 & 0.72 & 0.85 & 0.23\\
& Bin 5    & 1.38 &  0.87 & 1.39 & 0.99 & 0.84 & 0.67 & 0.26\\
& Bin 6    & 1.46 &  0.47 & 0.09 & 0.51 & 0.50 & 0.43 & 0.49\\
& Redshift bin AC vs CC & 1.88 & 0.18 & 0.07 & 0.06 & 0.00 & 0.23 & 0.48\\
& Scale  & 1.48 &  0.63 & 0.07 & 0.69 & 0.62 & 0.37 & 0.38\\
\hline
Band powers & Scale & 0.40 & 1.49 & 0.16 & 1.04 & 0.84 & 0.51 & 0.88\\
2PCFs & Scale  & -2.03 & 3.69 & 0.30 & 1.81 & 1.63 & 0.34 & 0.20\\
\hline
\end{tabular}
\tablefoot{The first column indicates the summary statistic and the second column lists the split applied to the data vector. The third and fourth columns report the tier 1 evidence-based metric of the Bayes ratio and the tension level inferred from the suspiciousness, respectively. The next three columns present the results of the tier 2 multi-dimensional parameter metric test for $\Omega_{\rm m}$, $S_8$, and their combination. The final two columns shows the tension level arising from the tier 3 PPD metric test in terms of the $p$-value for data vector predictions for the data subset listed in the second column inferred from the PPD of the other subset (column 8) and vice versa (column 9).}
\end{table*} 
\subsubsection{Redshift bin split}      
\label{sec:zbin}
The first split at the data vector level is designed to test the internal consistency between the six tomographic redshift bins. In this way, we can probe the data for any errors in our redshift calibration and redshift-dependent modelling effects, such as the impact of baryon feedback or the effect of IA of galaxies, which has a larger relative contribution to the total signal compared to the lensing signal at low redshifts. For each bin, we divided the theory vector into one subset containing the autocorrelation of the specific bin and its cross-correlation with the remaining redshift bins. The second subset consisted of all auto-correlations of the remaining redshift bins and their cross-correlation. This split is analogous to the consistency test between KiDS-1000 redshift bins presented in \citet{Asgari21}. A complementary method of testing the consistency between redshift bins is the entire removal of single tomographic bins from the analysis. This test is demonstrated in \citetalias{Wright25} and is commonly applied in weak lensing studies \citep[see for example][]{Amon22,LiHSC23}.

The first six panels of Fig. \ref{fig:datavectorlevel_cosebis} show the marginalised posterior distribution of the split cosmological analysis for the split by redshift bin. Each panel represents the constraints from a single redshift bin and its cross-correlation with the other bins in yellow and the constraints from the auto- and cross-correlations between the remaining bins in red. For reference, we visualize the constraints from the fiducial cosmic shear analysis with the black dashed line. A visual inspection of the contours indicates a good agreement between the split and fiducial analyses. As expected, low redshift bins only yield loose constraints on cosmological parameters, which however are in good agreement with the remaining tomographic bins. The largest shift between contours is observed in the split of the second bin. We note that the previous consistency analysis of KiDS-1000 data \citep[see Appendix B in][]{Asgari21} also showed the largest discrepancy in the second tomographic bin. However, we emphasise that along with the inclusion of additional survey area and calibration data and the re-reduction of previously released data in KiDS-DR5, the definition of tomographic bins changed from equidistant binning in photometric redshift in KiDS-1000 to equi-populated binning in KiDS-Legacy, which \citet{Sipp21} recommend as a better choice for the reduction of statistical errors. Therefore, any direct comparison between the consistency analysis of tomographic bins in KiDS-1000 and KiDS-Legacy should be made with caution.

The consistency levels for the redshift bin splits are listed in Table \ref{tab:datavectorlevel}. The first two columns show the results of the tier 1 test with evidence-based metrics. The Bayes' ratio indicates preference for the single cosmological model, ranging from 'barely worth mentioning' in bin 2 to 'strong' in bin 6 according to Jeffreys' scale. In terms of the suspiciousness, all bins are found to be in agreement with $N_{\sigma,S}<1$ except for bins 2 and 4, which show slightly larger values at $1.16\sigma$ and $1.39\sigma$, respectively. The good consistency between redshift bins is further confirmed by the tier 2 multi-parameter metric test. All redshift bins show agreement with $N_{\sigma}\leq1.39$, which is found when considering a shift in $\Omega_{\rm m}$ in the fifth bin. The final two columns list the consistency level from the tier 3 PPD test in terms of the $p$-value for data vector predictions for the data subset listed in the second column, inferred from the TPD of the other subset (column 8) and vice versa (column 9). Here, all $p$-values pass our threshold of $p>0.01$. Overall, we highlight that all consistency metrics indicate a good internal consistency for the split by redshift bin. In particular, while the previous KiDS-1000 analysis showed an internal inconsistency between redshift bins of up to $3\sigma$, we find the KiDS-Legacy data to be in better internal agreement with consistency levels better than $1.39\sigma$ and compatible with typical statistical fluctuations.
\subsubsection{Auto-correlation versus cross-correlation}
\label{sec:accc}
In addition to the redshift bin split, we applied a split of the data vector between auto- and cross-correlation signals of the tomographic bins. This split allowed us to probe the data for systematic effects that affect the two types of correlation signals through different processes. In particular, the individual IA contributions to the cosmic shear signal, given in Eq. \eqref{eq:IA}, can be attributed to either the auto- or the cross-correlation signal. The II term is generated through the alignment of physically close galaxies due to tidal forces of the nearby large-scale structure. Therefore, it predominantly affects the autocorrelation of tomographic bins. The GI term on the other hand is induced by the large-scale structure causing both the intrinsic alignment of nearby galaxies and the lensing of distant galaxies, which leads to an anti-correlation between the shapes of galaxy pairs that are separated in redshift. Therefore, this effect manifests itself primarily in the cross-correlation signal between tomographic bins. We divided the theory vector into one subset consisting of the auto-correlation signal of all six tomographic bins and the second subset containing all cross-correlations between bins.

The bottom right panel of Fig. \ref{fig:datavectorlevel_cosebis} shows the posterior from the consistency test between the autocorrelation group of all tomographic bins in yellow and the cross-correlation group in red. The constraints from the fiducial cosmic shear analysis are visualised with the black dashed line. The consistency metrics, listed in Table \ref{tab:datavectorlevel} signify an agreement between both groups in all tests. This finding is in agreement with the almost indistinguishable posteriors shown in the last panel of Fig. \ref{fig:datavectorlevel_cosebis}. At first glance, it may be surprising that the auto-correlation and cross-correlation parts of the data vector have the same constraining power as each other, and the fiducial data vector. Typically the GI contribution to the cross-correlation signal provides the main information to constrain the IA amplitude, and as such a cosmic shear analysis that excludes the cross-terms decreases the overall constraining power. The fact that we have found the same constraining power can be understood by recognising that the NLA-$M$ intrinsic alignment model parameters in our analysis are prior dominated and also shared between the two data splits with the cross-correlation data informing the auto-correlation IA model. If this was not the case then we would expect to see a degradation in constraining power when analysing the auto- and cross-correlations separately.
\subsubsection{Scales}
\label{sec:angular}
\begin{figure*}
\includegraphics[width=\textwidth]{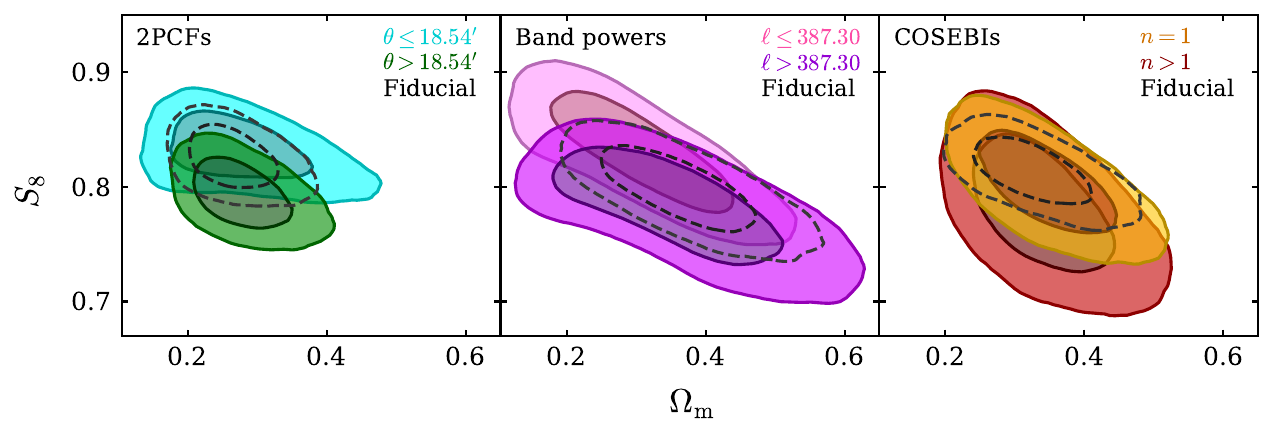}
\caption{Posterior distribution of the two instances of cosmological parameters in a split by scale for 2PCFs (left), band powers (middle), and COSEBIs (right) in comparison to the fiducial analysis with each summary statistic, illustrated by the black dashed lines. The inner and outer contours of the marginalised posteriors correspond to the 68\% and 95\% credible intervals, respectively.}
\label{fig:datavectorlevel_angular}
\end{figure*}
When selecting which scales to include in a cosmic shear analysis there is a trade-off between the desire to maximise signal-to-noise by including as much data as possible, and minimising the impact of unaccounted systematic errors that are expected to contaminate the smallest angular scales. Our fiducial cosmological constraints analyse data collected over the angular range $\theta \in [2\arcmin,300\arcmin]$, and in this section we assess the scale consistency by separating small and large scale data using all three statistics, 2PCFs, band power spectra and COSEBIs.

For the 2PCFs, we split the nine logarithmically spaced angular bins at an angle of $\theta\approx18.54\arcmin$, with the first and second set consisting of four and five angular bins, respectively, ensuring comparable signal-to-noise in both sub-samples\footnote{Following previous KiDS analyses, we limit the $\xi_{-}$ correlation function to $\theta>4\arcmin$, removing the first $\theta$-bin from the analysis.}. For this comparison, we found tension at $3.69\sigma$ between small and large scales when considering the suspiciousness test. This can be observed in the left panel of Fig. \ref{fig:datavectorlevel_angular}, which shows a preference for lower values of $S_8$ at large scales and a preference for higher values at small scales, with a $1.81\sigma$ shift in $S_8$ between the two. The PPD-based test, however, concludes that the data vectors are consistent. We note that \citetalias{Wright25} do not include 2PCFs in the fiducial cosmic shear analysis, only providing 2PCF constraints for completeness and comparison with previous works. This is because the 2PCF is particularly sensitive to the effect of baryon feedback which is challenging to model \citep[see for example][]{Asgari20}. As we have not optimised the angular scales for a 2PCF KiDS-Legacy analysis \citep[see for example][]{Krause21} we conclude that the tension reported by the suspicious tier 1 test is likely to be caused by an imperfect modelling of baryonic effects for the fiducial $\theta\in[2\arcmin,300\arcmin]$ angular scale range. As such we do not expect the cosmological constraints from 2PCFs to be as reliable as those inferred with our fiducial COSEBIs statistic, which restrict the range of physical scales entering the analysis (see for example fig. 1 in \citetalias{Wright25}), making them less susceptible to scale-dependent effects. We further note that on large angular scales, the assumption of a Gaussian likelihood does not hold \citep[see for example][]{Sellentin18,Louca20,Joachimi21,Oehl24}, which particularly affects the $\xi_+$ correlation function, leading to a potential bias in the inferred $S_8$.

With our band power spectra we can test the consistency between small and large multipoles in Fourier space (see fig. 1 in \citetalias{Wright25} for the band power filter functions that are compact in $\ell$). We divide the eight logarithmically spaced bands between $\ell \in \left[100,1500\right]$ at a limit of $\ell\approx387$, creating two subsets consisting of four bands each. The middle panel of Fig. \ref{fig:datavectorlevel_angular} shows the cosmological constraints from this split analysis with the second to last row of Table \ref{tab:datavectorlevel} reporting the suite of consistency metrics. We found an agreement between the low and high $\ell$ band power measurements for all our tests.

It is hard to define a data split for a scale-sensitivity analysis with COSEBIs as each $E_n$ mode is sensitive to a range of $\ell$-scales, varying only in the way those scales are combined (see for example fig. 1 in \citetalias{Wright25}). We therefore chose to conduct a data-split analysis of the COSEBI $n=1$ mode, which carries the majority of the constraining power, versus the other $n=2$ to $n=6$ modes. The result is visualised in the right panel of Fig. \ref{fig:datavectorlevel_angular} with the consistency metrics listed in Table \ref{tab:datavectorlevel}, demonstrating an agreement between the COSEBIs modes in all tests.
\subsection{Splits at the catalogue level}
\label{sec:results_catalogue}
\begin{figure*}[h!]
\centering
\includegraphics[width=\textwidth]{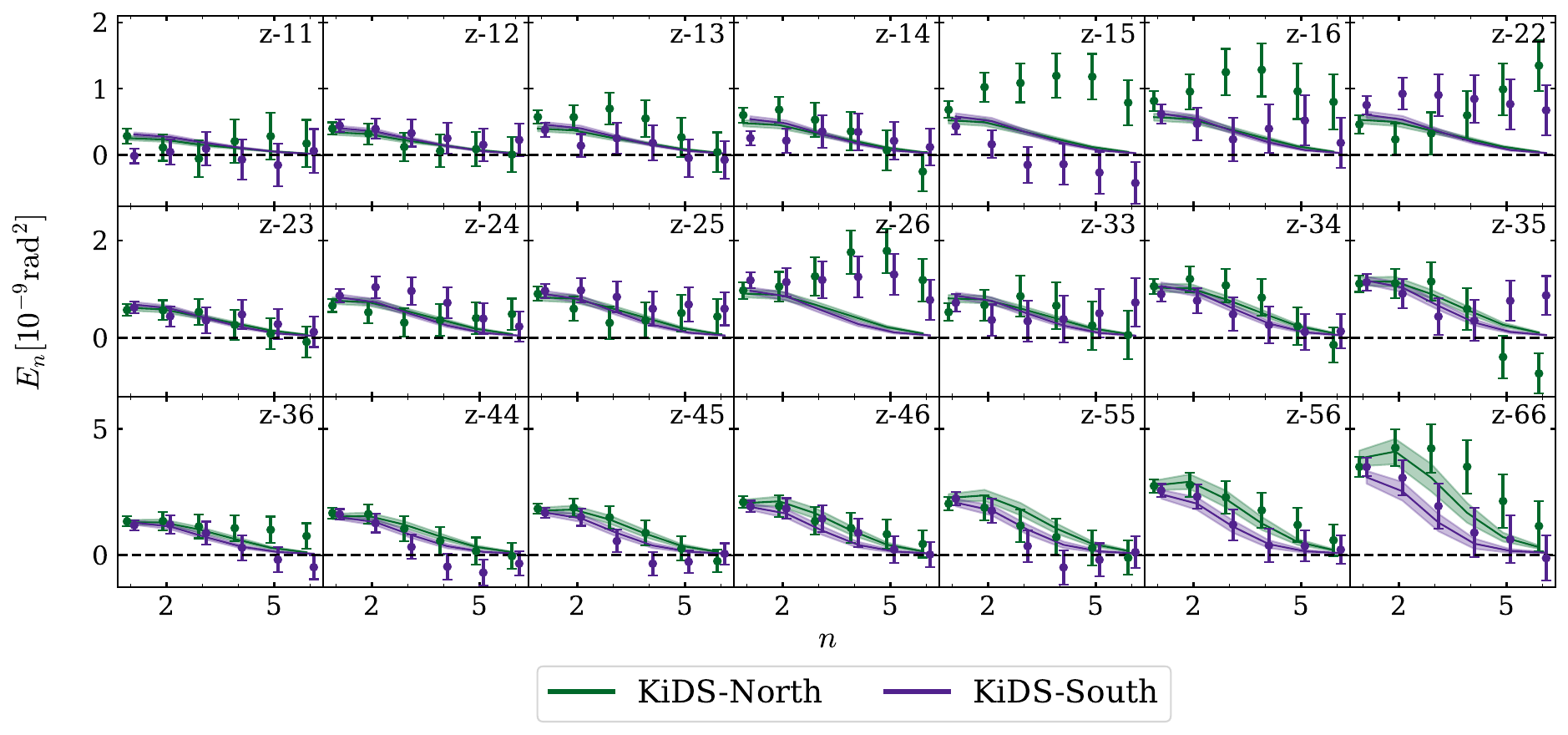}
\caption{COSEBIs E-mode measurements and their best-fitting model for a split cosmological analysis of the North-South split catalogue. The green and purple data points show the measurements of the KiDS-North and KiDS-South sample, respectively. The best-fitting theoretical predictions are given by the solid lines, and the $1\sigma$ interval of the TPDs are illustrated by the shaded regions. Each panel represents auto- or cross-correlation between tomographic bins. For visualisation purposes, we display the discrete $n$ modes with an offset on the x-axis.We note that the E-mode signals are highly correlated within a tomographic bin and advise against a so-called \enquote*{$\chi$-by-eye}.}
\label{fig:datavector_northsouth}
\end{figure*}
In this section, we split the data on the catalogue level, analysing the two distinct sections of the KiDS footprint, KiDS-North, and KiDS-South, and splitting the galaxy sample by colour. When conducting the likelihood analysis, we construct a joint data vector and covariance matrix of both subsets, doubling the dimensionality of the data vector with respect to the fiducial analysis. In Appendix \ref{ap:catlevel}, we report the data properties of each catalogue split with calibrated redshifts and shear measurements for each sample along with the B-mode signal of each subset, which we found to be consistent with the null hypothesis.
\subsubsection{North-South split}
\label{sec:hemisphere}
\begin{table*}
\caption{Consistency metrics for catalogue-level splits of KiDS-Legacy data.}
\label{tab:catlevel}
\centering
\begin{tabular}{lcc|ccc|cc}
\hline\hline
&\multicolumn{2}{c}{Tier 1}&\multicolumn{3}{c}{Tier 2}&\multicolumn{2}{c}{Tier 3}\\
Split&  $\log_{10}R$ & $N_{\sigma,S}$ & $N_{\sigma}(\Delta\Omega_{\rm m})$ & $N_{\sigma}(\Delta S_{\rm 8})$ & $N_{\sigma}(\Delta(\Omega_{\rm m}, S_{\rm 8}))$ & $p(A|B)$ & $ p(B|A)$\\
\hline
North vs South              & 0.83 &  0.71 & 1.09 & 0.36 & 0.60 & 0.36 & 0.39\\
Red vs  blue, $T_{\rm B}=3.0$ & -0.44 &  2.61 & 1.08 & 0.08 & 0.74 & 0.76 & 0.40\\
Red vs blue ,$T_{\rm B}=1.9$ & -1.29 &  2.84 & 0.70 & 0.46 & 0.25 & 0.64 & 0.35\\
\hline
\end{tabular}
\tablefoot{The second and third columns report the Bayes ratio and the tension level inferred from the suspiciousness, respectively. The next three columns present the results of the tier 2 test for $\Omega_{\rm m}$, $S_8$, and their combination. The final two columns shows the tension level arising from the tier 3 test in terms of the $p$-value for data vector predictions for the first subset of the catalogue inferred from the TPD of the other subset (column 7) and vice versa (column 8).}
\end{table*}
As discussed in Sect. \ref{sec:data}, the KiDS observations were taken on two distinct patches on sky: one at the celestial equator, dubbed KiDS-North, and one at the South Galactic Pole, dubbed KiDS-South. The two patches, which are shown in fig. 2 in \citetalias{Wright25}, cover a similar area on sky with approximately $496\,{\rm deg}^2$ of post-masking data in KiDS-North and $472\,{\rm deg}^2$ in KiDS-South. As a result, both patches contain a comparable number of sources per tomographic bin. In the fiducial analysis pipeline, we combined independent measurements of the cosmic shear 2PCF per patch into a single measurement, from which we computed the COSEBIs data vector. In the North-South split, we kept the 2PCF measurements in each patch separate in order to test their consistency. By doing so, we obtained one data vector per hemisphere. Since the two patches are separated on sky, we did not expect a cross-correlation signal between patches. Therefore, the combined covariance matrix only consists of two non-zero blocks, each containing the covariance for North and South, respectively.  

\begin{figure*}
\centering
\includegraphics[width=\textwidth]{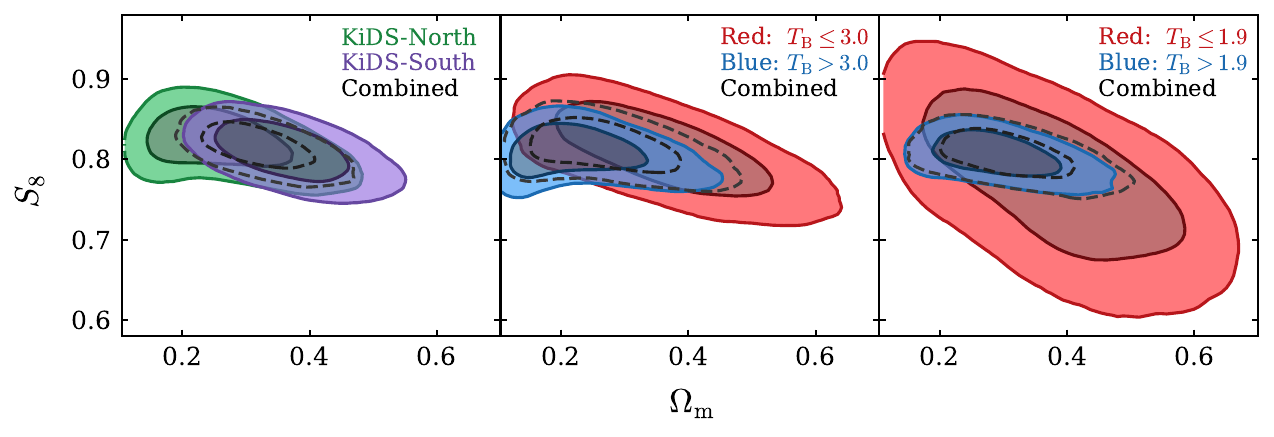}
\caption{Posterior distribution of parameter duplicates in the $\Omega_{\rm m}-S_8$ plane for catalogue-level splits for COSEBIs. Left panel: North-South split. Middle panel: Red-blue split defined via a cut on the spectral type of $T_{\rm B}=3.0$. Right panel: Red-blue split defined via a cut on the spectral type of $T_{\rm B}=1.9$. For reference, the black dashed contours show constraints from the analysis with a single set of parameters modelling both data subsets. The inner and outer contours of the marginalised posteriors correspond to the 68\% and 95\% credible intervals, respectively.}
\label{fig:catlevel_cosebis}
\end{figure*}

In Fig. \ref{fig:datavector_northsouth}, we present the COSEBIs data vector per hemisphere along with the corresponding theory prediction from the best-fitting model with uncertainties inferred from the TPDs. As the data properties of the KiDS-North and KiDS-South patches are very similar, in terms of the redshift distributions, multiplicative shear and redshift calibration, and the ellipticity dispersion (see Appendix \ref{ap:catlevel}), we have chosen to use a single shared set of observational nuisance parameters in our likelihood analysis, shown in the left panel of Fig. \ref{fig:catlevel_cosebis}. Although the KiDS-North patch tends to favour a shift towards lower values of $\Omega_{\rm m}$ compared to our fiducial analysis, both patches are in good agreement with a value of $N_{\sigma}(\Delta\Omega_{\rm m})=1.09$ inferred in the tier 2 multi-parameter consistency test. Considering $S_8$, we found both patches to be in agreement, which is confirmed by the tier 1 evidence and tier 3 PPD tests tabulated in the first row in Table \ref{tab:catlevel}. We conclude that the cosmological constraints from observations in KiDS-North and KiDS-South are fully consistent.
\subsubsection{Red-blue split}
\label{sec:redblue}
Observational evidence shows that red early-type galaxies intrinsically align, in contrast to blue late-type galaxies, where intrinsic alignments have yet to be detected \citep{Hirata07,Joachimi11,Heymans13,Samuroff19,Georgiou19,Johnston19,Fortuna20,Tugendhat20,Samuroff24,Georgiou25}. We therefore chose to split the KiDS-Legacy galaxies into a sample of red and blue galaxies to study the impact of intrinsic galaxy alignments on our cosmic shear signal. This also allows us to explore a secondary effect where the more spherical shape of red galaxies changes the populations’ ellipticity distribution, leading to possible differences in the shear calibration correction \citep{Kannawadi19}.
\begin{figure*}
\centering
\includegraphics[width=\textwidth]{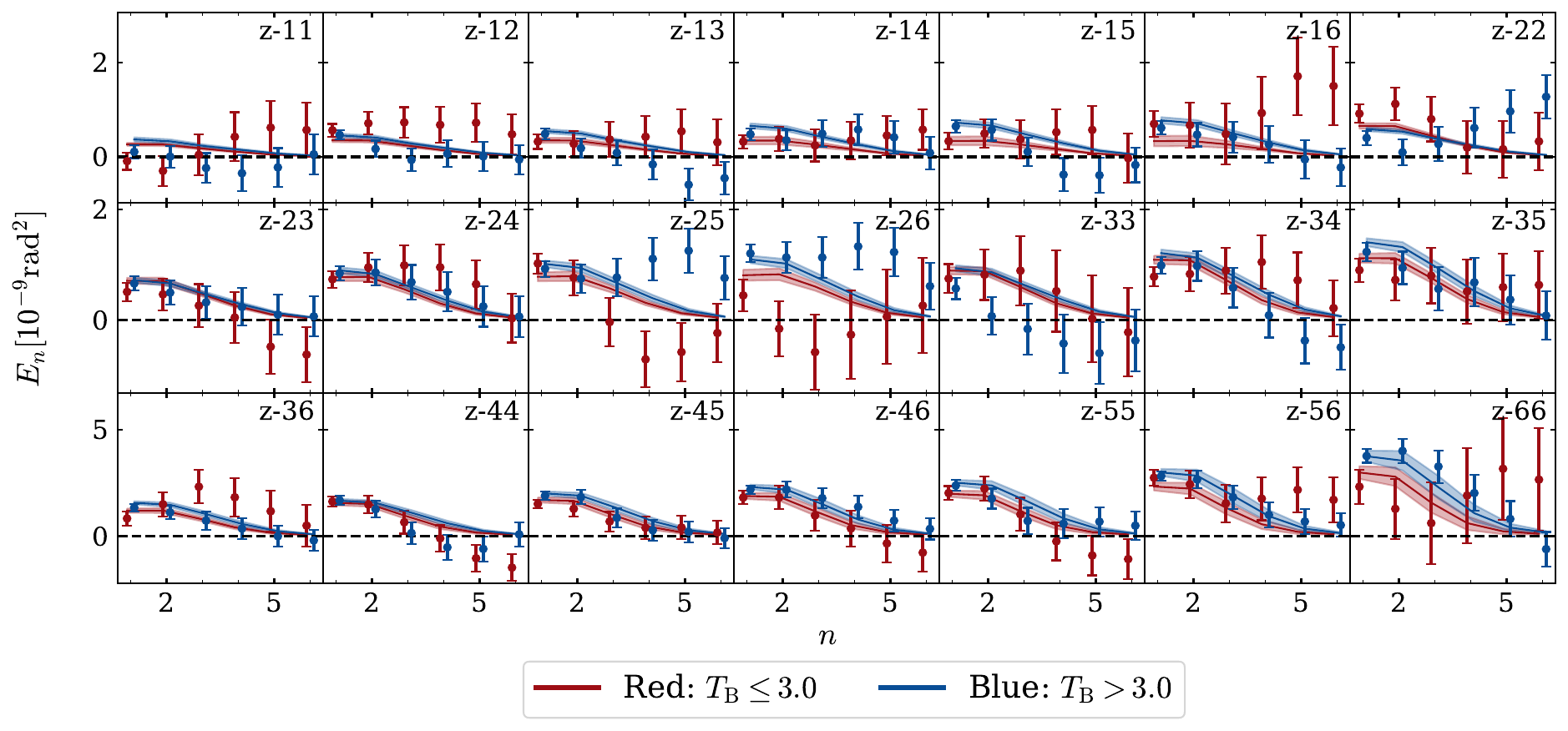}
\caption{COSEBIs E-mode measurements and their best-fitting model for a split cosmological analysis of the red-blue split catalogue, defined via a cut on the spectral type $T_{\rm B}=3.0$. The red and blue data points show the measurements of the red and blue sample, respectively. The best-fitting theoretical predictions are given by the solid lines, and the $1\sigma$ interval of the TPDs are illustrated by the shaded regions. Each panel represents auto- or cross-correlation between tomographic bins, as indicated by the label in the top right corner. For visualisation purposes, we display the discrete $n$ modes with an offset on the x-axis. We note that the E-mode signals are highly correlated within a tomographic bin and advise against a so-called \enquote*{$\chi$-by-eye}.}
\label{fig:datavector_redblue_3p0}
\end{figure*}
\begin{figure*}
\sidecaption
\includegraphics[width=12cm]{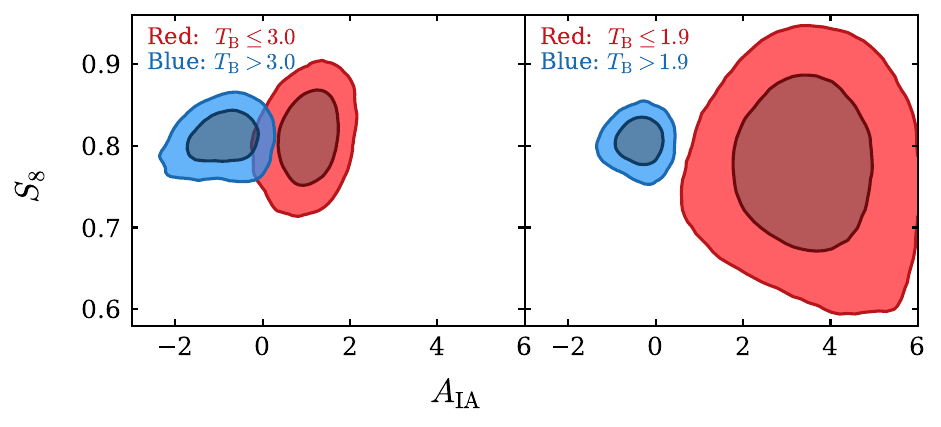}
\caption{Constraints on $S_8$ and $A_{\rm IA}$ for colour-based splits of the catalogue. Left panel: Red-blue split defined via a cut on the spectral type of $T_{\rm B}=3.0$. Right panel: Red-blue split defined via a cut on the spectral type of $T_{\rm B}=1.9$. For this threshold, the catalogue only contains very few red galaxies with $z_{\rm B}>1.14$. Therefore, the red galaxy sample only encompasses the first five tomographic bins. The inner and outer contours of the marginalised posteriors correspond to the 68\% and 95\% credible intervals, respectively.}
\label{fig:redblue_S8-AIA}
\end{figure*}

We followed the methodology of \citet{Li21}, dividing the KiDS-Legacy sample into two subsets based on the spectral type $T_{\rm B}$ reported by the {\sc bpz} code. \citet{Li21} defined blue galaxies via a threshold of $T_{\rm B}>3.0$ to provide a similar constraining power per data subset. This is in contrast to appendix B in \citetalias{Wright25}, where red galaxies are selected with a threshold of $T_{\rm B}\leq1.9$. We present an analysis of each threshold\footnote{We note another alternative, that we do not explore here, is to split a catalogue by colour as proposed by \citet{McCullough24}, applying a SOM-based selection on $r-z$ colour in order to derive a high-purity sample of blue galaxies.}, where dividing the sample at a threshold of $T_{\rm B}=3.0$ results in the red sample containing approximately one third of the galaxies, with the more stringent cut of $T_{\rm B}=1.9$ leaving $\sim 16\%$ of galaxies in the red sample. In Appendix \ref{ap:catlevel}, we present the redshift distributions of each sample and the separate redshift nuisance parameters that are marginalised over in our analysis. In contrast to the division of galaxies by hemisphere, the red and blue galaxy samples are expected to be highly correlated and these correlations are taken into account via the cross-covariance between cosmic shear measurements of the red and blue samples computed with the {\sc OneCovariance} code.

Our fiducial cosmic shear analysis adopts the NLA-$M$ IA model, which sets any blue galaxy alignment to zero for $T_{\rm B}>1.9$. Thus, this model is not applicable when considering a split between red and blue galaxies at a threshold of $T_{\rm B}=3.0$, which requires a recalibration of the IA mass scaling. Since this is beyond the scope of this work, we revert to the standard NLA model in this analysis for both $T_{\rm B}$ thresholds with a separate amplitude parameter, $A_{\rm IA}$, for each galaxy sample. We note that as discussed in \citetalias{Wright25}, the cosmological constraints are highly consistent between different IA models. Therefore, we do not expect the choice of IA model to make an impact on the internal consistency test in cosmological parameter space.

The measured COSEBIs E modes for the colour-based split with a threshold $T_{\rm B}=3.0$ and the TPDs from the best-fitting theory model are illustrated in Fig. \ref{fig:datavector_redblue_3p0}. Here, each colour represents measurements of the auto- and cross-correlation between tomographic bins of the given sample. We chose not to include cross-correlation measurements between the red and blue bins as these signals mix contributions from theoretical predictions for the red and blue signals that cannot be easily modelled with our current pipeline. The cross-correlation between the red and blue data points is, however, taken into account in our consistency analysis via the joint covariance matrix. 

The posterior distribution for both instances of $\Omega_{\rm m}$ and $S_8$ in the split cosmological analysis are shown in the middle and right panels of Fig. \ref{fig:catlevel_cosebis}. As expected, the red galaxy sample yields weaker constraints on $S_8$ and $\Omega_{\rm m}$ than the blue sample due to the higher number density of blue galaxies. This is particularly true for the $T_{\rm B}=1.9$ selected sample where the marginalised posteriors are almost unconstrained by the red galaxy sample. Nevertheless, the cosmological parameter posterior distributions for both red-blue splits are in good agreement. Looking at the consistency metrics listed in Table \ref{tab:catlevel} for the $T_{\rm B}=3.0$ ($T_{\rm B}=1.9$) selected samples, we found that the suspiciousness test shows a preference for the split cosmological model at $2.61\sigma$ ($2.84\sigma$). In contrast, the tier 2 test yields an agreement between parameters with $N_\sigma\leq1.08$ ($N_\sigma\leq 0.70$) when considering $S_8$, $\Omega_{\rm m}$, and their combination. This is confirmed by the tier 3 tests, which suggests a good agreement between the TPDs and the observed data in both subsets. 

The tier 1 evidence-based preference for the split cosmological model can be explained by the additional freedom in the modelling of intrinsic galaxy alignments. While the combined galaxy sample analysis assumes a shared IA amplitude for both red and blue galaxies, the split analysis models intrinsic alignments with two independent parameters. As shown in Fig. \ref{fig:redblue_S8-AIA}, there is a significant difference between the IA amplitudes for the two samples with a marginal mode and the 1D 68\% highest posterior density interval of $A^{\rm blue}_{\rm IA} = -0.75^{+0.43}_{-0.62}$ and $A^{\rm red}_{\rm IA} = 1.08^{+0.46}_{-0.44}$ for the $T_{\rm B}=3.0$ sample split. For the $T_{\rm B}=1.9$ split, we found the IA amplitude of blue galaxies to be $A^{\rm blue}_{\rm IA} = -0.32^{+0.33}_{-0.36}$, while the red galaxy sample yields $A^{\rm red}_{\rm IA} = 3.32^{+1.13}_{-0.97}$. This corresponds to a difference in their posterior distribution at $N_\sigma(\Delta A_{\rm IA})=2.81$ for $T_{\rm B}=3.0$ and $N_\sigma(\Delta A_{\rm IA})=2.57$ for $T_{\rm B}=1.9$. These results are compatible with the central assumption of the NLA-$M$ model used in our fiducial cosmic shear analysis, which assumes zero alignment of blue galaxies. We therefore do not interpret the tier 1 result as an indication of internal inconsistency given that the physical mechanism behind red and blue galaxy alignment is expected to differ. As the evidence-based tier 1 consistency test compresses the full posterior into a single statistic, the physical difference between red and blue galaxies is reflected as a preference for the split cosmological model. Based on the other consistency metrics reported in Table \ref{tab:catlevel}, we can therefore conclude there is internal consistency between our samples of red and blue galaxies.
\subsection{Consistency between summary statistics}
\label{sec:statlevel}
\begin{figure}
\centering
\includegraphics[width=0.49\textwidth]{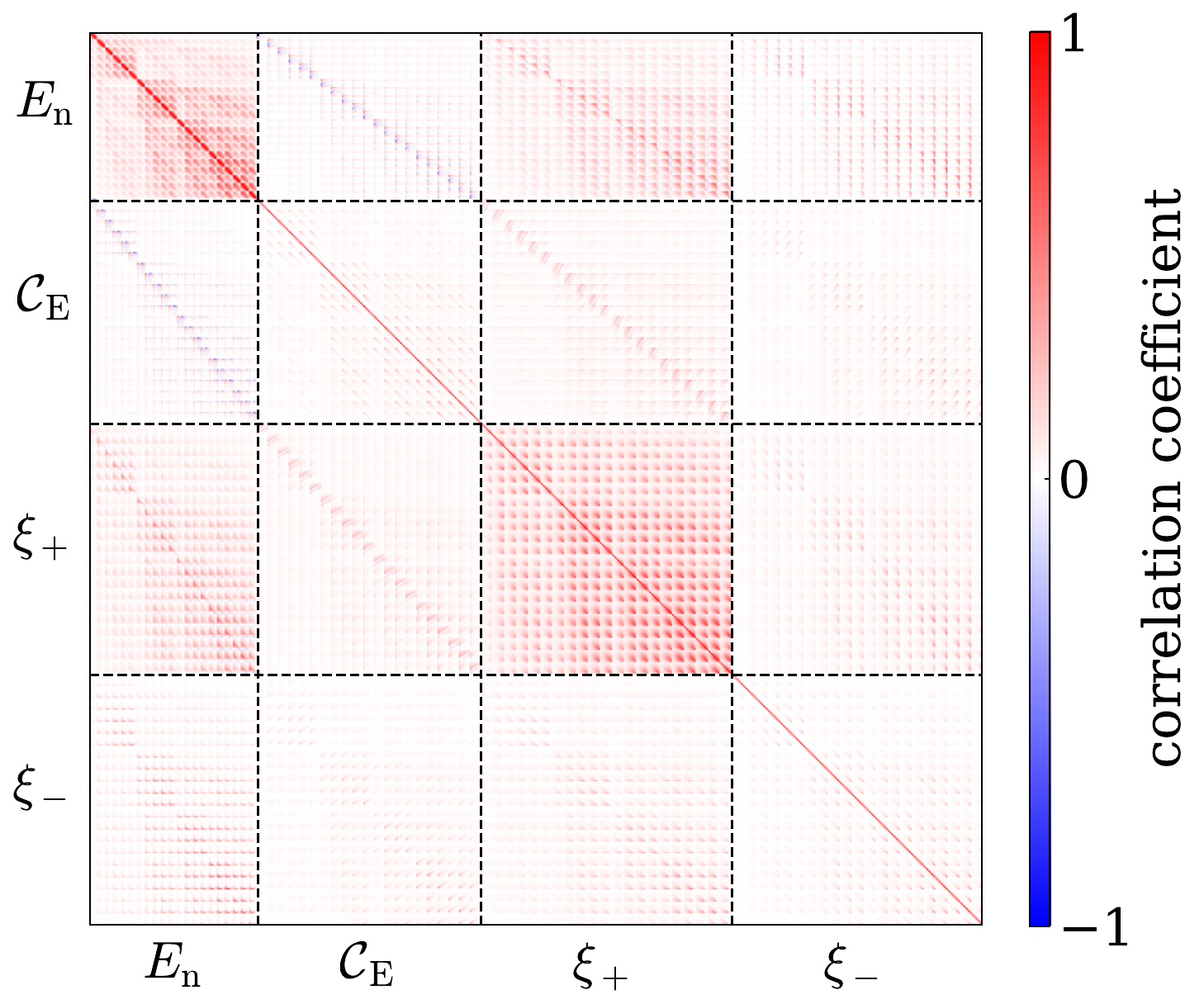}
\caption{Correlation matrix between measurements of COSEBIs, band powers, and 2PCFs.}
\label{fig:covariance}
\end{figure}
\begin{figure*}
\centering
 \includegraphics[width=\textwidth]{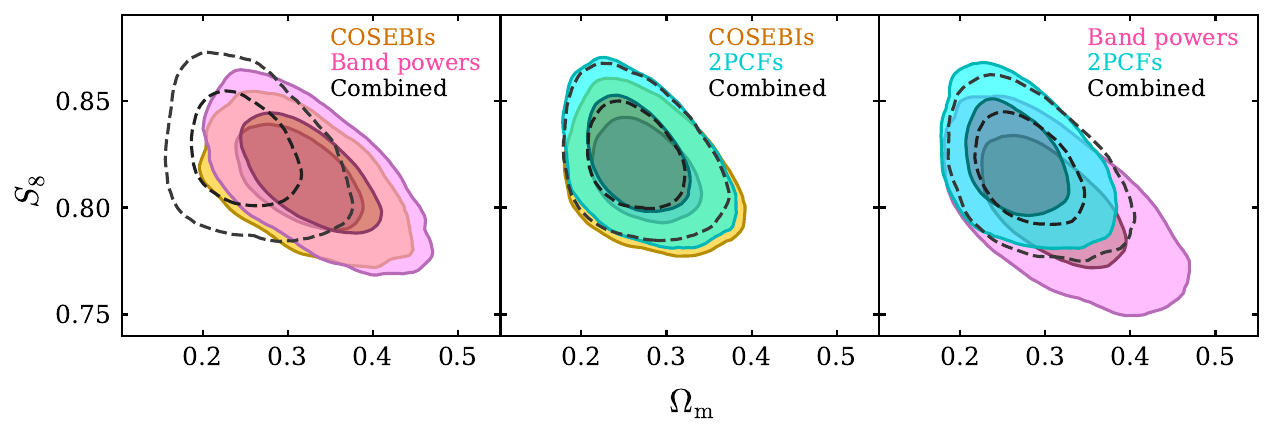}
\caption{Posterior distribution of parameter duplicates in the $\Omega_{\rm m}-S_8$ plane for split cosmological analyses with two summary statistics. For reference, the black dashed contours show constraints from the analysis with a single set of parameters modelling both data subsets. The inner and outer contours of the marginalised posteriors correspond to the 68\% and 95\% credible intervals, respectively.}
\label{fig:statlevel_cosebis}
\end{figure*}
While the two primary summary statistics in the fiducial analysis, COSEBIs and band powers, as well as the additional statistic, binned 2PCFs, are derived from the same cosmic shear 2PCF measurements, they differ in their sensitivity to spatial scales as well as systematic and modelling effects. Thus, we quantified the consistency between all combinations of two summary statistics. Although this does not correspond to a split of the data vector or the catalogue since each statistic originates from the same measurements of the cosmic shear 2PCFs, we can nevertheless use the same methodology to quantify whether or not the statistics prefer different cosmologies. In practice, we construct three data vectors that each combine measurements of two summary statistics and model the theoretical prediction with one set of parameters per statistic. As depicted in Eq. \eqref{eq:summarystatistic}, they differ only in the corresponding weight function, which vary in their sensitivity to different angular scales. Therefore, we expect the statistics to be highly correlated, since the different summary statistics are derived from the same two-point correlation function measurements. Thus, it is particularly important to derive a robust estimate of the covariance between summary statistics, which is enabled by the {\sc OneCovariance} code. The resulting correlation matrix is displayed in Fig. \ref{fig:covariance}. As discussed in \citet{Reischke23}, the full covariance matrix between all summary statistics can be non-positive definite due to numerical noise. However, the sub-covariance matrices between two summary statistics are still positive definite and invertible. 

\begin{figure}
\includegraphics[width=0.5\textwidth]{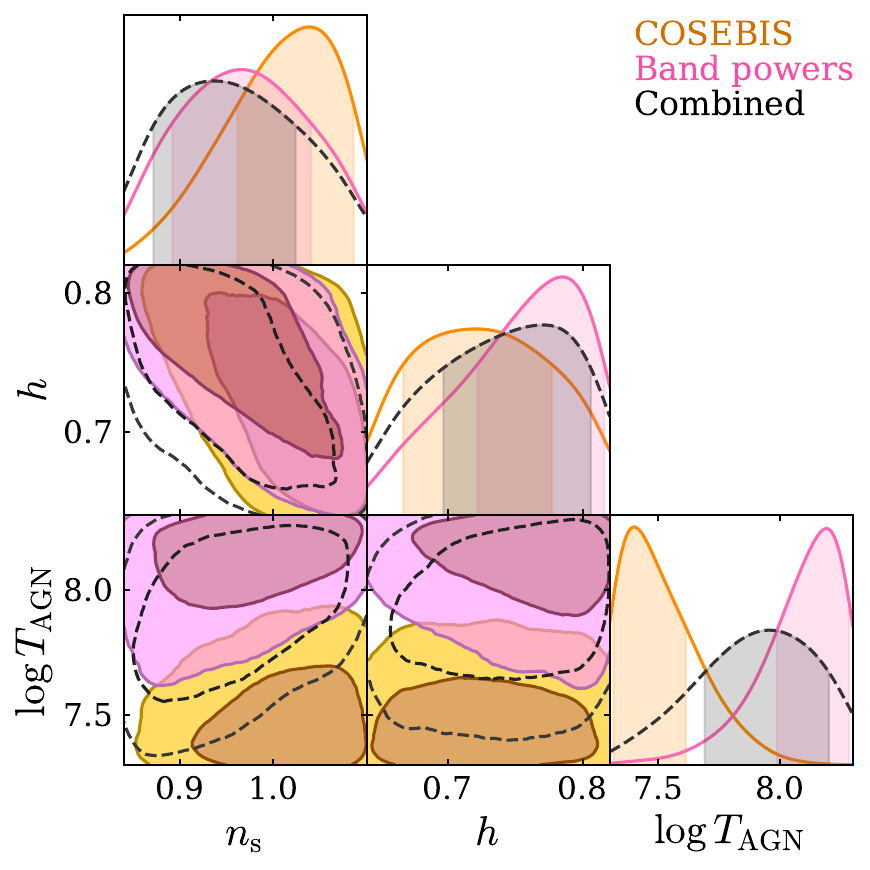}
\caption{Constraints on $n_{\rm s}$, $h$, and $\log T_{\rm AGN}$ in a split cosmological analysis with COSEBIs (yellow) and band powers (pink). For reference, the black dashed contours show constraints from the analysis with a single set of parameters modelling both datasets. The inner and outer contours of the marginalised posteriors correspond to the 68\% and 95\% credible intervals, respectively.}
\label{fig:cosebis_bandpowers}
\end{figure}

The marginalised posterior distributions for $\Omega_{\rm m}$ and $S_8$ are displayed in Fig. \ref{fig:statlevel_cosebis}. Here, the solid contours refer to the analysis with the split cosmological model and the black dashed lines show constraints from the analysis of the combined data vector of two summary statistics with a single set of parameters. For the combination of COSEBIs and 2PCFs as well as band powers and 2PCFs, we found the constraints from the combined analysis to be in agreement with the split analyses. However, for the combination of COSEBIs and band powers we observe a preference for lower values of $\Omega_{\rm m}$ in the combined analysis. We further investigated the origin of this feature by decomposing the likelihood into the contribution from the auto-correlation of COSEBIs and band powers and the contribution from their cross-correlation. This analysis shows that the shift towards low $\Omega_{\rm m}$ is driven by the cross-covariance terms between COSEBIs and band powers. Additionally, we inspected the posterior distribution of the remaining split parameters, displayed in Fig. \ref{fig:cosebis_bandpowers}. We found that the two instances of the baryon feedback parameter, $\log T_{\rm AGN}$, show a preference for different amounts of baryonic feedback. While the COSEBIs posterior peaks at a low value of $\log T_{\rm AGN}$, corresponding to a dark matter only scenario, the posterior from band powers tends towards the upper edge of the prior. The origin of this feature most likely lies in their varying response to different scales. In the combined analysis with a single set of parameters, in which both statistics share the same baryonic feedback parameter, this most likely causes the shift of the posterior towards low $\Omega_{\rm m}$. As a consequence, the evidence-based consistency metric, provided in Table \ref{tab:statlevel}, reports a preference for the split cosmological model with $N_{\sigma,S}=3.78$. In the tier 2 parameter space and tier 3 data space metrics, however, we find both statistics to be in agreement. We note that our default consistency test in parameter space focusses on the two parameters that are mostly constrained by our cosmic shear data, $\Omega_{\rm m}$ and $S_8$. Considering the apparent discrepancy in the baryon feedback parameter, we computed the significance of the shift in $\log T_{\rm AGN}$, which we found to be $N_{\sigma}(\Delta\log T_{\rm AGN})=2.54$. Furthermore, we note that the posterior in the combined analysis of COSEBIs and band powers closely resembles the posterior of an analysis with 2PCFs (see appendix F in \citetalias{Wright25}). As can be inferred from the corresponding window functions, the combination of COSEBIs and band powers covers approximately the same range of scales that is probed by 2PCFs, which provides an explanation for the similarity between their posteriors.

For the remaining combinations of summary statistics, our consistency analysis finds an agreement in all tests. Therefore, we conclude that for the parameters of interest, we find the three summary statistics to be in agreement. This confirms the result of the fiducial cosmic shear analysis \citepalias{Wright25}, which reports a good agreement between marginalised $S_8$ constraints inferred individually with the three statistics. However, we note that a combined analysis of two summary statistics proves to be challenging due to the high degree of correlation between the statistics, as showcased in our combined analysis of COSEBIs and band powers. 
\begin{table*}
\caption{Consistency metrics for the combination of summary statistics.}
\label{tab:statlevel}
\centering
\begin{tabular}{lcc|ccc|cc}
\hline\hline
&\multicolumn{2}{c}{Tier 1}&\multicolumn{3}{c}{Tier 2}&\multicolumn{2}{c}{Tier 3}\\
Statistic&  $\log_{10}R$ & $N_{\sigma,S}$ & $N_{\sigma}(\Delta\Omega_{\rm m})$ & $N_{\sigma}(\Delta S_{\rm 8})$ & $N_{\sigma}(\Delta(\Omega_{\rm m}, S_{\rm 8}))$ & $p(A|B)$ & $ p(B|A)$\\
\hline
COSEBIS vs Bandpowers & 0.09 &  3.78 & 0.87 & 0.06 & 0.85 & 0.24 & 0.85 \\
COSEBIS vs $\xi_\pm$ & 2.75 &  0.21 & 0.24 & 0.45 & 0.15 & 0.45 & 0.09 \\
Bandpowers vs $\xi_\pm$ & 1.46 &  1.31 & 0.82 & 1.09 & 0.57 & 0.87 & 0.08 \\
\hline
\end{tabular}
\tablefoot{The second and third columns report the Bayes ratio and the tension level inferred from the suspiciousness, respectively. The next three columns present the results of the tier 2 test for $\Omega_{\rm m}$, $S_8$, and their combination. The final two columns shows the tension level arising from the tier 3 test in terms of the $p$-value for data vector predictions for the first statistic inferred from the TPD of the other statistic (column 7) and vice versa (column 8).}
\end{table*}
\section{Combination with external data}
\label{sec:results_external}
Weak lensing on its own only is only equipped to constrain a nearly degenerate combination of $\sigma_8$ and $\Omega_{\rm m}$. This is commonly described in terms of the $S_8$ parameter, where the width of the degeneracy is reflected in the uncertainty on $S_8$. Combining our KiDS-Legacy cosmic shear measurements with external datasets allows us to break the degeneracy between the two parameters. In particular, spectroscopic galaxy surveys provide complementary constraints on the expansion history of the Universe through measurements of the BAO feature over a range of redshifts. While the BAO feature resides in the quasi-linear clustering regime, RSD measurements probe the growth of structure. We adopted BAO measurements from DESI DR1 \citep{DESI24} as well as BAO and RSD measurements from eBOSS DR16 \citep{Alam21}, which provide a tight constraint on $\Omega_{\rm m}$. Additionally, observations of SN Ia provide an alternative method of constraining the expansion history of the Universe. This method relies on measurements of the luminosity distances as a function of redshift, which provide an independent constraint on the matter density. Here, we employed SN Ia measurements from the Pantheon+ compilation \citep{Scolnic22,Brout22}.
\begin{table*}
\caption{Consistency metrics for the combination of KiDS-Legacy with external data.}
\label{tab:external}
\centering
\begin{tabular}{lcc|ccc|cc}
\hline\hline
&\multicolumn{2}{c}{Tier 1}&\multicolumn{3}{c}{Tier 2}&\multicolumn{2}{c}{Tier 3}\\
External dataset&  $\log_{10}R$ & $N_{\sigma,S}$ & $N_{\sigma}(\Delta\Omega_{\rm m})$ & $N_{\sigma}(\Delta S_{\rm 8})$ & $N_{\sigma}(\Delta(\Omega_{\rm m}, S_{\rm 8}))$ & $p(A|B)$ & $ p(B|A)$\\
\hline
DESI Y1 BAO             & 0.24 & 0.26 & 0.45 & 0.55 & 0.35 & 0.46 & 0.49\\
eBOSS DR16              & 0.28 & 0.29 & 0.22 & 0.41 & 0.19 & 0.47 & 0.59\\
Pantheon+               & 0.40 & 0.44 & 0.29 & 0.23 & 0.13 & 0.47 & 0.80\\
DESI Y1 BAO + Pantheon+ & 0.32 & 0.45 & 0.06 & 0.90 & 0.10 & 0.47 & 0.77\\
eBOSS DR16 + Pantheon+  & 0.34 & 0.44 & 0.06 & 0.44 & 0.05 & 0.47 & 0.79\\
DES Y3                  & 1.14 & 0.31 & 0.81 & 0.58 & 0.05 & 0.35 & 0.30\\
{\it Planck} 2018       & 0.99 & 0.77 & 0.17 & 0.61 & 0.15 & 0.47 & 0.57\\
\hline
\end{tabular}
\tablefoot{The third and fourth columns report the Bayes ratio and the tension level inferred from the suspiciousness, respectively. The next three columns present the results of the tier 2 test for $\Omega_{\rm m}$, $S_8$, and their combination. The final two columns shows the tension level arising from the tier 3 test in terms of the $p$-value for data vector predictions for the external data dataset listed in the second column inferred from the KiDS-Legacy TPD (column 8) and vice versa (column 9).}
\end{table*}

\begin{figure}
\centering
 \includegraphics[width=0.5\textwidth]{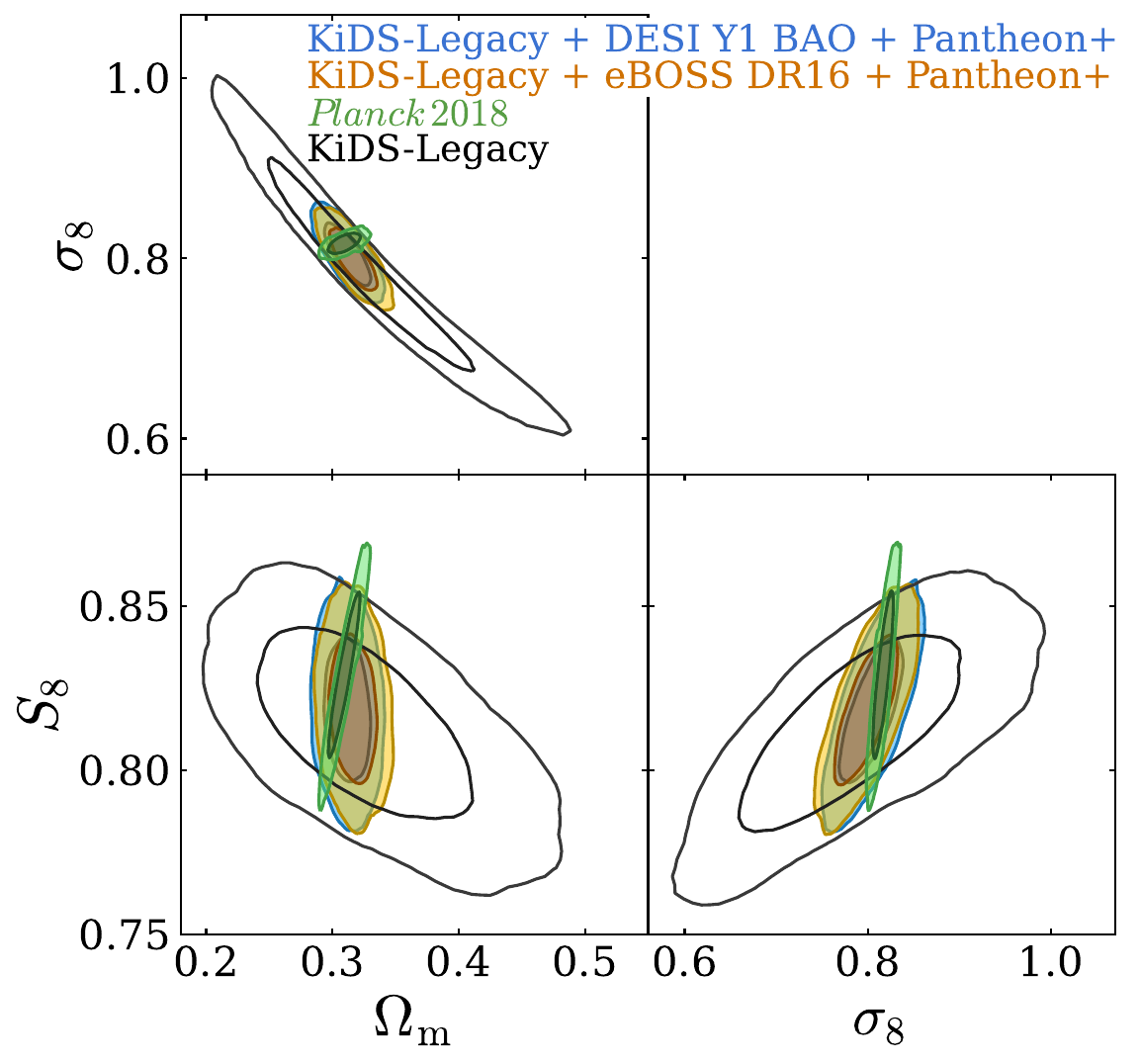}
\caption{Marginalised constraints for the joint distributions of $\Omega_{\rm m}$, $\sigma_8$, and $S_8$ from KiDS-Legacy cosmic shear data (black), its combination with Pantheon+ SN Ia data and DESI Y1 BAO data (blue), and its combination with Pantheon+ SN Ia data and eBOSS DR16 BAO and RSD data (orange). These results can be compared to CMB constraints (green) inferred with the compressed {\it Planck} likelihood by \cite{Prince19}. The inner and outer contours of the marginalised posteriors correspond to the 68\% and 95\% credible intervals, respectively.}
\label{fig:external_cosebis}
\end{figure}

We treated the KiDS data and the external data vectors as independent. Therefore, we computed the joint likelihood by multiplying the individual likelihoods of each experiment. Additionally, we assume independence between the BAO and SN Ia measurements and conduct a joint analysis of KiDS and Pantheon+ data in combination with DESI Y1 BAO and eBOSS DR16, respectively. In Table \ref{tab:external}, we quantify the consistency between KiDS-Legacy and the external datasets. As can be observed in Fig. \ref{fig:external_cosebis}, BAO, RSD, and SN Ia measurements put a tight constraint on the matter density. Additionally, BAO measurements constrain the Hubble parameter by incorporating an external calibration of the absolute BAO scale. Since the external datasets are sensitive to parameters that are mostly unconstrained by cosmic shear, we found a good agreement between KiDS-Legacy and DESI, eBOSS, and Pantheon+ in all tests. In the joint analyses, we found the marginal mode and the highest posterior density interval to be
\begin{equation}
\begin{aligned}
    &\text{KiDS + DESI + Pantheon+:\;}&S_8&=\,0.818^{+0.015}_{-0.014,}\\
    &&\sigma_8&=\,0.803^{+0.024}_{-0.021,}\\
    &&\Omega_{\rm m}&=\,0.311^{+0.011}_{-0.012,}\\
    &\text{KiDS + eBOSS + Pantheon+:\;}&S_8&=\,0.819^{+0.014}_{-0.015,}\\
    && \sigma_8&=\,0.798^{+0.023}_{-0.022,}\\
    && \Omega_{\rm m}&=\,0.315^{+0.012}_{-0.013}\;.
\end{aligned}
\end{equation}
The breaking of the $\sigma_8$-$\Omega_{\rm m}$ degeneracy results in a reduction in the uncertainty on $\sigma_8$ by about 72\% compared to the fiducial constraint of \citetalias{Wright25}. In terms of $S_8$, this corresponds to a 22\% uncertainty reduction. We note, however, that the preferred degeneracy direction for KiDS-Legacy in the more general $\Sigma_8=\sigma_8(\Omega_{\rm m}/0.3)^\alpha$ parameterisation differs from the $\alpha=0.5$ assumed in the definition of $S_8$, as discussed in sect. 5.1 in \citetalias{Wright25}. Using the preferred $\alpha=0.58$, we find $\Sigma_8 = 0.819^{+0.015}_{-0.013}$ in the combined analysis of KiDS-Legacy + DESI Y1 BAO + Pantheon+, which is consistent with the results of \citetalias{Wright25}.

\begin{figure}
\centering
\includegraphics[width=0.5\textwidth]{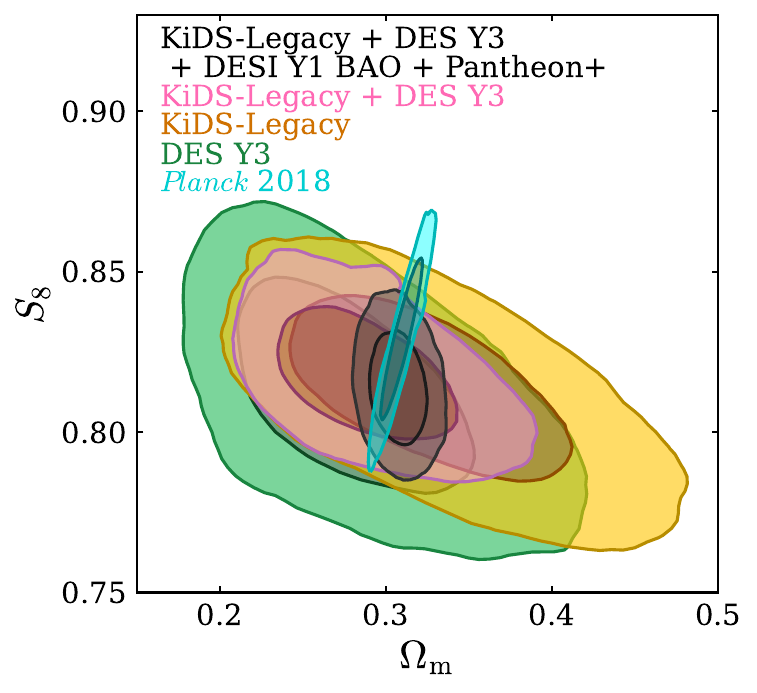}
\caption{Marginalised cosmic shear constraints for the joint distribution of $\Omega_{\rm m}$ and $S_8$ from KiDS-Legacy (yellow), DES-Y3 (green), and their combination (pink). These results can be compared to the CMB posterior inferred with the compressed {\it Planck} likelihood by \cite{Prince19}. The black contour shows constraints from a joint analysis of KiDS-Legacy with DES Y3, DESI Y1 BAO, and Pantheon+ data. The inner and outer contours of the marginalised posteriors correspond to the 68\% and 95\% credible intervals, respectively.}
\label{fig:kids_des}
\end{figure}

Additionally, we present a joint analysis between cosmic shear measurements of DES and KiDS. This analysis was previously conducted in \citetalias{KiDS+DES}, who combined cosmic shear data from DES Y3 and KiDS-1000 data and \citet{Garcia24}, who reanalysed DES Y3, KiDS-1000, and HSC DR1 data with a common harmonic-space pipeline. Here, we adopted the \enquote*{KiDS-excised} DES Y3 2PCF data vector of \citetalias{KiDS+DES} and update the combined analysis with our KiDS-Legacy COSEBIs measurements. We sampled the parameter space with our KiDS-Legacy pipeline within our fiducial prior space and adopted the independent Gaussian priors for incorporating the uncertainty on the shear and redshift calibration in DES \citepalias[see Table 1 of][]{KiDS+DES}. We note that the DES Y3 measurements excluded the overlap region between both surveys, which are therefore considered to be independent. Following the methodology of \citetalias{KiDS+DES}, we adopted independent IA parameters for both surveys. We employed the fiducial NLA-$M$ model for KiDS IA in our combined analysis. Since the prior on the NLA-$M$ parameters is survey-dependent and an application of the NLA-$M$ model to DES data is beyond the scope of this work, we modelled DES IA with the NLA-$z$ model, which is the fiducial IA model in the Hybrid analysis pipeline of \citetalias{KiDS+DES}. However, as shown in \citetalias{Wright25}, changes in the IA modelling only have a minor impact on the cosmological constraints in KiDS-Legacy. Therefore, we did not expect the choice of IA model to make a significant impact on the consistency analysis between KiDS and DES. The respective posteriors in the combined parameter space of $S_8$ and $\Omega_{\rm m}$ for KiDS-Legacy (yellow), DES Y3 (green), and their combination (pink) are illustrated in Fig. \ref{fig:kids_des} and their consistency is quantified in Table \ref{tab:external}. Overall, we found an agreement between both surveys up to $N_{\sigma} = 0.81$, which is reported by the parameter space test in $\Omega_{\rm m}$. Thus, we considered the cosmological constraints from both surveys to be consistent, which is in agreement with the earlier study conducted with KiDS-1000 data. The joint analysis of KiDS-Legacy + DES Y3 yields
\begin{equation}
\begin{aligned}
    S_8&=0.818^{+0.012}_{-0.014,}\\
    \sigma_8&=0.838^{+0.059}_{-0.060,}\\
    \Omega_{\rm m}&=0.277^{+0.042}_{-0.028}\;.
\end{aligned}
\end{equation}
Compared to the earlier study of joint KiDS + DES data, this corresponds to a 26\% reduction in uncertainty on $S_8$ and a shift towards higher $S_8$ by about $1\sigma$, which can be attributed to the preference for higher $S_8$ in the KiDS-Legacy dataset.

The consistency analysis between KiDS-Legacy and various probes of the low-redshift Universe individually signifies a good agreement. Therefore, this enables a joint analysis with our compilation of external datasets. We note that we did not consider a joint analysis with both DESI Y1 BAO and eBOSS DR16 data given the overlapping survey footprints. Therefore, we conduct a joint analysis of KiDS-Legacy with DES Y3, DESI Y1 BAO, and Pantheon+ data, yielding
\begin{equation}
\begin{aligned}
    S_8 &= 0.814^{+0.011}_{-0.012,}\\
    \sigma_8 &= 0.802^{+0.022}_{-0.018,}\\
    \Omega_{\rm m} &= 0.307^{+0.011}_{-0.011}.\;
\end{aligned}
\end{equation}
This represents a 38\% improvement in constraining power on $S_8$ and a 75\% improvement in $\sigma_8$ compared to the fiducial constraints with KiDS-Legacy.

In light of the apparent $S_8$ tension that was reported in earlier cosmic shear studies, we quantify the consistency between our KiDS-Legacy cosmic shear constraints and CMB measurements from {\it Planck}. We conducted a consistency analysis, assuming independence between both surveys, and quantify the consistency in Table \ref{tab:external}. For modelling the CMB power spectra we employed the {\sc CosmoPower} emulator \citep{SpurioMancini22}, which was shown to reproduce the fiducial {\it Planck} parameter constraints with similar accuracy to common Boltzmann solvers. We found no evidence for a disagreement between KiDS-Legacy cosmic shear measurements and {\it Planck} CMB data with $N_\sigma=0.77$ inferred with the suspiciousness statistic. Moreover, we found $S_8$ to be in agreement at $0.61\sigma$ between both surveys. This removes the $S_8$ tension, which was found to be significant at the $\sim2\sigma$ level in earlier KiDS studies. Combined analyses of probes of the early Universe, such as CMB measurements, and probes of the late Universe, such as cosmic shear, are commonly employed in studies of extended cosmological models beyond $\Lambda$CDM. A detailed study of extended cosmological models is beyond the scope of this work, and we therefore leave the combined analysis of KiDS-Legacy and {\it Planck} data for a forthcoming publication.
\section{Summary and conclusions}
\label{sec:conclusions}
In this analysis, we have demonstrated that the KiDS-Legacy cosmic shear data exhibits a high level of internal consistency. We find an agreement between the cosmological constraints from different tomographic redshift bins, the auto- and cross-correlation measurements, the COSEBIs, band power spectra and 2PCF statistics, the North and South regions within the KiDS footprint, and the red and blue galaxies. Our three tiers of consistency metrics use Bayesian evidence, measured shifts in multi-dimensional parameter space and TPDs to quantify consistency between different data splits. From our range of consistency tests, it is worth highlighting two key results. In a red-blue galaxy split analysis, we confirm the results of previous studies that find strong intrinsic galaxy alignments between red early-type galaxies and no significant alignment in the blue galaxy population. This supports our decision to adopt a new colour-dependent NLA-M model in our primary KiDS-Legacy analysis. 
In our analysis of the cosmic shear signal measured across the six tomographic redshift bins, we find a $N_{\sigma}<1.39$ consistency. This represents a marked improvement over previous KiDS analyses where the second tomographic bin, covering a range of $0.3 < z_{\rm B} \leq 0.5$, was identified as a significant outlier. We credit this improvement to advances in redshift calibration methodology and the enhanced spectroscopic dataset adopted for KiDS-Legacy \citep[see][for details]{Wright23_redshifts}.

In \citet{Wright25}, we present our fiducial cosmic shear analysis, reporting a result of $S_8 = 0.815^{+0.016}_{-0.021}$. We used our three-tier analysis to demonstrate external consistency between this result and BAO constraints\footnote{Similar results are found when combining KiDS-Legacy with BAO and RSD galaxy clustering data from eBOSS DR16, where $S_8=0.819^{+0.014}_{-0.015}$.} from DESI DR1 combined with the Pantheon+ SN Ia compilation. We inferred a joint constraint of $S_8=0.818^{+0.015}_{-0.014}$, representing a 22\% reduction in uncertainty on this parameter over our fiducial result. KiDS-Legacy is also shown to be consistent with cosmic shear data from DES Y3, where a joint analysis of the two surveys finds $S_8=0.818^{+0.012}_{-0.014}$. Combining the KiDS, DES, DESI, and Pantheon datasets, we were able to deliver a 1.4\% precision measurement of $S_8 = 0.814^{+0.011}_{-0.012}$. This result is consistent with $S_8$ measurements of the cosmic microwave background by {\it Planck}, which is in $0.77\sigma$ agreement with KiDS-Legacy.

The KiDS-Legacy cosmic shear data exhibits a high level of internal and external consistency resulting from significant improvements in the data reduction and scientific analyses of KiDS since its inception over a decade ago. Upcoming cosmology experiments will be required to pass the three tiers of stringent internal consistency analyses presented here and, as such, this analysis provides a useful blueprint for future studies.
\begin{acknowledgements}
We thank the anonymous referee for their constructive comments,
which helped to improve the manuscript.
BS, ZY, and CH acknowledge support from the Max Planck Society and the Alexander von Humboldt Foundation in the framework of the Max Planck-Humboldt Research Award endowed by the Federal Ministry of Education and Research. AHW is supported by the Deutsches Zentrum für Luft- und Raumfahrt (DLR), made possible by the Bundesministerium für Wirtschaft und Klimaschutz, and acknowledges funding from the German Science Foundation DFG, via the Collaborative Research Center SFB1491 "Cosmic Interacting Matters - From Source to Signal". PB acknowledges financial support from the Canadian Space Agency (Grant No. 23EXPROSS1) and the Waterloo Centre for Astrophysics. LL is supported by the Austrian Science Fund (FWF) [ESP 357-N]. SJ acknowledges the Ramón y Cajal Fellowship (RYC2022-036431-I) from the Spanish Ministry of Science and the Dennis Sciama Fellowship at the University of Portsmouth. HHo, SSL, and MY acknowledge support from the European Research Council (ERC) under the European Union’s Horizon 2020 research and innovation program with Grant agreement No. 101053992. LM acknowledges the financial contribution from the grant PRIN-MUR 2022 20227RNLY3 “The concordance cosmological model: stress-tests with galaxy clusters” supported by Next Generation EU and from the grant ASI n. 2024-10-HH.0 “Attività scientifiche per la missione Euclid – fase E”. MvWK acknowledges the support by the UKSA and STFC (grant no. ST/X001075/1). AD, HHi, CM, and RR are supported by an ERC Consolidator Grant (No. 770935). HHi is supported by a DFG Heisenberg grant (Hi 1495/5-1), the DFG Collaborative Research Center SFB1491, and the DLR project 50QE2305. MB, PJ, and AJW are supported by the Polish National Science Center through grant no. 2020/38/E/ST9/00395. MB is also supported by the Polish National Science Center through grant no. 2020/39/B/ST9/03494. LP acknowledges support from the DLR grant 50QE2002. TT acknowledges funding from the Swiss National Science Foundation under the Ambizione project PZ00P2\_193352. YZ acknowledges the studentship from the UK Science and Technology Facilities Council (STFC). BJ acknowledges support by the ERC-selected UKRI Frontier Research Grant EP/Y03015X/1 and by STFC Consolidated Grant ST/V000780/1. KK acknowledges support from the Royal Society and Imperial College. SSL is receiving funding from the programme \enquote{Netzwerke 2021}, an initiative of the Ministry of Culture and Science of the State of Northrhine Westphalia. CG is funded by the MICINN project PID2022-141079NB-C32. CH acknowledges support from the UK Science and Technology Facilities Council (STFC) under grant ST/V000594/1. MA is supported by the UK Science and Technology Facilities Council (STFC) under grant number ST/Y002652/1 and the Royal Society under grant numbers RGSR2222268 and ICAR1231094. CM acknowledges support from the Beecroft Trust, the Spanish Ministry of Science under the grant number PID2021-128338NB-I00. JHD acknowledges support from an STFC Ernest Rutherford Fellowship (project reference ST/S004858/1). MR acknowledges financial support from the INAF grant 2022". AL acknowledges support from the research project grant `Understanding the Dynamic Universe' funded by the Knut and Alice Wallenberg Foundation under Dnr KAW 2018.0067. NRN acknowledges financial support from the National Science Foundation of China, Research Fund for Excellent International Scholars (grant n. 12150710511), and from the research grant from China Manned Space Project n. CMS-CSST-2021-A01.
\\
{\it Kilo-Degree Survey:} Based on observations made with ESO Telescopes at the La Silla Paranal Observatory under programme IDs 179.A-2004, 177.A-3016, 177.A-3017, 177.A-3018, 298.A-5015.
\\
{\it Dark Energy Spectroscopic Instrument:} This research used data obtained with the Dark Energy Spectroscopic Instrument (DESI). DESI construction and operations is managed by the Lawrence Berkeley National Laboratory. This material is based upon work supported by the U.S. Department of Energy, Office of Science, Office of High-Energy Physics, under Contract No. DE–AC02–05CH11231, and by the National Energy Research Scientific Computing Center, a DOE Office of Science User Facility under the same contract. Additional support for DESI was provided by the U.S. National Science Foundation (NSF), Division of Astronomical Sciences under Contract No. AST-0950945 to the NSF’s National Optical-Infrared Astronomy Research Laboratory; the Science and Technology Facilities Council of the United Kingdom; the Gordon and Betty Moore Foundation; the Heising-Simons Foundation; the French Alternative Energies and Atomic Energy Commission (CEA); the National Council of Science and Technology of Mexico (CONACYT); the Ministry of Science and Innovation of Spain (MICINN), and by the DESI Member Institutions: www.desi.lbl.gov/collaborating-institutions. The DESI collaboration is honored to be permitted to conduct scientific research on Iolkam Du’ag (Kitt Peak), a mountain with particular significance to the Tohono O’odham Nation. Any opinions, findings, and conclusions or recommendations expressed in this material are those of the author(s) and do not necessarily reflect the views of the U.S. National Science Foundation, the U.S. Department of Energy, or any of the listed funding agencies.
\\
{\it SDSS-IV:} Funding for the Sloan Digital Sky 
Survey IV has been provided by the 
Alfred P. Sloan Foundation, the U.S. 
Department of Energy Office of 
Science, and the Participating 
Institutions. 
SDSS-IV acknowledges support and 
resources from the Center for High 
Performance Computing at the 
University of Utah. The SDSS 
website is www.sdss4.org.
SDSS-IV is managed by the 
Astrophysical Research Consortium 
for the Participating Institutions 
of the SDSS Collaboration including 
the Brazilian Participation Group, 
the Carnegie Institution for Science, 
Carnegie Mellon University, Center for 
Astrophysics | Harvard \& 
Smithsonian, the Chilean Participation 
Group, the French Participation Group, 
Instituto de Astrof\'isica de 
Canarias, The Johns Hopkins 
University, Kavli Institute for the 
Physics and Mathematics of the 
Universe (IPMU) / University of 
Tokyo, the Korean Participation Group, 
Lawrence Berkeley National Laboratory, 
Leibniz Institut f\"ur Astrophysik 
Potsdam (AIP), Max-Planck-Institut 
f\"ur Astronomie (MPIA Heidelberg), 
Max-Planck-Institut f\"ur 
Astrophysik (MPA Garching), 
Max-Planck-Institut f\"ur 
Extraterrestrische Physik (MPE), 
National Astronomical Observatories of 
China, New Mexico State University, 
New York University, University of 
Notre Dame, Observat\'ario 
Nacional / MCTI, The Ohio State 
University, Pennsylvania State 
University, Shanghai 
Astronomical Observatory, United 
Kingdom Participation Group, 
Universidad Nacional Aut\'onoma 
de M\'exico, University of Arizona, 
University of Colorado Boulder, 
University of Oxford, University of 
Portsmouth, University of Utah, 
University of Virginia, University 
of Washington, University of 
Wisconsin, Vanderbilt University, 
and Yale University.
\\
{\it Dark Energy Survey:} This project used public archival data from the Dark Energy Survey (DES). Funding for the DES Projects has been provided by the U.S. Department of Energy, the U.S. National Science Foundation, the Ministry of Science and Education of Spain, the Science and Technology FacilitiesCouncil of the United Kingdom, the Higher Education Funding Council for England, the National Center for Supercomputing Applications at the University of Illinois at Urbana-Champaign, the Kavli Institute of Cosmological Physics at the University of Chicago, the Center for Cosmology and Astro-Particle Physics at the Ohio State University, the Mitchell Institute for Fundamental Physics and Astronomy at Texas A\&M University, Financiadora de Estudos e Projetos, Funda{\c c}{\~a}o Carlos Chagas Filho de Amparo {\`a} Pesquisa do Estado do Rio de Janeiro, Conselho Nacional de Desenvolvimento Cient{\'i}fico e Tecnol{\'o}gico and the Minist{\'e}rio da Ci{\^e}ncia, Tecnologia e Inova{\c c}{\~a}o, the Deutsche Forschungsgemeinschaft, and the Collaborating Institutions in the Dark Energy Survey.
The Collaborating Institutions are Argonne National Laboratory, the University of California at Santa Cruz, the University of Cambridge, Centro de Investigaciones Energ{\'e}ticas, Medioambientales y Tecnol{\'o}gicas-Madrid, the University of Chicago, University College London, the DES-Brazil Consortium, the University of Edinburgh, the Eidgen{\"o}ssische Technische Hochschule (ETH) Z{\"u}rich, Fermi National Accelerator Laboratory, the University of Illinois at Urbana-Champaign, the Institut de Ci{\`e}ncies de l'Espai (IEEC/CSIC), the Institut de F{\'i}sica d'Altes Energies, Lawrence Berkeley National Laboratory, the Ludwig-Maximilians Universit{\"a}t M{\"u}nchen and the associated Excellence Cluster Universe, the University of Michigan, the National Optical Astronomy Observatory, the University of Nottingham, The Ohio State University, the OzDES Membership Consortium, the University of Pennsylvania, the University of Portsmouth, SLAC National Accelerator Laboratory, Stanford University, the University of Sussex, and Texas A\&M University.
Based in part on observations at Cerro Tololo Inter-American Observatory, National Optical Astronomy Observatory, which is operated by the Association of Universities for Research in Astronomy (AURA) under a cooperative agreement with the National Science Foundation. BG acknowledges support from the UKRI Stephen Hawking Fellowship (grant reference EP/Y017137/1).
\\
{\it Planck:} Based on observations obtained with Planck (http://www.esa.int/Planck), an ESA science mission with instruments and contributions directly funded by ESA Member States, NASA, and Canada.
\\
{\it Software:} The figures in this work were created with {\sc matplotlib} \citep{Matplotlib} and {\sc ChainConsumer} \citep{Chainconsumer}, making use of the {\sc NumPy} \citep{Numpy}, {\sc SciPy} \citep{Scipy}, {\sc pandas} \citep{Pandas}, {\sc Cosmosis} \citep{Zuntz15}, {\sc Nautilus} \citep{Lange23}, {\sc camb} \citep{Lewis00,Howlett12}, and {\sc CosmoPower} \citep{SpurioMancini22} software packages.
\\
{\it Author Contributions:} All authors contributed to the development and writing of this paper. The authorship list is given in three groups: the lead authors (BS,AHW), followed by two alphabetical groups. The first alphabetical group includes those who are key contributors to both the scientific analysis and the data products of this manuscript and release. The second group covers those who have either made a significant contribution to the preparation of data products or to the scientific analyses of KiDS since its inception.
\end{acknowledgements}
\bibliographystyle{aa}
\bibliography{bibliography.bib}

\begin{thebibliography}{118}
\expandafter\ifx\csname natexlab\endcsname\relax\def\natexlab#1{#1}\fi

\bibitem[{{Abdalla} {et~al.}(2022){Abdalla}, {Abell{\'a}n}, {Aboubrahim},
  {Agnello}, {Akarsu}, {Akrami}, {Alestas}, {Aloni}, {Amendola}, {Anchordoqui},
  {Anderson}, {Arendse}, {Asgari}, {Ballardini}, {Barger}, {Basilakos},
  {Batista}, {Battistelli}, {Battye}, {Benetti}, {Benisty}, {Berlin}, {de
  Bernardis}, {Berti}, {Bidenko}, {Birrer}, {Blakeslee}, {Boddy}, {Bom},
  {Bonilla}, {Borghi}, {Bouchet}, {Braglia}, {Buchert}, {Buckley-Geer},
  {Calabrese}, {Caldwell}, {Camarena}, {Capozziello}, {Casertano}, {Chen},
  {Chluba}, {Chen}, {Chen}, {Chudaykin}, {Cicoli}, {Copi}, {Courbin},
  {Cyr-Racine}, {Czerny}, {Dainotti}, {D'Amico}, {Davis}, {de Cruz P{\'e}rez},
  {de Haro}, {Delabrouille}, {Denton}, {Dhawan}, {Dienes}, {Di Valentino},
  {Du}, {Eckert}, {Escamilla-Rivera}, {Fert{\'e}}, {Finelli}, {Fosalba},
  {Freedman}, {Frusciante}, {Gazta{\~n}aga}, {Giar{\`e}}, {Giusarma},
  {G{\'o}mez-Valent}, {Handley}, {Harrison}, {Hart}, {Hazra}, {Heavens},
  {Heinesen}, {Hildebrandt}, {Hill}, {Hogg}, {Holz}, {Hooper}, {Hosseininejad},
  {Huterer}, {Ishak}, {Ivanov}, {Jaffe}, {Jang}, {Jedamzik}, {Jimenez},
  {Joseph}, {Joudaki}, {Kamionkowski}, {Karwal}, {Kazantzidis}, {Keeley},
  {Klasen}, {Komatsu}, {Koopmans}, {Kumar}, {Lamagna}, {Lazkoz}, {Lee},
  {Lesgourgues}, {Levi Said}, {Lewis}, {L'Huillier}, {Lucca}, {Maartens},
  {Macri}, {Marfatia}, {Marra}, {Martins}, {Masi}, {Matarrese}, {Mazumdar},
  {Melchiorri}, {Mena}, {Mersini-Houghton}, {Mertens}, {Milakovi{\'c}},
  {Minami}, {Miranda}, {Moreno-Pulido}, {Moresco}, {Mota}, {Mottola}, {Mozzon},
  {Muir}, {Mukherjee}, {Mukherjee}, {Naselsky}, {Nath}, {Nesseris},
  {Niedermann}, {Notari}, {Nunes}, {{\'O} Colg{\'a}in}, {Owens},
  {{\"O}z{\"u}lker}, {Pace}, {Paliathanasis}, {Palmese}, {Pan}, {Paoletti},
  {Perez Bergliaffa}, {Perivolaropoulos}, {Pesce}, {Pettorino}, {Philcox},
  {Pogosian}, {Poulin}, {Poulot}, {Raveri}, {Reid}, {Renzi}, {Riess}, {Sabla},
  {Salucci}, {Salzano}, {Saridakis}, {Sathyaprakash}, {Schmaltz},
  {Sch{\"o}neberg}, {Scolnic}, {Sen}, {Sehgal}, {Shafieloo}, {Sheikh-Jabbari},
  {Silk}, {Silvestri}, {Skara}, {Sloth}, {Soares-Santos}, {Sol{\`a} Peracaula},
  {Songsheng}, {Soriano}, {Staicova}, {Starkman}, {Szapudi}, {Teixeira},
  {Thomas}, {Treu}, {Trott}, {van de Bruck}, {Vazquez}, {Verde}, {Visinelli},
  {Wang}, {Wang}, {Wang}, {Watkins}, {Watson}, {Webb}, {Weiner}, {Weltman},
  {Witte}, {Wojtak}, \& {Yadav}}]{Abdalla22}
{Abdalla}, E., {Abell{\'a}n}, G.~F., {Aboubrahim}, A., {et~al.} 2022,
  \href{http://dx.doi.org/10.1016/j.jheap.2022.04.002}{\color{black}Journal of
  High Energy Astrophysics},
  \href{https://ui.adsabs.harvard.edu/abs/2022JHEAp..34...49A}{34, 49}

\bibitem[{{Adame} {et~al.}(2025){Adame}, {Aguilar}, {Ahlen}, {Alam},
  {Alexander}, {Alvarez}, {Alves}, {Anand}, {Andrade}, {Armengaud}, {Avila},
  {Aviles}, {Awan}, {Bahr-Kalus}, {Bailey}, {Baltay}, {Bault}, {Behera},
  {BenZvi}, {Bera}, {Beutler}, {Bianchi}, {Blake}, {Blum}, {Brieden},
  {Brodzeller}, {Brooks}, {Buckley-Geer}, {Burtin}, {Calderon}, {Canning},
  {Carnero Rosell}, {Cereskaite}, {Cervantes-Cota}, {Chabanier}, {Chaussidon},
  {Chaves-Montero}, {Chen}, {Chen}, {Claybaugh}, {Cole}, {Cuceu}, {Davis},
  {Dawson}, {de la Macorra}, {de Mattia}, {Deiosso}, {Dey}, {Dey}, {Ding},
  {Doel}, {Edelstein}, {Eftekharzadeh}, {Eisenstein}, {Elliott}, {Fagrelius},
  {Fanning}, {Ferraro}, {Ereza}, {Findlay}, {Flaugher}, {Font-Ribera},
  {Forero-S{\'a}nchez}, {Forero-Romero}, {Frenk}, {Garcia-Quintero},
  {Gazta{\~n}aga}, {Gil-Mar{\'\i}n}, {Gontcho a Gontcho}, {Gonzalez-Morales},
  {Gonzalez-Perez}, {Gordon}, {Green}, {Gruen}, {Gsponer}, {Gutierrez}, {Guy},
  {Hadzhiyska}, {Hahn}, {Hanif}, {Herrera-Alcantar}, {Honscheid}, {Howlett},
  {Huterer}, {Ir{\v{s}}i{\v{c}}}, {Ishak}, {Juneau}, {Kara{\c{c}}ayl{\i}},
  {Kehoe}, {Kent}, {Kirkby}, {Kremin}, {Krolewski}, {Lai}, {Lan}, {Landriau},
  {Lang}, {Lasker}, {Le Goff}, {Le Guillou}, {Leauthaud}, {Levi}, {Li},
  {Linder}, {Lodha}, {Magneville}, {Manera}, {Margala}, {Martini}, {Maus},
  {McDonald}, {Medina-Varela}, {Meisner}, {Mena-Fern{\'a}ndez}, {Miquel},
  {Moon}, {Moore}, {Moustakas}, {Mueller}, {Mu{\~n}oz-Guti{\'e}rrez}, {Myers},
  {Nadathur}, {Napolitano}, {Neveux}, {Newman}, {Nguyen}, {Nie}, {Niz},
  {Noriega}, {Padmanabhan}, {Paillas}, {Palanque-Delabrouille}, {Pan},
  {Penmetsa}, {Percival}, {Pieri}, {Pinon}, {Poppett}, {Porredon}, {Prada},
  {P{\'e}rez-Fern{\'a}ndez}, {P{\'e}rez-R{\`a}fols}, {Rabinowitz}, {Raichoor},
  {Ram{\'\i}rez-P{\'e}rez}, {Ramirez-Solano}, {Rashkovetskyi}, {Ravoux},
  {Rezaie}, {Rich}, {Rocher}, {Rockosi}, {Roe}, {Rosado-Marin}, {Ross},
  {Rossi}, {Ruggeri}, {Ruhlmann-Kleider}, {Samushia}, {Sanchez}, {Saulder},
  {Schlafly}, {Schlegel}, {Schubnell}, {Seo}, {Shafieloo}, {Sharples},
  {Silber}, {Slosar}, {Smith}, {Sprayberry}, {Tan}, {Tarl{\'e}}, {Taylor},
  {Trusov}, {Ure{\~n}a-L{\'o}pez}, {Vaisakh}, {Valcin}, {Valdes},
  {Vargas-Maga{\~n}a}, {Verde}, {Walther}, {Wang}, {Wang}, {Weaver},
  {Weaverdyck}, {Wechsler}, {Weinberg}, {White}, {Yu}, {Yu}, {Yuan},
  {Y{\`e}che}, {Zaborowski}, {Zarrouk}, {Zhang}, {Zhao}, {Zhao}, {Zhou}, \&
  {Zhuang}}]{DESI24}
{Adame}, A.~G., {Aguilar}, J., {Ahlen}, S., {et~al.} 2025,
  \href{http://dx.doi.org/10.1088/1475-7516/2025/02/021}{\color{black}\jcap},
  \href{https://ui.adsabs.harvard.edu/abs/2025JCAP...02..021A}{2025, 021}

\bibitem[{{Aihara} {et~al.}(2018){Aihara}, {Arimoto}, {Armstrong}, {Arnouts},
  {Bahcall}, {Bickerton}, {Bosch}, {Bundy}, {Capak}, {Chan}, {Chiba}, {Coupon},
  {Egami}, {Enoki}, {Finet}, {Fujimori}, {Fujimoto}, {Furusawa}, {Furusawa},
  {Goto}, {Goulding}, {Greco}, {Greene}, {Gunn}, {Hamana}, {Harikane},
  {Hashimoto}, {Hattori}, {Hayashi}, {Hayashi}, {He{\l}miniak}, {Higuchi},
  {Hikage}, {Ho}, {Hsieh}, {Huang}, {Huang}, {Ikeda}, {Imanishi}, {Inoue},
  {Iwasawa}, {Iwata}, {Jaelani}, {Jian}, {Kamata}, {Karoji}, {Kashikawa},
  {Katayama}, {Kawanomoto}, {Kayo}, {Koda}, {Koike}, {Kojima}, {Komiyama},
  {Konno}, {Koshida}, {Koyama}, {Kusakabe}, {Leauthaud}, {Lee}, {Lin}, {Lin},
  {Lupton}, {Mand elbaum}, {Matsuoka}, {Medezinski}, {Mineo}, {Miyama},
  {Miyatake}, {Miyazaki}, {Momose}, {More}, {More}, {Moritani}, {Moriya},
  {Morokuma}, {Mukae}, {Murata}, {Murayama}, {Nagao}, {Nakata}, {Niida},
  {Niikura}, {Nishizawa}, {Obuchi}, {Oguri}, {Oishi}, {Okabe}, {Okamoto},
  {Okura}, {Ono}, {Onodera}, {Onoue}, {Osato}, {Ouchi}, {Price}, {Pyo}, {Sako},
  {Sawicki}, {Shibuya}, {Shimasaku}, {Shimono}, {Shirasaki}, {Silverman},
  {Simet}, {Speagle}, {Spergel}, {Strauss}, {Sugahara}, {Sugiyama}, {Suto},
  {Suyu}, {Suzuki}, {Tait}, {Takada}, {Takata}, {Tamura}, {Tanaka}, {Tanaka},
  {Tanaka}, {Tanaka}, {Terai}, {Terashima}, {Toba}, {Tominaga}, {Toshikawa},
  {Turner}, {Uchida}, {Uchiyama}, {Umetsu}, {Uraguchi}, {Urata}, {Usuda},
  {Utsumi}, {Wang}, {Wang}, {Wong}, {Yabe}, {Yamada}, {Yamanoi}, {Yasuda},
  {Yeh}, {Yonehara}, \& {Yuma}}]{HSC1}
{Aihara}, H., {Arimoto}, N., {Armstrong}, R., {et~al.} 2018,
  \href{http://dx.doi.org/10.1093/pasj/psx066}{\color{black}\pasj},
  \href{https://ui.adsabs.harvard.edu/abs/2018PASJ...70S...4A}{70, S4}

\bibitem[{{Alam} {et~al.}(2017){Alam}, {Ata}, {Bailey}, {Beutler}, {Bizyaev},
  {Blazek}, {Bolton}, {Brownstein}, {Burden}, {Chuang}, {Comparat}, {Cuesta},
  {Dawson}, {Eisenstein}, {Escoffier}, {Gil-Mar{\'\i}n}, {Grieb}, {Hand}, {Ho},
  {Kinemuchi}, {Kirkby}, {Kitaura}, {Malanushenko}, {Malanushenko}, {Maraston},
  {McBride}, {Nichol}, {Olmstead}, {Oravetz}, {Padmanabhan},
  {Palanque-Delabrouille}, {Pan}, {Pellejero-Ibanez}, {Percival}, {Petitjean},
  {Prada}, {Price-Whelan}, {Reid}, {Rodr{\'\i}guez-Torres}, {Roe}, {Ross},
  {Ross}, {Rossi}, {Rubi{\~n}o-Mart{\'\i}n}, {Saito}, {Salazar-Albornoz},
  {Samushia}, {S{\'a}nchez}, {Satpathy}, {Schlegel}, {Schneider},
  {Sc{\'o}ccola}, {Seo}, {Sheldon}, {Simmons}, {Slosar}, {Strauss}, {Swanson},
  {Thomas}, {Tinker}, {Tojeiro}, {Maga{\~n}a}, {Vazquez}, {Verde}, {Wake},
  {Wang}, {Weinberg}, {White}, {Wood-Vasey}, {Y{\`e}che}, {Zehavi}, {Zhai}, \&
  {Zhao}}]{Alam17}
{Alam}, S., {Ata}, M., {Bailey}, S., {et~al.} 2017,
  \href{http://dx.doi.org/10.1093/mnras/stx721}{\color{black}\mnras},
  \href{https://ui.adsabs.harvard.edu/abs/2017MNRAS.470.2617A}{470, 2617}

\bibitem[{{Alam} {et~al.}(2021){Alam}, {Aubert}, {Avila}, {Balland},
  {Bautista}, {Bershady}, {Bizyaev}, {Blanton}, {Bolton}, {Bovy}, {Brinkmann},
  {Brownstein}, {Burtin}, {Chabanier}, {Chapman}, {Choi}, {Chuang}, {Comparat},
  {Cousinou}, {Cuceu}, {Dawson}, {de la Torre}, {de Mattia}, {Agathe}, {des
  Bourboux}, {Escoffier}, {Etourneau}, {Farr}, {Font-Ribera}, {Frinchaboy},
  {Fromenteau}, {Gil-Mar{\'\i}n}, {Le Goff}, {Gonzalez-Morales},
  {Gonzalez-Perez}, {Grabowski}, {Guy}, {Hawken}, {Hou}, {Kong}, {Parker},
  {Klaene}, {Kneib}, {Lin}, {Long}, {Lyke}, {de la Macorra}, {Martini},
  {Masters}, {Mohammad}, {Moon}, {Mueller}, {Mu{\~n}oz-Guti{\'e}rrez}, {Myers},
  {Nadathur}, {Neveux}, {Newman}, {Noterdaeme}, {Oravetz}, {Oravetz},
  {Palanque-Delabrouille}, {Pan}, {Paviot}, {Percival}, {P{\'e}rez-R{\`a}fols},
  {Petitjean}, {Pieri}, {Prakash}, {Raichoor}, {Ravoux}, {Rezaie}, {Rich},
  {Ross}, {Rossi}, {Ruggeri}, {Ruhlmann-Kleider}, {S{\'a}nchez}, {S{\'a}nchez},
  {S{\'a}nchez-Gallego}, {Sayres}, {Schneider}, {Seo}, {Shafieloo}, {Slosar},
  {Smith}, {Stermer}, {Tamone}, {Tinker}, {Tojeiro}, {Vargas-Maga{\~n}a},
  {Variu}, {Wang}, {Weaver}, {Weijmans}, {Y{\`e}che}, {Zarrouk}, {Zhao},
  {Zhao}, \& {Zheng}}]{Alam21}
{Alam}, S., {Aubert}, M., {Avila}, S., {et~al.} 2021,
  \href{http://dx.doi.org/10.1103/PhysRevD.103.083533}{\color{black}\prd},
  \href{https://ui.adsabs.harvard.edu/abs/2021PhRvD.103h3533A}{103, 083533}

\bibitem[{{Amon} \& {Efstathiou}(2022)}]{Amon22b}
{Amon}, A. \& {Efstathiou}, G. 2022,
  \href{http://dx.doi.org/10.1093/mnras/stac2429}{\color{black}\mnras},
  \href{https://ui.adsabs.harvard.edu/abs/2022MNRAS.516.5355A}{516, 5355}

\bibitem[{{Amon} {et~al.}(2022){Amon}, {Gruen}, {Troxel}, {MacCrann},
  {Dodelson}, {Choi}, {Doux}, {Secco}, {Samuroff}, {Krause}, {Cordero},
  {Myles}, {DeRose}, {Wechsler}, {Gatti}, {Navarro-Alsina}, {Bernstein},
  {Jain}, {Blazek}, {Alarcon}, {Fert{\'e}}, {Lemos}, {Raveri}, {Campos},
  {Prat}, {S{\'a}nchez}, {Jarvis}, {Alves}, {Andrade-Oliveira}, {Baxter},
  {Bechtol}, {Becker}, {Bridle}, {Camacho}, {Carnero Rosell}, {Carrasco Kind},
  {Cawthon}, {Chang}, {Chen}, {Chintalapati}, {Crocce}, {Davis}, {Diehl},
  {Drlica-Wagner}, {Eckert}, {Eifler}, {Elvin-Poole}, {Everett}, {Fang},
  {Fosalba}, {Friedrich}, {Gaztanaga}, {Giannini}, {Gruendl}, {Harrison},
  {Hartley}, {Herner}, {Huang}, {Huff}, {Huterer}, {Kuropatkin}, {Leget},
  {Liddle}, {McCullough}, {Muir}, {Pandey}, {Park}, {Porredon}, {Refregier},
  {Rollins}, {Roodman}, {Rosenfeld}, {Ross}, {Rykoff}, {Sanchez},
  {Sevilla-Noarbe}, {Sheldon}, {Shin}, {Troja}, {Tutusaus}, {Tutusaus},
  {Varga}, {Weaverdyck}, {Yanny}, {Yin}, {Zhang}, {Zuntz}, {Aguena}, {Allam},
  {Annis}, {Bacon}, {Bertin}, {Bhargava}, {Brooks}, {Buckley-Geer}, {Burke},
  {Carretero}, {Costanzi}, {da Costa}, {Pereira}, {De Vicente}, {Desai},
  {Dietrich}, {Doel}, {Ferrero}, {Flaugher}, {Frieman}, {Garc{\'\i}a-Bellido},
  {Gaztanaga}, {Gerdes}, {Giannantonio}, {Gschwend}, {Gutierrez}, {Hinton},
  {Hollowood}, {Honscheid}, {Hoyle}, {James}, {Kron}, {Kuehn}, {Lahav}, {Lima},
  {Lin}, {Maia}, {Marshall}, {Martini}, {Melchior}, {Menanteau}, {Miquel},
  {Mohr}, {Morgan}, {Ogando}, {Palmese}, {Paz-Chinch{\'o}n}, {Petravick},
  {Pieres}, {Romer}, {Sanchez}, {Scarpine}, {Schubnell}, {Serrano}, {Smith},
  {Soares-Santos}, {Tarle}, {Thomas}, {To}, {Weller}, \& {DES
  Collaboration}}]{Amon22}
{Amon}, A., {Gruen}, D., {Troxel}, M.~A., {et~al.} 2022,
  \href{http://dx.doi.org/10.1103/PhysRevD.105.023514}{\color{black}\prd},
  \href{https://ui.adsabs.harvard.edu/abs/2022PhRvD.105b3514A}{105, 023514}

\bibitem[{{Amon} {et~al.}(2023){Amon}, {Robertson}, {Miyatake}, {Heymans},
  {White}, {DeRose}, {Yuan}, {Wechsler}, {Varga}, {Bocquet}, {Dvornik}, {More},
  {Ross}, {Hoekstra}, {Alarcon}, {Asgari}, {Blazek}, {Campos}, {Chen}, {Choi},
  {Crocce}, {Diehl}, {Doux}, {Eckert}, {Elvin-Poole}, {Everett}, {Fert{\'e}},
  {Gatti}, {Giannini}, {Gruen}, {Gruendl}, {Hartley}, {Herner}, {Hildebrandt},
  {Huang}, {Huff}, {Joachimi}, {Lee}, {MacCrann}, {Myles}, {Navarro-Alsina},
  {Nishimichi}, {Prat}, {Secco}, {Sevilla-Noarbe}, {Sheldon}, {Shin},
  {Tr{\"o}ster}, {Troxel}, {Tutusaus}, {Wright}, {Yin}, {Aguena}, {Allam},
  {Annis}, {Bacon}, {Bilicki}, {Brooks}, {Burke}, {Carnero Rosell},
  {Carretero}, {Castander}, {Cawthon}, {Costanzi}, {da Costa}, {Pereira}, {de
  Jong}, {De Vicente}, {Desai}, {Dietrich}, {Doel}, {Ferrero}, {Frieman},
  {Garc{\'\i}a-Bellido}, {Gerdes}, {Gschwend}, {Gutierrez}, {Hinton},
  {Hollowood}, {Honscheid}, {Huterer}, {Kannawadi}, {Kuehn}, {Kuropatkin},
  {Lahav}, {Lima}, {Maia}, {Marshall}, {Menanteau}, {Miquel}, {Mohr}, {Morgan},
  {Muir}, {Paz-Chinch{\'o}n}, {Pieres}, {Plazas Malag{\'o}n}, {Porredon},
  {Rodriguez-Monroy}, {Roodman}, {Sanchez}, {Serrano}, {Shan}, {Suchyta},
  {Swanson}, {Tarle}, {Thomas}, {To}, \& {Zhang}}]{Amon23}
{Amon}, A., {Robertson}, N.~C., {Miyatake}, H., {et~al.} 2023,
  \href{http://dx.doi.org/10.1093/mnras/stac2938}{\color{black}\mnras},
  \href{https://ui.adsabs.harvard.edu/abs/2023MNRAS.518..477A}{518, 477}

\bibitem[{{Asgari} {et~al.}(2021){Asgari}, {Lin}, {Joachimi}, {Giblin},
  {Heymans}, {Hildebrandt}, {Kannawadi}, {St{\"o}lzner}, {Tr{\"o}ster}, {van
  den Busch}, {Wright}, {Bilicki}, {Blake}, {de Jong}, {Dvornik}, {Erben},
  {Getman}, {Hoekstra}, {K{\"o}hlinger}, {Kuijken}, {Miller}, {Radovich},
  {Schneider}, {Shan}, \& {Valentijn}}]{Asgari21}
{Asgari}, M., {Lin}, C.-A., {Joachimi}, B., {et~al.} 2021,
  \href{http://dx.doi.org/10.1051/0004-6361/202039070}{\color{black}\aap},
  \href{https://ui.adsabs.harvard.edu/abs/2021A&A...645A.104A}{645, A104}

\bibitem[{{Asgari} {et~al.}(2012){Asgari}, {Schneider}, \& {Simon}}]{Asgari12}
{Asgari}, M., {Schneider}, P., \& {Simon}, P. 2012,
  \href{http://dx.doi.org/10.1051/0004-6361/201218828}{\color{black}\aap},
  \href{https://ui.adsabs.harvard.edu/abs/2012A&A...542A.122A}{542, A122}

\bibitem[{{Asgari} {et~al.}(2020){Asgari}, {Tr{\"o}ster}, {Heymans},
  {Hildebrandt}, {van den Busch}, {Wright}, {Choi}, {Erben}, {Joachimi},
  {Joudaki}, {Kannawadi}, {Kuijken}, {Lin}, {Schneider}, \& {Zuntz}}]{Asgari20}
{Asgari}, M., {Tr{\"o}ster}, T., {Heymans}, C., {et~al.} 2020,
  \href{http://dx.doi.org/10.1051/0004-6361/201936512}{\color{black}\aap},
  \href{https://ui.adsabs.harvard.edu/abs/2020A&A...634A.127A}{634, A127}

\bibitem[{{Bacon} {et~al.}(2000){Bacon}, {Refregier}, \& {Ellis}}]{Bacon00}
{Bacon}, D.~J., {Refregier}, A.~R., \& {Ellis}, R.~S. 2000,
  \href{http://dx.doi.org/10.1046/j.1365-8711.2000.03851.x}{\color{black}\mnras},
  \href{https://ui.adsabs.harvard.edu/abs/2000MNRAS.318..625B}{318, 625}

\bibitem[{{Bautista} {et~al.}(2021){Bautista}, {Paviot}, {Vargas Maga{\~n}a},
  {de la Torre}, {Fromenteau}, {Gil-Mar{\'\i}n}, {Ross}, {Burtin}, {Dawson},
  {Hou}, {Kneib}, {de Mattia}, {Percival}, {Rossi}, {Tojeiro}, {Zhao}, {Zhao},
  {Alam}, {Brownstein}, {Chapman}, {Choi}, {Chuang}, {Escoffier}, {de la
  Macorra}, {du Mas des Bourboux}, {Mohammad}, {Moon}, {M{\"u}ller},
  {Nadathur}, {Newman}, {Schneider}, {Seo}, \& {Wang}}]{Bautista21}
{Bautista}, J.~E., {Paviot}, R., {Vargas Maga{\~n}a}, M., {et~al.} 2021,
  \href{http://dx.doi.org/10.1093/mnras/staa2800}{\color{black}\mnras},
  \href{https://ui.adsabs.harvard.edu/abs/2021MNRAS.500..736B}{500, 736}

\bibitem[{{Becker} \& {Rozo}(2016)}]{Becker16}
{Becker}, M.~R. \& {Rozo}, E. 2016,
  \href{http://dx.doi.org/10.1093/mnras/stv3018}{\color{black}\mnras},
  \href{https://ui.adsabs.harvard.edu/abs/2016MNRAS.457..304B}{457, 304}

\bibitem[{{Ben{\'\i}tez}(2000)}]{Benitez00}
{Ben{\'\i}tez}, N. 2000,
  \href{http://dx.doi.org/10.1086/308947}{\color{black}\apj},
  \href{https://ui.adsabs.harvard.edu/abs/2000ApJ...536..571B}{536, 571}

\bibitem[{Bishop(2006)}]{Bishop}
Bishop, C.~M. 2006, {Pattern Recognition and Machine Learning}, Information
  science and statistics (Springer), 738

\bibitem[{{Bridle} \& {King}(2007)}]{Bridle07}
{Bridle}, S. \& {King}, L. 2007,
  \href{http://dx.doi.org/10.1088/1367-2630/9/12/444}{\color{black}New Journal
  of Physics}, \href{https://ui.adsabs.harvard.edu/abs/2007NJPh....9..444B}{9,
  444}

\bibitem[{{Brout} {et~al.}(2022){Brout}, {Scolnic}, {Popovic}, {Riess}, {Carr},
  {Zuntz}, {Kessler}, {Davis}, {Hinton}, {Jones}, {Kenworthy}, {Peterson},
  {Said}, {Taylor}, {Ali}, {Armstrong}, {Charvu}, {Dwomoh}, {Meldorf},
  {Palmese}, {Qu}, {Rose}, {Sanchez}, {Stubbs}, {Vincenzi}, {Wood}, {Brown},
  {Chen}, {Chambers}, {Coulter}, {Dai}, {Dimitriadis}, {Filippenko}, {Foley},
  {Jha}, {Kelsey}, {Kirshner}, {M{\"o}ller}, {Muir}, {Nadathur}, {Pan}, {Rest},
  {Rojas-Bravo}, {Sako}, {Siebert}, {Smith}, {Stahl}, \& {Wiseman}}]{Brout22}
{Brout}, D., {Scolnic}, D., {Popovic}, B., {et~al.} 2022,
  \href{http://dx.doi.org/10.3847/1538-4357/ac8e04}{\color{black}\apj},
  \href{https://ui.adsabs.harvard.edu/abs/2022ApJ...938..110B}{938, 110}

\bibitem[{{Chaussidon} {et~al.}(2023){Chaussidon}, {Y{\`e}che},
  {Palanque-Delabrouille}, {Alexander}, {Yang}, {Ahlen}, {Bailey}, {Brooks},
  {Cai}, {Chabanier}, {Davis}, {Dawson}, {de laMacorra}, {Dey}, {Dey},
  {Eftekharzadeh}, {Eisenstein}, {Fanning}, {Font-Ribera}, {Gazta{\~n}aga}, {A
  Gontcho}, {Gonzalez-Morales}, {Guy}, {Herrera-Alcantar}, {Honscheid},
  {Ishak}, {Jiang}, {Juneau}, {Kehoe}, {Kisner}, {Kov{\'a}cs}, {Kremin}, {Lan},
  {Landriau}, {Le Guillou}, {Levi}, {Magneville}, {Martini}, {Meisner},
  {Moustakas}, {Mu{\~n}oz-Guti{\'e}rrez}, {Myers}, {Newman}, {Nie}, {Percival},
  {Poppett}, {Prada}, {Raichoor}, {Ravoux}, {Ross}, {Schlafly}, {Schlegel},
  {Tan}, {Tarl{\'e}}, {Zhou}, {Zhou}, \& {Zou}}]{Chaussidon23}
{Chaussidon}, E., {Y{\`e}che}, C., {Palanque-Delabrouille}, N., {et~al.} 2023,
  \href{http://dx.doi.org/10.3847/1538-4357/acb3c2}{\color{black}\apj},
  \href{https://ui.adsabs.harvard.edu/abs/2023ApJ...944..107C}{944, 107}

\bibitem[{{Dalal} {et~al.}(2023){Dalal}, {Li}, {Nicola}, {Zuntz}, {Strauss},
  {Sugiyama}, {Zhang}, {Rau}, {Mandelbaum}, {Takada}, {More}, {Miyatake},
  {Kannawadi}, {Shirasaki}, {Taniguchi}, {Takahashi}, {Osato}, {Hamana},
  {Oguri}, {Nishizawa}, {Malag{\'o}n}, {Sunayama}, {Alonso}, {Slosar}, {Luo},
  {Armstrong}, {Bosch}, {Hsieh}, {Komiyama}, {Lupton}, {Lust}, {MacArthur},
  {Miyazaki}, {Murayama}, {Nishimichi}, {Okura}, {Price}, {Tait}, {Tanaka}, \&
  {Wang}}]{Dalal23}
{Dalal}, R., {Li}, X., {Nicola}, A., {et~al.} 2023,
  \href{http://dx.doi.org/10.1103/PhysRevD.108.123519}{\color{black}\prd},
  \href{https://ui.adsabs.harvard.edu/abs/2023PhRvD.108l3519D}{108, 123519}

\bibitem[{{Dark Energy Survey and Kilo-Degree Survey Collaboration}
  {et~al.}(2023){Dark Energy Survey and Kilo-Degree Survey Collaboration},
  {Abbott}, {Aguena}, {Alarcon}, {Alves}, {Amon}, {Andrade-Oliveira}, {Asgari},
  {Avila}, {Bacon}, {Bechtol}, {Becker}, {Bernstein}, {Bertin}, {Bilicki},
  {Blazek}, {Bocquet}, {Brooks}, {Burger}, {Burke}, {Camacho}, {Campos},
  {Carnero Rosell}, {Carrasco Kind}, {Carretero}, {Castander}, {Cawthon},
  {Chang}, {Chen}, {Choi}, {Conselice}, {Cordero}, {Crocce}, {da Costa}, {da
  Silva Pereira}, {Dalal}, {Davis}, {de Jong}, {DeRose}, {Desai}, {Diehl},
  {Dodelson}, {Doel}, {Doux}, {Drlica-Wagner}, {Dvornik}, {Eckert}, {Eifler},
  {Elvin-Poole}, {Everett}, {Fang}, {Ferrero}, {Fert{\'e}}, {Flaugher},
  {Friedrich}, {Frieman}, {Garc{\'\i}a-Bellido}, {Gatti}, {Giannini}, {Giblin},
  {Gruen}, {Gruendl}, {Gutierrez}, {Harrison}, {Hartley}, {Herner}, {Heymans},
  {Hildebrandt}, {Hinton}, {Hoekstra}, {Hollowood}, {Honscheid}, {Huang},
  {Huff}, {Huterer}, {James}, {Jarvis}, {Jeffrey}, {Jeltema}, {Joachimi},
  {Joudaki}, {Kannawadi}, {Krause}, {Kuehn}, {Kuijken}, {Kuropatkin}, {Lahav},
  {Leget}, {Lemos}, {Li}, {Li}, {Liddle}, {Lima}, {Lin}, {Lin}, {MacCrann},
  {Mahony}, {Marshall}, {McCullough}, {Mena-Fern{\'a}ndez}, {Menanteau},
  {Miquel}, {Mohr}, {Muir}, {Myles}, {Napolitano}, {Navarro-Alsina}, {Ogando},
  {Palmese}, {Pandey}, {Park}, {Paterno}, {Peacock}, {Petravick}, {Pieres},
  {Plazas Malag{\'o}n}, {Porredon}, {Prat}, {Radovich}, {Raveri}, {Reischke},
  {Robertson}, {Rollins}, {Romer}, {Roodman}, {Rykoff}, {Samuroff},
  {S{\'a}nchez}, {Sanchez}, {Sanchez}, {Schneider}, {Secco}, {Sevilla-Noarbe},
  {Shan}, {Sheldon}, {Shin}, {Sif{\'o}n}, {Smith}, {Soares-Santos},
  {St{\"o}lzner}, {Suchyta}, {Swanson}, {Tarle}, {Thomas}, {To}, {Troxel},
  {Tr{\"o}ster}, {Tutusaus}, {van den Busch}, {Varga}, {Walker}, {Weaverdyck},
  {Wechsler}, {Weller}, {Wiseman}, {Wright}, {Yanny}, {Yin}, {Yoon}, {Zhang},
  \& {Zuntz}}]{KiDS+DES}
{Dark Energy Survey and Kilo-Degree Survey Collaboration}, {Abbott}, T.~M.~C.,
  {Aguena}, M., {et~al.} 2023,
  \href{http://dx.doi.org/10.21105/astro.2305.17173}{\color{black}The Open
  Journal of Astrophysics},
  \href{https://ui.adsabs.harvard.edu/abs/2023OJAp....6E..36D}{6, 36}

\bibitem[{{Dark Energy Survey Collaboration} {et~al.}(2016){Dark Energy Survey
  Collaboration}, {Abbott}, {Abdalla}, {Aleksi{\'c}}, {Allam}, {Amara},
  {Bacon}, {Balbinot}, {Banerji}, {Bechtol}, {Benoit-L{\'e}vy}, {Bernstein},
  {Bertin}, {Blazek}, {Bonnett}, {Bridle}, {Brooks}, {Brunner}, {Buckley-Geer},
  {Burke}, {Caminha}, {Capozzi}, {Carlsen}, {Carnero-Rosell}, {Carollo},
  {Carrasco-Kind}, {Carretero}, {Castander}, {Clerkin}, {Collett}, {Conselice},
  {Crocce}, {Cunha}, {D'Andrea}, {da Costa}, {Davis}, {Desai}, {Diehl},
  {Dietrich}, {Dodelson}, {Doel}, {Drlica-Wagner}, {Estrada}, {Etherington},
  {Evrard}, {Fabbri}, {Finley}, {Flaugher}, {Foley}, {Fosalba}, {Frieman},
  {Garc{\'\i}a-Bellido}, {Gaztanaga}, {Gerdes}, {Giannantonio}, {Goldstein},
  {Gruen}, {Gruendl}, {Guarnieri}, {Gutierrez}, {Hartley}, {Honscheid}, {Jain},
  {James}, {Jeltema}, {Jouvel}, {Kessler}, {King}, {Kirk}, {Kron}, {Kuehn},
  {Kuropatkin}, {Lahav}, {Li}, {Lima}, {Lin}, {Maia}, {Makler}, {Manera},
  {Maraston}, {Marshall}, {Martini}, {McMahon}, {Melchior}, {Merson}, {Miller},
  {Miquel}, {Mohr}, {Morice-Atkinson}, {Naidoo}, {Neilsen}, {Nichol}, {Nord},
  {Ogando}, {Ostrovski}, {Palmese}, {Papadopoulos}, {Peiris}, {Peoples},
  {Percival}, {Plazas}, {Reed}, {Refregier}, {Romer}, {Roodman}, {Ross},
  {Rozo}, {Rykoff}, {Sadeh}, {Sako}, {S{\'a}nchez}, {Sanchez}, {Santiago},
  {Scarpine}, {Schubnell}, {Sevilla-Noarbe}, {Sheldon}, {Smith}, {Smith},
  {Soares-Santos}, {Sobreira}, {Soumagnac}, {Suchyta}, {Sullivan}, {Swanson},
  {Tarle}, {Thaler}, {Thomas}, {Thomas}, {Tucker}, {Vieira}, {Vikram},
  {Walker}, {Wechsler}, {Weller}, {Wester}, {Whiteway}, {Wilcox}, {Yanny},
  {Zhang}, \& {Zuntz}}]{DES6}
{Dark Energy Survey Collaboration}, {Abbott}, T., {Abdalla}, F.~B., {et~al.}
  2016, \href{http://dx.doi.org/10.1093/mnras/stw641}{\color{black}\mnras},
  \href{https://ui.adsabs.harvard.edu/abs/2016MNRAS.460.1270D}{460, 1270}

\bibitem[{{de Jong} {et~al.}(2015){de Jong}, {Verdoes Kleijn}, {Boxhoorn},
  {Buddelmeijer}, {Capaccioli}, {Getman}, {Grado}, {Helmich}, {Huang},
  {Irisarri}, {Kuijken}, {La Barbera}, {McFarland}, {Napolitano}, {Radovich},
  {Sikkema}, {Valentijn}, {Begeman}, {Brescia}, {Cavuoti}, {Choi}, {Cordes},
  {Covone}, {Dall'Ora}, {Hildebrandt}, {Longo}, {Nakajima}, {Paolillo},
  {Puddu}, {Rifatto}, {Tortora}, {van Uitert}, {Buddendiek},
  {Harnois-D{\'e}raps}, {Erben}, {Eriksen}, {Heymans}, {Hoekstra}, {Joachimi},
  {Kitching}, {Klaes}, {Koopmans}, {K{\"o}hlinger}, {Roy}, {Sif{\'o}n},
  {Schneider}, {Sutherland}, {Viola}, \& {Vriend}}]{Dejong15}
{de Jong}, J. T.~A., {Verdoes Kleijn}, G.~A., {Boxhoorn}, D.~R., {et~al.} 2015,
  \href{http://dx.doi.org/10.1051/0004-6361/201526601}{\color{black}\aap},
  \href{https://ui.adsabs.harvard.edu/abs/2015A&A...582A..62D}{582, A62}

\bibitem[{{de Jong} {et~al.}(2017){de Jong}, {Verdoes Kleijn}, {Erben},
  {Hildebrandt}, {Kuijken}, {Sikkema}, {Brescia}, {Bilicki}, {Napolitano},
  {Amaro}, {Begeman}, {Boxhoorn}, {Buddelmeijer}, {Cavuoti}, {Getman}, {Grado},
  {Helmich}, {Huang}, {Irisarri}, {La Barbera}, {Longo}, {McFarland},
  {Nakajima}, {Paolillo}, {Puddu}, {Radovich}, {Rifatto}, {Tortora},
  {Valentijn}, {Vellucci}, {Vriend}, {Amon}, {Blake}, {Choi}, {Conti}, {Gwyn},
  {Herbonnet}, {Heymans}, {Hoekstra}, {Klaes}, {Merten}, {Miller}, {Schneider},
  \& {Viola}}]{Dejong17}
{de Jong}, J. T.~A., {Verdoes Kleijn}, G.~A., {Erben}, T., {et~al.} 2017,
  \href{http://dx.doi.org/10.1051/0004-6361/201730747}{\color{black}\aap},
  \href{https://ui.adsabs.harvard.edu/abs/2017A&A...604A.134D}{604, A134}

\bibitem[{{de Mattia} {et~al.}(2021){de Mattia}, {Ruhlmann-Kleider},
  {Raichoor}, {Ross}, {Tamone}, {Zhao}, {Alam}, {Avila}, {Burtin}, {Bautista},
  {Beutler}, {Brinkmann}, {Brownstein}, {Chapman}, {Chuang}, {Comparat}, {du
  Mas des Bourboux}, {Dawson}, {de la Macorra}, {Gil-Mar{\'\i}n},
  {Gonzalez-Perez}, {Gorgoni}, {Hou}, {Kong}, {Lin}, {Nadathur}, {Newman},
  {Mueller}, {Percival}, {Rezaie}, {Rossi}, {Schneider}, {Tiwari}, {Vivek},
  {Wang}, \& {Zhao}}]{deMattia21}
{de Mattia}, A., {Ruhlmann-Kleider}, V., {Raichoor}, A., {et~al.} 2021,
  \href{http://dx.doi.org/10.1093/mnras/staa3891}{\color{black}\mnras},
  \href{https://ui.adsabs.harvard.edu/abs/2021MNRAS.501.5616D}{501, 5616}

\bibitem[{{DESI Collaboration} {et~al.}(2022){DESI Collaboration}, {Abareshi},
  {Aguilar}, {Ahlen}, {Alam}, {Alexander}, {Alfarsy}, {Allen}, {Allende
  Prieto}, {Alves}, {Ameel}, {Armengaud}, {Asorey}, {Aviles}, {Bailey},
  {Balaguera-Antol{\'\i}nez}, {Ballester}, {Baltay}, {Bault}, {Beltran},
  {Benavides}, {BenZvi}, {Berti}, {Besuner}, {Beutler}, {Bianchi}, {Blake},
  {Blanc}, {Blum}, {Bolton}, {Bose}, {Bramall}, {Brieden}, {Brodzeller},
  {Brooks}, {Brownewell}, {Buckley-Geer}, {Cahn}, {Cai}, {Canning}, {Capasso},
  {Carnero Rosell}, {Carton}, {Casas}, {Castander}, {Cervantes-Cota},
  {Chabanier}, {Chaussidon}, {Chuang}, {Circosta}, {Cole}, {Cooper}, {da
  Costa}, {Cousinou}, {Cuceu}, {Davis}, {Dawson}, {de la Cruz-Noriega}, {de la
  Macorra}, {de Mattia}, {Della Costa}, {Demmer}, {Derwent}, {Dey}, {Dey},
  {Dhungana}, {Ding}, {Dobson}, {Doel}, {Donald-McCann}, {Donaldson},
  {Douglass}, {Duan}, {Dunlop}, {Edelstein}, {Eftekharzadeh}, {Eisenstein},
  {Enriquez-Vargas}, {Escoffier}, {Evatt}, {Fagrelius}, {Fan}, {Fanning},
  {Fawcett}, {Ferraro}, {Ereza}, {Flaugher}, {Font-Ribera}, {Forero-Romero},
  {Frenk}, {Fromenteau}, {G{\"a}nsicke}, {Garcia-Quintero}, {Garrison},
  {Gazta{\~n}aga}, {Gerardi}, {Gil-Mar{\'\i}n}, {Gontcho A Gontcho},
  {Gonzalez-Morales}, {Gonzalez-de-Rivera}, {Gonzalez-Perez}, {Gordon},
  {Graur}, {Green}, {Grove}, {Gruen}, {Gutierrez}, {Guy}, {Hahn}, {Harris},
  {Herrera}, {Herrera-Alcantar}, {Honscheid}, {Howlett}, {Huterer},
  {Ir{\v{s}}i{\v{c}}}, {Ishak}, {Jelinsky}, {Jiang}, {Jimenez}, {Jing},
  {Joyce}, {Jullo}, {Juneau}, {Kara{\c{c}}ayl{\i}}, {Karamanis}, {Karcher},
  {Karim}, {Kehoe}, {Kent}, {Kirkby}, {Kisner}, {Kitaura}, {Koposov},
  {Kov{\'a}cs}, {Kremin}, {Krolewski}, {L'Huillier}, {Lahav}, {Lambert},
  {Lamman}, {Lan}, {Landriau}, {Lane}, {Lang}, {Lange}, {Lasker}, {Le Guillou},
  {Leauthaud}, {Le Van Suu}, {Levi}, {Li}, {Magneville}, {Manera}, {Manser},
  {Marshall}, {Martini}, {McCollam}, {McDonald}, {Meisner},
  {Mena-Fern{\'a}ndez}, {Meneses-Rizo}, {Mezcua}, {Miller}, {Miquel},
  {Montero-Camacho}, {Moon}, {Moustakas}, {Mueller}, {Mu{\~n}oz-Guti{\'e}rrez},
  {Myers}, {Nadathur}, {Najita}, {Napolitano}, {Neilsen}, {Newman}, {Nie},
  {Ning}, {Niz}, {Norberg}, {Noriega}, {O'Brien}, {Obuljen},
  {Palanque-Delabrouille}, {Palmese}, {Zhiwei}, {Pappalardo}, {PENG},
  {Percival}, {Perruchot}, {Pogge}, {Poppett}, {Porredon}, {Prada},
  {Prochaska}, {Pucha}, {P{\'e}rez-Fern{\'a}ndez}, {P{\'e}rez-R{\`a}fols},
  {Rabinowitz}, \& {Raichoor}}]{Abareshi22}
{DESI Collaboration}, {Abareshi}, B., {Aguilar}, J., {et~al.} 2022,
  \href{http://dx.doi.org/10.3847/1538-3881/ac882b}{\color{black}\aj},
  \href{https://ui.adsabs.harvard.edu/abs/2022AJ....164..207D}{164, 207}

\bibitem[{{DESI Collaboration} {et~al.}(2016){DESI Collaboration}, {Aghamousa},
  {Aguilar}, {Ahlen}, {Alam}, {Allen}, {Allende Prieto}, {Annis}, {Bailey},
  {Balland}, {Ballester}, {Baltay}, {Beaufore}, {Bebek}, {Beers}, {Bell},
  {Bernal}, {Besuner}, {Beutler}, {Blake}, {Bleuler}, {Blomqvist}, {Blum},
  {Bolton}, {Briceno}, {Brooks}, {Brownstein}, {Buckley-Geer}, {Burden},
  {Burtin}, {Busca}, {Cahn}, {Cai}, {Cardiel-Sas}, {Carlberg}, {Carton},
  {Casas}, {Castander}, {Cervantes-Cota}, {Claybaugh}, {Close}, {Coker},
  {Cole}, {Comparat}, {Cooper}, {Cousinou}, {Crocce}, {Cuby}, {Cunningham},
  {Davis}, {Dawson}, {de la Macorra}, {De Vicente}, {Delubac}, {Derwent},
  {Dey}, {Dhungana}, {Ding}, {Doel}, {Duan}, {Ealet}, {Edelstein},
  {Eftekharzadeh}, {Eisenstein}, {Elliott}, {Escoffier}, {Evatt}, {Fagrelius},
  {Fan}, {Fanning}, {Farahi}, {Farihi}, {Favole}, {Feng}, {Fernandez},
  {Findlay}, {Finkbeiner}, {Fitzpatrick}, {Flaugher}, {Flender}, {Font-Ribera},
  {Forero-Romero}, {Fosalba}, {Frenk}, {Fumagalli}, {Gaensicke}, {Gallo},
  {Garcia-Bellido}, {Gaztanaga}, {Pietro Gentile Fusillo}, {Gerard},
  {Gershkovich}, {Giannantonio}, {Gillet}, {Gonzalez-de-Rivera},
  {Gonzalez-Perez}, {Gott}, {Graur}, {Gutierrez}, {Guy}, {Habib}, {Heetderks},
  {Heetderks}, {Heitmann}, {Hellwing}, {Herrera}, {Ho}, {Holland}, {Honscheid},
  {Huff}, {Hutchinson}, {Huterer}, {Hwang}, {Illa Laguna}, {Ishikawa},
  {Jacobs}, {Jeffrey}, {Jelinsky}, {Jennings}, {Jiang}, {Jimenez}, {Johnson},
  {Joyce}, {Jullo}, {Juneau}, {Kama}, {Karcher}, {Karkar}, {Kehoe}, {Kennamer},
  {Kent}, {Kilbinger}, {Kim}, {Kirkby}, {Kisner}, {Kitanidis}, {Kneib},
  {Koposov}, {Kovacs}, {Koyama}, {Kremin}, {Kron}, {Kronig}, {Kueter-Young},
  {Lacey}, {Lafever}, {Lahav}, {Lambert}, {Lampton}, {Landriau}, {Lang},
  {Lauer}, {Le Goff}, {Le Guillou}, {Le Van Suu}, {Lee}, {Lee}, {Leitner},
  {Lesser}, {Levi}, {L'Huillier}, {Li}, {Liang}, {Lin}, {Linder}, {Loebman},
  {Luki{\'c}}, {Ma}, {MacCrann}, {Magneville}, {Makarem}, {Manera}, {Manser},
  {Marshall}, {Martini}, {Massey}, {Matheson}, {McCauley}, {McDonald},
  {McGreer}, {Meisner}, {Metcalfe}, {Miller}, {Miquel}, {Moustakas}, {Myers},
  {Naik}, {Newman}, {Nichol}, {Nicola}, {Nicolati da Costa}, {Nie}, {Niz},
  {Norberg}, {Nord}, {Norman}, {Nugent}, {O'Brien}, {Oh}, {Olsen}, {Padilla},
  {Padmanabhan}, {Padmanabhan}, {Palanque-Delabrouille}, {Palmese},
  {Pappalardo}, {P{\^a}ris}, {Park}, {Patej}, {Peacock}, {Peiris}, {Peng},
  {Percival}, {Perruchot}, {Pieri}, {Pogge}, {Pollack}, {Poppett}, {Prada},
  {Prakash}, {Probst}, {Rabinowitz}, {Raichoor}, {Ree}, {Refregier}, {Regal},
  {Reid}, {Reil}, {Rezaie}, {Rockosi}, {Roe}, {Ronayette}, {Roodman}, {Ross},
  {Ross}, {Rossi}, {Rozo}, {Ruhlmann-Kleider}, {Rykoff}, {Sabiu}, {Samushia},
  {Sanchez}, {Sanchez}, {Schlegel}, {Schneider}, {Schubnell}, {Secroun},
  {Seljak}, {Seo}, {Serrano}, {Shafieloo}, {Shan}, {Sharples}, {Sholl},
  {Shourt}, {Silber}, {Silva}, {Sirk}, {Slosar}, {Smith}, {Smoot}, {Som},
  {Song}, {Sprayberry}, {Staten}, {Stefanik}, {Tarle}, {Sien Tie}, {Tinker},
  {Tojeiro}, {Valdes}, {Valenzuela}, {Valluri}, {Vargas-Magana}, {Verde},
  {Walker}, {Wang}, {Wang}, {Weaver}, {Weaverdyck}, {Wechsler}, {Weinberg},
  {White}, {Yang}, {Yeche}, {Zhang}, {Zhao}, {Zheng}, {Zhou}, {Zhou}, {Zhu},
  {Zou}, \& {Zu}}]{Aghamousa16}
{DESI Collaboration}, {Aghamousa}, A., {Aguilar}, J., {et~al.} 2016,
  \href{https://ui.adsabs.harvard.edu/abs/2016arXiv161100036D}{arXiv e-prints,
  arXiv:1611.00036}

\bibitem[{{Doux} {et~al.}(2021){Doux}, {Baxter}, {Lemos}, {Chang}, {Alarcon},
  {Amon}, {Campos}, {Choi}, {Gatti}, {Gruen}, {Jarvis}, {MacCrann}, {Park},
  {Prat}, {Rau}, {Raveri}, {Samuroff}, {DeRose}, {Hartley}, {Hoyle}, {Troxel},
  {Zuntz}, {Abbott}, {Aguena}, {Allam}, {Annis}, {Avila}, {Bacon}, {Bertin},
  {Bhargava}, {Brooks}, {Burke}, {Carrasco Kind}, {Carretero}, {Cawthon},
  {Costanzi}, {da Costa}, {Pereira}, {Desai}, {Diehl}, {Dietrich}, {Doel},
  {Everett}, {Ferrero}, {Fosalba}, {Frieman}, {Garc{\'\i}a-Bellido}, {Gerdes},
  {Giannantonio}, {Gruendl}, {Gschwend}, {Gutierrez}, {Hinton}, {Hollowood},
  {Honscheid}, {Huff}, {Huterer}, {Jain}, {James}, {Krause}, {Kuehn},
  {Kuropatkin}, {Lahav}, {Lidman}, {Lima}, {Maia}, {Menanteau}, {Miquel},
  {Morgan}, {Muir}, {Ogando}, {Palmese}, {Paz-Chinch{\'o}n}, {Plazas},
  {Sanchez}, {Scarpine}, {Schubnell}, {Serrano}, {Sevilla-Noarbe}, {Smith},
  {Suchyta}, {Swanson}, {Tarle}, {To}, {Tucker}, {Varga}, {Weller}, \&
  {Wilkinson}}]{Doux21}
{Doux}, C., {Baxter}, E., {Lemos}, P., {et~al.} 2021,
  \href{http://dx.doi.org/10.1093/mnras/stab526}{\color{black}\mnras},
  \href{https://ui.adsabs.harvard.edu/abs/2021MNRAS.503.2688D}{503, 2688}

\bibitem[{{du Mas des Bourboux} {et~al.}(2020){du Mas des Bourboux}, {Rich},
  {Font-Ribera}, {de Sainte Agathe}, {Farr}, {Etourneau}, {Le Goff}, {Cuceu},
  {Balland}, {Bautista}, {Blomqvist}, {Brinkmann}, {Brownstein}, {Chabanier},
  {Chaussidon}, {Dawson}, {Gonz{\'a}lez-Morales}, {Guy}, {Lyke}, {de la
  Macorra}, {Mueller}, {Myers}, {Nitschelm}, {Mu{\~n}oz Guti{\'e}rrez},
  {Palanque-Delabrouille}, {Parker}, {Percival}, {P{\'e}rez-R{\`a}fols},
  {Petitjean}, {Pieri}, {Ravoux}, {Rossi}, {Schneider}, {Seo}, {Slosar},
  {Stermer}, {Vivek}, {Y{\`e}che}, \& {Youles}}]{Bourboux20}
{du Mas des Bourboux}, H., {Rich}, J., {Font-Ribera}, A., {et~al.} 2020,
  \href{http://dx.doi.org/10.3847/1538-4357/abb085}{\color{black}\apj},
  \href{https://ui.adsabs.harvard.edu/abs/2020ApJ...901..153D}{901, 153}

\bibitem[{{Edge} {et~al.}(2013){Edge}, {Sutherland}, {Kuijken}, {Driver},
  {McMahon}, {Eales}, \& {Emerson}}]{Edge13}
{Edge}, A., {Sutherland}, W., {Kuijken}, K., {et~al.} 2013, The Messenger,
  \href{https://ui.adsabs.harvard.edu/abs/2013Msngr.154...32E}{154, 32}

\bibitem[{{Efstathiou} \& {Lemos}(2018)}]{Efstathiou18}
{Efstathiou}, G. \& {Lemos}, P. 2018,
  \href{http://dx.doi.org/10.1093/mnras/sty099}{\color{black}\mnras},
  \href{https://ui.adsabs.harvard.edu/abs/2018MNRAS.476..151E}{476, 151}

\bibitem[{{Euclid Collaboration} {et~al.}(2025){Euclid Collaboration},
  {Mellier}, {Abdurro'uf}, {Acevedo Barroso}, {Ach{\'u}carro}, {Adamek},
  {Adam}, {Addison}, {Aghanim}, {Aguena}, {Ajani}, {Akrami}, {Al-Bahlawan},
  {Alavi}, {Albuquerque}, {Alestas}, {Alguero}, {Allaoui}, {Allen}, {Allevato},
  {Alonso-Tetilla}, {Altieri}, {Alvarez-Candal}, {Alvi}, {Amara}, {Amendola},
  {Amiaux}, {Andika}, {Andreon}, {Andrews}, {Angora}, {Angulo}, {Annibali},
  {Anselmi}, {Anselmi}, {Arcari}, {Archidiacono}, {Aric{\`o}}, {Arnaud},
  {Arnouts}, {Asgari}, {Asorey}, {Atayde}, {Atek}, {Atrio-Barandela}, {Aubert},
  {Aubourg}, {Auphan}, {Auricchio}, {Aussel}, {Aussel}, {Avelino},
  {Avgoustidis}, {Avila}, {Awan}, {Azzollini}, {Baccigalupi}, {Bachelet},
  {Bacon}, {Baes}, {Bagley}, {Bahr-Kalus}, {Balaguera-Antolinez}, {Balbinot},
  {Balcells}, {Baldi}, {Baldry}, {Balestra}, {Ballardini}, {Ballester},
  {Balogh}, {Ba{\~n}ados}, {Barbier}, {Bardelli}, {Baron}, {Barreiro},
  {Barrena}, {Barriere}, {Barros}, {Barthelemy}, {Bartolo}, {Basset},
  {Battaglia}, {Battisti}, {Baugh}, {Baumont}, {Bazzanini}, {Beaulieu},
  {Beckmann}, {Belikov}, {Bel}, {Bellagamba}, {Bella}, {Bellini}, {Benabed},
  {Bender}, {Benevento}, {Bennett}, {Benson}, {Bergamini}, {Bermejo-Climent},
  {Bernardeau}, {Bertacca}, {Berthe}, {Berthier}, {Bethermin}, {Beutler},
  {Bevillon}, {Bhargava}, {Bhatawdekar}, {Bianchi}, {Bisigello}, {Biviano},
  {Blake}, {Blanchard}, {Blazek}, {Blot}, {Bosco}, {Bodendorf}, {Boenke},
  {B{\"o}hringer}, {Boldrini}, {Bolzonella}, {Bonchi}, {Bonici}, {Bonino},
  {Bonino}, {Bonvin}, {Bon}, {Booth}, {Borgani}, {Borlaff}, {Borsato}, {Bose},
  {Botticella}, {Boucaud}, {Bouche}, {Boucher}, {Boutigny}, {Bouvard},
  {Bouwens}, {Bouy}, {Bowler}, {Bozza}, {Bozzo}, {Branchini}, {Brando},
  {Brau-Nogue}, {Brekke}, {Bremer}, {Brescia}, {Breton}, {Brinchmann},
  {Brinckmann}, {Brockley-Blatt}, {Brodwin}, {Brouard}, {Brown}, {Bruton},
  {Bucko}, {Buddelmeijer}, {Buenadicha}, {Buitrago}, {Burger}, {Burigana},
  {Busillo}, {Busonero}, {Cabanac}, {Cabayol-Garcia}, {Cagliari}, {Caillat},
  {Caillat}, {Calabrese}, {Calabro}, {Calderone}, {Calura}, {Camacho Quevedo},
  {Camera}, {Campos}, {Ca{\~n}as-Herrera}, {Candini}, {Cantiello},
  {Capobianco}, {Cappellaro}, {Cappelluti}, {Cappi}, {Caputi}, {Cara},
  {Carbone}, {Cardone}, {Carella}, {Carlberg}, {Carle}, {Carminati}, {Caro},
  {Carrasco}, {Carretero}, {Carrilho}, {Carron Duque}, \& {Carry}}]{Euclid24}
{Euclid Collaboration}, {Mellier}, Y., {Abdurro'uf}, {et~al.} 2025,
  \href{http://dx.doi.org/10.1051/0004-6361/202450810}{\color{black}\aap},
  \href{https://ui.adsabs.harvard.edu/abs/2025A&A...697A...1E}{697, A1}

\bibitem[{{Fenech Conti} {et~al.}(2017){Fenech Conti}, {Herbonnet}, {Hoekstra},
  {Merten}, {Miller}, \& {Viola}}]{Fenech17}
{Fenech Conti}, I., {Herbonnet}, R., {Hoekstra}, H., {et~al.} 2017,
  \href{http://dx.doi.org/10.1093/mnras/stx200}{\color{black}\mnras},
  \href{https://ui.adsabs.harvard.edu/abs/2017MNRAS.467.1627F}{467, 1627}

\bibitem[{{Flaugher} {et~al.}(2015){Flaugher}, {Diehl}, {Honscheid}, {Abbott},
  {Alvarez}, {Angstadt}, {Annis}, {Antonik}, {Ballester}, {Beaufore},
  {Bernstein}, {Bernstein}, {Bigelow}, {Bonati}, {Boprie}, {Brooks},
  {Buckley-Geer}, {Campa}, {Cardiel-Sas}, {Castander}, {Castilla}, {Cease},
  {Cela-Ruiz}, {Chappa}, {Chi}, {Cooper}, {da Costa}, {Dede}, {Derylo},
  {DePoy}, {de Vicente}, {Doel}, {Drlica-Wagner}, {Eiting}, {Elliott}, {Emes},
  {Estrada}, {Fausti Neto}, {Finley}, {Flores}, {Frieman}, {Gerdes},
  {Gladders}, {Gregory}, {Gutierrez}, {Hao}, {Holland}, {Holm}, {Huffman},
  {Jackson}, {James}, {Jonas}, {Karcher}, {Karliner}, {Kent}, {Kessler},
  {Kozlovsky}, {Kron}, {Kubik}, {Kuehn}, {Kuhlmann}, {Kuk}, {Lahav}, {Lathrop},
  {Lee}, {Levi}, {Lewis}, {Li}, {Mandrichenko}, {Marshall}, {Martinez},
  {Merritt}, {Miquel}, {Mu{\~n}oz}, {Neilsen}, {Nichol}, {Nord}, {Ogando},
  {Olsen}, {Palaio}, {Patton}, {Peoples}, {Plazas}, {Rauch}, {Reil}, {Rheault},
  {Roe}, {Rogers}, {Roodman}, {Sanchez}, {Scarpine}, {Schindler}, {Schmidt},
  {Schmitt}, {Schubnell}, {Schultz}, {Schurter}, {Scott}, {Serrano}, {Shaw},
  {Smith}, {Soares-Santos}, {Stefanik}, {Stuermer}, {Suchyta}, {Sypniewski},
  {Tarle}, {Thaler}, {Tighe}, {Tran}, {Tucker}, {Walker}, {Wang}, {Watson},
  {Weaverdyck}, {Wester}, {Woods}, {Yanny}, \& {DES Collaboration}}]{DES15}
{Flaugher}, B., {Diehl}, H.~T., {Honscheid}, K., {et~al.} 2015,
  \href{http://dx.doi.org/10.1088/0004-6256/150/5/150}{\color{black}\aj},
  \href{https://ui.adsabs.harvard.edu/abs/2015AJ....150..150F}{150, 150}

\bibitem[{{Fortuna} {et~al.}(2025){Fortuna}, {Dvornik}, {Hoekstra}, {Chisari},
  {Asgari}, {Bilicki}, {Heymans}, {Hildebrandt}, {Kuijken}, {Wright}, \&
  {Yao}}]{Fortuna24}
{Fortuna}, M.~C., {Dvornik}, A., {Hoekstra}, H., {et~al.} 2025,
  \href{http://dx.doi.org/10.1051/0004-6361/202452347}{\color{black}\aap},
  \href{https://ui.adsabs.harvard.edu/abs/2025A&A...694A.322C}{694, A322}

\bibitem[{{Fortuna} {et~al.}(2021){Fortuna}, {Hoekstra}, {Joachimi},
  {Johnston}, {Chisari}, {Georgiou}, \& {Mahony}}]{Fortuna20}
{Fortuna}, M.~C., {Hoekstra}, H., {Joachimi}, B., {et~al.} 2021,
  \href{http://dx.doi.org/10.1093/mnras/staa3802}{\color{black}\mnras},
  \href{https://ui.adsabs.harvard.edu/abs/2021MNRAS.501.2983F}{501, 2983}

\bibitem[{{Garc{\'\i}a-Garc{\'\i}a} {et~al.}(2024){Garc{\'\i}a-Garc{\'\i}a},
  {Zennaro}, {Aric{\`o}}, {Alonso}, \& {Angulo}}]{Garcia24}
{Garc{\'\i}a-Garc{\'\i}a}, C., {Zennaro}, M., {Aric{\`o}}, G., {Alonso}, D., \&
  {Angulo}, R.~E. 2024,
  \href{http://dx.doi.org/10.1088/1475-7516/2024/08/024}{\color{black}\jcap},
  \href{https://ui.adsabs.harvard.edu/abs/2024JCAP...08..024G}{2024, 024}

\bibitem[{{Georgiou} {et~al.}(2025){Georgiou}, {Chisari}, {Bilicki}, {La
  Barbera}, {Napolitano}, {Roy}, \& {Tortora}}]{Georgiou25}
{Georgiou}, C., {Chisari}, N.~E., {Bilicki}, M., {et~al.} 2025,
  \href{http://dx.doi.org/10.1051/0004-6361/202554134}{\color{black}\aap},
  \href{https://ui.adsabs.harvard.edu/abs/2025A&A...699A.252G}{699, A252}

\bibitem[{{Georgiou} {et~al.}(2019){Georgiou}, {Johnston}, {Hoekstra}, {Viola},
  {Kuijken}, {Joachimi}, {Chisari}, {Farrow}, {Hildebrandt}, {Holwerda}, \&
  {Kannawadi}}]{Georgiou19}
{Georgiou}, C., {Johnston}, H., {Hoekstra}, H., {et~al.} 2019,
  \href{http://dx.doi.org/10.1051/0004-6361/201834219}{\color{black}\aap},
  \href{https://ui.adsabs.harvard.edu/abs/2019A&A...622A..90G}{622, A90}

\bibitem[{{Gil-Mar{\'\i}n} {et~al.}(2020){Gil-Mar{\'\i}n}, {Bautista},
  {Paviot}, {Vargas-Maga{\~n}a}, {de la Torre}, {Fromenteau}, {Alam},
  {{\'A}vila}, {Burtin}, {Chuang}, {Dawson}, {Hou}, {de Mattia}, {Mohammad},
  {M{\"u}ller}, {Nadathur}, {Neveux}, {Percival}, {Raichoor}, {Rezaie}, {Ross},
  {Rossi}, {Ruhlmann-Kleider}, {Smith}, {Tamone}, {Tinker}, {Tojeiro}, {Wang},
  {Zhao}, {Zhao}, {Brinkmann}, {Brownstein}, {Choi}, {Escoffier}, {de la
  Macorra}, {Moon}, {Newman}, {Schneider}, {Seo}, \& {Vivek}}]{GilMarin20}
{Gil-Mar{\'\i}n}, H., {Bautista}, J.~E., {Paviot}, R., {et~al.} 2020,
  \href{http://dx.doi.org/10.1093/mnras/staa2455}{\color{black}\mnras},
  \href{https://ui.adsabs.harvard.edu/abs/2020MNRAS.498.2492G}{498, 2492}

\bibitem[{{Gong} {et~al.}(2019){Gong}, {Liu}, {Cao}, {Chen}, {Fan}, {Li}, {Li},
  {Li}, {Zhang}, \& {Zhan}}]{Gong19}
{Gong}, Y., {Liu}, X., {Cao}, Y., {et~al.} 2019,
  \href{http://dx.doi.org/10.3847/1538-4357/ab391e}{\color{black}\apj},
  \href{https://ui.adsabs.harvard.edu/abs/2019ApJ...883..203G}{883, 203}

\bibitem[{{Hahn} {et~al.}(2023){Hahn}, {Wilson}, {Ruiz-Macias}, {Cole},
  {Weinberg}, {Moustakas}, {Kremin}, {Tinker}, {Smith}, {Wechsler}, {Ahlen},
  {Alam}, {Bailey}, {Brooks}, {Cooper}, {Davis}, {Dawson}, {Dey}, {Dey},
  {Eftekharzadeh}, {Eisenstein}, {Fanning}, {Forero-Romero}, {Frenk},
  {Gazta{\~n}aga}, {A Gontcho}, {Guy}, {Honscheid}, {Ishak}, {Juneau}, {Kehoe},
  {Kisner}, {Lan}, {Landriau}, {Le Guillou}, {Levi}, {Magneville}, {Martini},
  {Meisner}, {Myers}, {Nie}, {Norberg}, {Palanque-Delabrouille}, {Percival},
  {Poppett}, {Prada}, {Raichoor}, {Ross}, {Gaines}, {Saulder}, {Schlafly},
  {Schlegel}, {Sierra-Porta}, {Tarle}, {Weaver}, {Y{\`e}che}, {Zarrouk},
  {Zhou}, {Zhou}, \& {Zou}}]{Hahn23}
{Hahn}, C., {Wilson}, M.~J., {Ruiz-Macias}, O., {et~al.} 2023,
  \href{http://dx.doi.org/10.3847/1538-3881/accff8}{\color{black}\aj},
  \href{https://ui.adsabs.harvard.edu/abs/2023AJ....165..253H}{165, 253}

\bibitem[{{Handley} \& {Lemos}(2019{\natexlab{a}})}]{handley19}
{Handley}, W. \& {Lemos}, P. 2019{\natexlab{a}},
  \href{http://dx.doi.org/10.1103/PhysRevD.100.023512}{\color{black}\prd},
  \href{https://ui.adsabs.harvard.edu/abs/2019PhRvD.100b3512H}{100, 023512}

\bibitem[{{Handley} \& {Lemos}(2019{\natexlab{b}})}]{Handley19b}
{Handley}, W. \& {Lemos}, P. 2019{\natexlab{b}},
  \href{http://dx.doi.org/10.1103/PhysRevD.100.043504}{\color{black}\prd},
  \href{https://ui.adsabs.harvard.edu/abs/2019PhRvD.100d3504H}{100, 043504}

\bibitem[{Harris {et~al.}(2020)Harris, Millman, van~der Walt, Gommers,
  Virtanen, Cournapeau, Wieser, Taylor, Berg, Smith, Kern, Picus, Hoyer, van
  Kerkwijk, Brett, Haldane, del R{\'{i}}o, Wiebe, Peterson,
  G{\'{e}}rard-Marchant, Sheppard, Reddy, Weckesser, Abbasi, Gohlke, \&
  Oliphant}]{Numpy}
Harris, C.~R., Millman, K.~J., van~der Walt, S.~J., {et~al.} 2020,
  \href{http://dx.doi.org/10.1038/s41586-020-2649-2}{\color{black}Nature}, 585,
  357

\bibitem[{{Heymans} {et~al.}(2013){Heymans}, {Grocutt}, {Heavens}, {Kilbinger},
  {Kitching}, {Simpson}, {Benjamin}, {Erben}, {Hildebrandt}, {Hoekstra},
  {Mellier}, {Miller}, {Van Waerbeke}, {Brown}, {Coupon}, {Fu},
  {Harnois-D{\'e}raps}, {Hudson}, {Kuijken}, {Rowe}, {Schrabback}, {Semboloni},
  {Vafaei}, \& {Velander}}]{Heymans13}
{Heymans}, C., {Grocutt}, E., {Heavens}, A., {et~al.} 2013,
  \href{http://dx.doi.org/10.1093/mnras/stt601}{\color{black}\mnras},
  \href{https://ui.adsabs.harvard.edu/abs/2013MNRAS.432.2433H}{432, 2433}

\bibitem[{{Heymans} {et~al.}(2021){Heymans}, {Tr{\"o}ster}, {Asgari}, {Blake},
  {Hildebrandt}, {Joachimi}, {Kuijken}, {Lin}, {S{\'a}nchez}, {van den Busch},
  {Wright}, {Amon}, {Bilicki}, {de Jong}, {Crocce}, {Dvornik}, {Erben},
  {Fortuna}, {Getman}, {Giblin}, {Glazebrook}, {Hoekstra}, {Joudaki},
  {Kannawadi}, {K{\"o}hlinger}, {Lidman}, {Miller}, {Napolitano}, {Parkinson},
  {Schneider}, {Shan}, {Valentijn}, {Verdoes Kleijn}, \& {Wolf}}]{Heymans21}
{Heymans}, C., {Tr{\"o}ster}, T., {Asgari}, M., {et~al.} 2021,
  \href{http://dx.doi.org/10.1051/0004-6361/202039063}{\color{black}\aap},
  \href{https://ui.adsabs.harvard.edu/abs/2021A&A...646A.140H}{646, A140}

\bibitem[{{Hildebrandt} {et~al.}(2021){Hildebrandt}, {van den Busch}, {Wright},
  {Blake}, {Joachimi}, {Kuijken}, {Tr{\"o}ster}, {Asgari}, {Bilicki}, {de
  Jong}, {Dvornik}, {Erben}, {Getman}, {Giblin}, {Heymans}, {Kannawadi}, {Lin},
  \& {Shan}}]{Hildebrandt21}
{Hildebrandt}, H., {van den Busch}, J.~L., {Wright}, A.~H., {et~al.} 2021,
  \href{http://dx.doi.org/10.1051/0004-6361/202039018}{\color{black}\aap},
  \href{https://ui.adsabs.harvard.edu/abs/2021A&A...647A.124H}{647, A124}

\bibitem[{{Hinton}(2016)}]{Chainconsumer}
{Hinton}, S.~R. 2016,
  \href{http://dx.doi.org/10.21105/joss.00045}{\color{black}The Journal of Open
  Source Software},
  \href{https://ui.adsabs.harvard.edu/abs/2016JOSS....1...45H}{1, 00045}

\bibitem[{{Hirata} {et~al.}(2007){Hirata}, {Mandelbaum}, {Ishak}, {Seljak},
  {Nichol}, {Pimbblet}, {Ross}, \& {Wake}}]{Hirata07}
{Hirata}, C.~M., {Mandelbaum}, R., {Ishak}, M., {et~al.} 2007,
  \href{http://dx.doi.org/10.1111/j.1365-2966.2007.12312.x}{\color{black}\mnras},
  \href{https://ui.adsabs.harvard.edu/abs/2007MNRAS.381.1197H}{381, 1197}

\bibitem[{{Hou} {et~al.}(2021){Hou}, {S{\'a}nchez}, {Ross}, {Smith}, {Neveux},
  {Bautista}, {Burtin}, {Zhao}, {Scoccimarro}, {Dawson}, {de Mattia}, {de la
  Macorra}, {du Mas des Bourboux}, {Eisenstein}, {Gil-Mar{\'\i}n}, {Lyke},
  {Mohammad}, {Mueller}, {Percival}, {Rossi}, {Vargas Maga{\~n}a}, {Zarrouk},
  {Zhao}, {Brinkmann}, {Brownstein}, {Chuang}, {Myers}, {Newman}, {Schneider},
  \& {Vivek}}]{Hou21}
{Hou}, J., {S{\'a}nchez}, A.~G., {Ross}, A.~J., {et~al.} 2021,
  \href{http://dx.doi.org/10.1093/mnras/staa3234}{\color{black}\mnras},
  \href{https://ui.adsabs.harvard.edu/abs/2021MNRAS.500.1201H}{500, 1201}

\bibitem[{{Howlett} {et~al.}(2012){Howlett}, {Lewis}, {Hall}, \&
  {Challinor}}]{Howlett12}
{Howlett}, C., {Lewis}, A., {Hall}, A., \& {Challinor}, A. 2012,
  \href{http://dx.doi.org/10.1088/1475-7516/2012/04/027}{\color{black}\jcap},
  \href{https://ui.adsabs.harvard.edu/abs/2012JCAP...04..027H}{2012, 027}

\bibitem[{{Howlett} {et~al.}(2015){Howlett}, {Ross}, {Samushia}, {Percival}, \&
  {Manera}}]{Howlett15}
{Howlett}, C., {Ross}, A.~J., {Samushia}, L., {Percival}, W.~J., \& {Manera},
  M. 2015, \href{http://dx.doi.org/10.1093/mnras/stu2693}{\color{black}\mnras},
  \href{https://ui.adsabs.harvard.edu/abs/2015MNRAS.449..848H}{449, 848}

\bibitem[{{Hu}(1999)}]{Hu99}
{Hu}, W. 1999, \href{http://dx.doi.org/10.1086/312210}{\color{black}\apjl},
  \href{https://ui.adsabs.harvard.edu/abs/1999ApJ...522L..21H}{522, L21}

\bibitem[{Hunter(2007)}]{Matplotlib}
Hunter, J.~D. 2007,
  \href{http://dx.doi.org/10.1109/MCSE.2007.55}{\color{black}Computing in
  Science \& Engineering}, 9, 90

\bibitem[{{Ivezi{\'c}} {et~al.}(2019){Ivezi{\'c}}, {Kahn}, {Tyson}, {Abel},
  {Acosta}, {Allsman}, {Alonso}, {AlSayyad}, {Anderson}, {Andrew}, {Angel},
  {Angeli}, {Ansari}, {Antilogus}, {Araujo}, {Armstrong}, {Arndt}, {Astier},
  {Aubourg}, {Auza}, {Axelrod}, {Bard}, {Barr}, {Barrau}, {Bartlett}, {Bauer},
  {Bauman}, {Baumont}, {Bechtol}, {Bechtol}, {Becker}, {Becla}, {Beldica},
  {Bellavia}, {Bianco}, {Biswas}, {Blanc}, {Blazek}, {Blandford}, {Bloom},
  {Bogart}, {Bond}, {Booth}, {Borgland}, {Borne}, {Bosch}, {Boutigny},
  {Brackett}, {Bradshaw}, {Brandt}, {Brown}, {Bullock}, {Burchat}, {Burke},
  {Cagnoli}, {Calabrese}, {Callahan}, {Callen}, {Carlin}, {Carlson},
  {Chandrasekharan}, {Charles-Emerson}, {Chesley}, {Cheu}, {Chiang}, {Chiang},
  {Chirino}, {Chow}, {Ciardi}, {Claver}, {Cohen-Tanugi}, {Cockrum}, {Coles},
  {Connolly}, {Cook}, {Cooray}, {Covey}, {Cribbs}, {Cui}, {Cutri}, {Daly},
  {Daniel}, {Daruich}, {Daubard}, {Daues}, {Dawson}, {Delgado}, {Dellapenna},
  {de Peyster}, {de Val-Borro}, {Digel}, {Doherty}, {Dubois},
  {Dubois-Felsmann}, {Durech}, {Economou}, {Eifler}, {Eracleous}, {Emmons},
  {Fausti Neto}, {Ferguson}, {Figueroa}, {Fisher-Levine}, {Focke}, {Foss},
  {Frank}, {Freemon}, {Gangler}, {Gawiser}, {Geary}, {Gee}, {Geha}, {Gessner},
  {Gibson}, {Gilmore}, {Glanzman}, {Glick}, {Goldina}, {Goldstein}, {Goodenow},
  {Graham}, {Gressler}, {Gris}, {Guy}, {Guyonnet}, {Haller}, {Harris},
  {Hascall}, {Haupt}, {Hernandez}, {Herrmann}, {Hileman}, {Hoblitt}, {Hodgson},
  {Hogan}, {Howard}, {Huang}, {Huffer}, {Ingraham}, {Innes}, {Jacoby}, {Jain},
  {Jammes}, {Jee}, {Jenness}, {Jernigan}, {Jevremovi{\'c}}, {Johns}, {Johnson},
  {Johnson}, {Jones}, {Juramy-Gilles}, {Juri{\'c}}, {Kalirai}, {Kallivayalil},
  {Kalmbach}, {Kantor}, {Karst}, {Kasliwal}, {Kelly}, {Kessler}, {Kinnison},
  {Kirkby}, {Knox}, {Kotov}, {Krabbendam}, {Krughoff}, {Kub{\'a}nek},
  {Kuczewski}, {Kulkarni}, {Ku}, {Kurita}, {Lage}, {Lambert}, {Lange},
  {Langton}, {Le Guillou}, {Levine}, {Liang}, {Lim}, {Lintott}, {Long},
  {Lopez}, {Lotz}, {Lupton}, {Lust}, {MacArthur}, {Mahabal}, {Mandelbaum},
  {Markiewicz}, {Marsh}, {Marshall}, {Marshall}, {May}, {McKercher}, {McQueen},
  {Meyers}, {Migliore}, {Miller}, {Mills}, {Miraval}, {Moeyens}, {Moolekamp},
  {Monet}, {Moniez}, {Monkewitz}, {Montgomery}, {Morrison}, {Mueller},
  {Muller}, {Mu{\~n}oz Arancibia}, {Neill}, {Newbry}, {Nief}, {Nomerotski},
  {Nordby}, {O'Connor}, {Oliver}, {Olivier}, {Olsen}, {O'Mullane}, {Ortiz},
  {Osier}, {Owen}, {Pain}, {Palecek}, {Parejko}, {Parsons}, {Pease},
  {Peterson}, {Peterson}, {Petravick}, {Libby Petrick}, {Petry},
  {Pierfederici}, {Pietrowicz}, {Pike}, {Pinto}, {Plante}, {Plate}, {Plutchak},
  {Price}, {Prouza}, {Radeka}, {Rajagopal}, {Rasmussen}, {Regnault}, {Reil},
  {Reiss}, {Reuter}, {Ridgway}, {Riot}, {Ritz}, {Robinson}, {Roby}, {Roodman},
  {Rosing}, {Roucelle}, {Rumore}, {Russo}, {Saha}, {Sassolas}, {Schalk},
  {Schellart}, {Schindler}, {Schmidt}, {Schneider}, {Schneider}, {Schoening},
  {Schumacher}, {Schwamb}, {Sebag}, {Selvy}, {Sembroski}, {Seppala}, {Serio},
  {Serrano}, {Shaw}, {Shipsey}, {Sick}, {Silvestri}, {Slater}, {Smith},
  {Smith}, {Sobhani}, {Soldahl}, {Storrie-Lombardi}, {Stover}, {Strauss},
  {Street}, {Stubbs}, {Sullivan}, {Sweeney}, {Swinbank}, {Szalay}, {Takacs},
  {Tether}, {Thaler}, {Thayer}, {Thomas}, {Thornton}, {Thukral}, {Tice},
  {Trilling}, {Turri}, {Van Berg}, {Vanden Berk}, {Vetter}, {Virieux},
  {Vucina}, {Wahl}, {Walkowicz}, {Walsh}, {Walter}, {Wang}, {Wang}, {Warner},
  {Wiecha}, {Willman}, {Winters}, {Wittman}, {Wolff}, {Wood-Vasey}, {Wu},
  {Xin}, {Yoachim}, \& {Zhan}}]{LSST}
{Ivezi{\'c}}, {\v{Z}}., {Kahn}, S.~M., {Tyson}, J.~A., {et~al.} 2019,
  \href{http://dx.doi.org/10.3847/1538-4357/ab042c}{\color{black}\apj},
  \href{https://ui.adsabs.harvard.edu/abs/2019ApJ...873..111I}{873, 111}

\bibitem[{Jeffreys(1939)}]{Jeffreys39}
Jeffreys, H. 1939, {The Theory of Probability}, Oxford Classic Texts in the
  Physical Sciences (OUP Oxford)

\bibitem[{{Joachimi} {et~al.}(2021){Joachimi}, {Lin}, {Asgari}, {Tr{\"o}ster},
  {Heymans}, {Hildebrandt}, {K{\"o}hlinger}, {S{\'a}nchez}, {Wright},
  {Bilicki}, {Blake}, {van den Busch}, {Crocce}, {Dvornik}, {Erben}, {Getman},
  {Giblin}, {Hoekstra}, {Kannawadi}, {Kuijken}, {Napolitano}, {Schneider},
  {Scoccimarro}, {Sellentin}, {Shan}, {von Wietersheim-Kramsta}, \&
  {Zuntz}}]{Joachimi21}
{Joachimi}, B., {Lin}, C.~A., {Asgari}, M., {et~al.} 2021,
  \href{http://dx.doi.org/10.1051/0004-6361/202038831}{\color{black}\aap},
  \href{https://ui.adsabs.harvard.edu/abs/2021A&A...646A.129J}{646, A129}

\bibitem[{{Joachimi} {et~al.}(2011){Joachimi}, {Mandelbaum}, {Abdalla}, \&
  {Bridle}}]{Joachimi11}
{Joachimi}, B., {Mandelbaum}, R., {Abdalla}, F.~B., \& {Bridle}, S.~L. 2011,
  \href{http://dx.doi.org/10.1051/0004-6361/201015621}{\color{black}\aap},
  \href{https://ui.adsabs.harvard.edu/abs/2011A&A...527A..26J}{527, A26}

\bibitem[{{Johnston} {et~al.}(2019){Johnston}, {Georgiou}, {Joachimi},
  {Hoekstra}, {Chisari}, {Farrow}, {Fortuna}, {Heymans}, {Joudaki}, {Kuijken},
  \& {Wright}}]{Johnston19}
{Johnston}, H., {Georgiou}, C., {Joachimi}, B., {et~al.} 2019,
  \href{http://dx.doi.org/10.1051/0004-6361/201834714}{\color{black}\aap},
  \href{https://ui.adsabs.harvard.edu/abs/2019A&A...624A..30J}{624, A30}

\bibitem[{{Kaiser}(1992)}]{Kaiser92}
{Kaiser}, N. 1992, \href{http://dx.doi.org/10.1086/171151}{\color{black}\apj},
  \href{https://ui.adsabs.harvard.edu/abs/1992ApJ...388..272K}{388, 272}

\bibitem[{{Kaiser} {et~al.}(2000){Kaiser}, {Wilson}, \& {Luppino}}]{Kaiser00}
{Kaiser}, N., {Wilson}, G., \& {Luppino}, G.~A. 2000,
  \href{https://ui.adsabs.harvard.edu/abs/2000astro.ph..3338K}{arXiv e-prints,
  astro}

\bibitem[{{Kannawadi} {et~al.}(2019){Kannawadi}, {Hoekstra}, {Miller}, {Viola},
  {Fenech Conti}, {Herbonnet}, {Erben}, {Heymans}, {Hildebrandt}, {Kuijken},
  {Vakili}, \& {Wright}}]{Kannawadi19}
{Kannawadi}, A., {Hoekstra}, H., {Miller}, L., {et~al.} 2019,
  \href{http://dx.doi.org/10.1051/0004-6361/201834819}{\color{black}\aap},
  \href{https://ui.adsabs.harvard.edu/abs/2019A&A...624A..92K}{624, A92}

\bibitem[{{Kilbinger} {et~al.}(2017){Kilbinger}, {Heymans}, {Asgari},
  {Joudaki}, {Schneider}, {Simon}, {Van Waerbeke}, {Harnois-D{\'e}raps},
  {Hildebrandt}, {K{\"o}hlinger}, {Kuijken}, \& {Viola}}]{Kilbinger17}
{Kilbinger}, M., {Heymans}, C., {Asgari}, M., {et~al.} 2017,
  \href{http://dx.doi.org/10.1093/mnras/stx2082}{\color{black}\mnras},
  \href{https://ui.adsabs.harvard.edu/abs/2017MNRAS.472.2126K}{472, 2126}

\bibitem[{{K{\"o}hlinger} {et~al.}(2019){K{\"o}hlinger}, {Joachimi}, {Asgari},
  {Viola}, {Joudaki}, \& {Tr{\"o}ster}}]{Koehlinger19}
{K{\"o}hlinger}, F., {Joachimi}, B., {Asgari}, M., {et~al.} 2019,
  \href{http://dx.doi.org/10.1093/mnras/stz132}{\color{black}\mnras},
  \href{https://ui.adsabs.harvard.edu/abs/2019MNRAS.484.3126K}{484, 3126}

\bibitem[{{Krause} {et~al.}(2021){Krause}, {Fang}, {Pandey}, {Secco}, {Alves},
  {Huang}, {Blazek}, {Prat}, {Zuntz}, {Eifler}, {MacCrann}, {DeRose}, {Crocce},
  {Porredon}, {Jain}, {Troxel}, {Dodelson}, {Huterer}, {Liddle}, {Leonard},
  {Amon}, {Chen}, {Elvin-Poole}, {Fert{\'e}}, {Muir}, {Park}, {Samuroff},
  {Brandao-Souza}, {Weaverdyck}, {Zacharegkas}, {Rosenfeld}, {Campos},
  {Chintalapati}, {Choi}, {Di Valentino}, {Doux}, {Herner}, {Lemos},
  {Mena-Fern{\'a}ndez}, {Omori}, {Paterno}, {Rodriguez-Monroy}, {Rogozenski},
  {Rollins}, {Troja}, {Tutusaus}, {Wechsler}, {Abbott}, {Aguena}, {Allam},
  {Andrade-Oliveira}, {Annis}, {Bacon}, {Baxter}, {Bechtol}, {Bernstein},
  {Brooks}, {Buckley-Geer}, {Burke}, {Carnero Rosell}, {Carrasco Kind},
  {Carretero}, {Castander}, {Cawthon}, {Chang}, {Costanzi}, {da Costa},
  {Pereira}, {De Vicente}, {Desai}, {Diehl}, {Doel}, {Everett}, {Evrard},
  {Ferrero}, {Flaugher}, {Fosalba}, {Frieman}, {Garc{\'\i}a-Bellido},
  {Gaztanaga}, {Gerdes}, {Giannantonio}, {Gruen}, {Gruendl}, {Gschwend},
  {Gutierrez}, {Hartley}, {Hinton}, {Hollowood}, {Honscheid}, {Hoyle}, {Huff},
  {James}, {Kuehn}, {Kuropatkin}, {Lahav}, {Lima}, {Maia}, {Marshall},
  {Martini}, {Melchior}, {Menanteau}, {Miquel}, {Mohr}, {Morgan}, {Myles},
  {Palmese}, {Paz-Chinch{\'o}n}, {Petravick}, {Pieres}, {Plazas Malag{\'o}n},
  {Sanchez}, {Scarpine}, {Schubnell}, {Serrano}, {Sevilla-Noarbe}, {Smith},
  {Soares-Santos}, {Suchyta}, {Tarle}, {Thomas}, {To}, {Varga}, \&
  {Weller}}]{Krause21}
{Krause}, E., {Fang}, X., {Pandey}, S., {et~al.} 2021,
  \href{https://ui.adsabs.harvard.edu/abs/2021arXiv210513548K}{\href{http://dx.doi.org/10.48550/arXiv.2105.13548}{\color{black}arXiv
  e-prints}, arXiv:2105.13548}

\bibitem[{{Kuijken} {et~al.}(2019){Kuijken}, {Heymans}, {Dvornik},
  {Hildebrandt}, {de Jong}, {Wright}, {Erben}, {Bilicki}, {Giblin}, {Shan},
  {Getman}, {Grado}, {Hoekstra}, {Miller}, {Napolitano}, {Paolilo}, {Radovich},
  {Schneider}, {Sutherland }, {Tewes}, {Tortora}, {Valentijn}, \& {Verdoes
  Kleijn}}]{Kuijken19}
{Kuijken}, K., {Heymans}, C., {Dvornik}, A., {et~al.} 2019,
  \href{http://dx.doi.org/10.1051/0004-6361/201834918}{\color{black}\aap},
  \href{https://ui.adsabs.harvard.edu/abs/2019A&A...625A...2K}{625, A2}

\bibitem[{{Kuijken} {et~al.}(2015){Kuijken}, {Heymans}, {Hildebrandt},
  {Nakajima}, {Erben}, {de Jong}, {Viola}, {Choi}, {Hoekstra}, {Miller}, {van
  Uitert}, {Amon}, {Blake}, {Brouwer}, {Buddendiek}, {Conti}, {Eriksen},
  {Grado}, {Harnois-D{\'e}raps}, {Helmich}, {Herbonnet}, {Irisarri},
  {Kitching}, {Klaes}, {La Barbera}, {Napolitano}, {Radovich}, {Schneider},
  {Sif{\'o}n}, {Sikkema}, {Simon}, {Tudorica}, {Valentijn}, {Verdoes Kleijn},
  \& {van Waerbeke}}]{Kuijken15}
{Kuijken}, K., {Heymans}, C., {Hildebrandt}, H., {et~al.} 2015,
  \href{http://dx.doi.org/10.1093/mnras/stv2140}{\color{black}\mnras},
  \href{https://ui.adsabs.harvard.edu/abs/2015MNRAS.454.3500K}{454, 3500}

\bibitem[{Kullback \& Leibler(1951)}]{Kullback51}
Kullback, S. \& Leibler, R.~A. 1951,
  \href{http://dx.doi.org/10.1214/aoms/1177729694}{\color{black}The Annals of
  Mathematical Statistics}, 22, 79

\bibitem[{{Lange}(2023)}]{Lange23}
{Lange}, J.~U. 2023,
  \href{http://dx.doi.org/10.1093/mnras/stad2441}{\color{black}\mnras},
  \href{https://ui.adsabs.harvard.edu/abs/2023MNRAS.525.3181L}{525, 3181}

\bibitem[{{Lewis} {et~al.}(2000){Lewis}, {Challinor}, \& {Lasenby}}]{Lewis00}
{Lewis}, A., {Challinor}, A., \& {Lasenby}, A. 2000,
  \href{http://dx.doi.org/10.1086/309179}{\color{black}\apj},
  \href{https://ui.adsabs.harvard.edu/abs/2000ApJ...538..473L}{538, 473}

\bibitem[{{Li} {et~al.}(2021){Li}, {Kuijken}, {Hoekstra}, {Hildebrandt},
  {Joachimi}, \& {Kannawadi}}]{Li21}
{Li}, S.-S., {Kuijken}, K., {Hoekstra}, H., {et~al.} 2021,
  \href{http://dx.doi.org/10.1051/0004-6361/202039254}{\color{black}\aap},
  \href{https://ui.adsabs.harvard.edu/abs/2021A&A...646A.175L}{646, A175}

\bibitem[{{Li} {et~al.}(2023{\natexlab{a}}){Li}, {Kuijken}, {Hoekstra},
  {Miller}, {Heymans}, {Hildebrandt}, {van den Busch}, {Wright}, {Yoon},
  {Bilicki}, {Bravo}, \& {Lagos}}]{Li23b}
{Li}, S.-S., {Kuijken}, K., {Hoekstra}, H., {et~al.} 2023{\natexlab{a}},
  \href{http://dx.doi.org/10.1051/0004-6361/202245210}{\color{black}\aap},
  \href{https://ui.adsabs.harvard.edu/abs/2023A&A...670A.100L}{670, A100}

\bibitem[{{Li} {et~al.}(2023{\natexlab{b}}){Li}, {Zhang}, {Sugiyama}, {Dalal},
  {Terasawa}, {Rau}, {Mandelbaum}, {Takada}, {More}, {Strauss}, {Miyatake},
  {Shirasaki}, {Hamana}, {Oguri}, {Luo}, {Nishizawa}, {Takahashi}, {Nicola},
  {Osato}, {Kannawadi}, {Sunayama}, {Armstrong}, {Bosch}, {Komiyama}, {Lupton},
  {Lust}, {MacArthur}, {Miyazaki}, {Murayama}, {Nishimichi}, {Okura}, {Price},
  {Tait}, {Tanaka}, \& {Wang}}]{LiHSC23}
{Li}, X., {Zhang}, T., {Sugiyama}, S., {et~al.} 2023{\natexlab{b}},
  \href{http://dx.doi.org/10.1103/PhysRevD.108.123518}{\color{black}\prd},
  \href{https://ui.adsabs.harvard.edu/abs/2023PhRvD.108l3518L}{108, 123518}

\bibitem[{{Lima} {et~al.}(2008){Lima}, {Cunha}, {Oyaizu}, {Frieman}, {Lin}, \&
  {Sheldon}}]{Lima08}
{Lima}, M., {Cunha}, C.~E., {Oyaizu}, H., {et~al.} 2008,
  \href{http://dx.doi.org/10.1111/j.1365-2966.2008.13510.x}{\color{black}\mnras},
  \href{https://ui.adsabs.harvard.edu/abs/2008MNRAS.390..118L}{390, 118}

\bibitem[{{Longley} {et~al.}(2023){Longley}, {Chang}, {Walter}, {Zuntz},
  {Ishak}, {Mandelbaum}, {Miyatake}, {Nicola}, {Pedersen}, {Pereira}, {Prat},
  {S{\'a}nchez}, {Secco}, {Tr{\"o}ster}, {Troxel}, {Wright}, \& {LSST Dark
  Energy Science Collaboration}}]{Longley23}
{Longley}, E.~P., {Chang}, C., {Walter}, C.~W., {et~al.} 2023,
  \href{http://dx.doi.org/10.1093/mnras/stad246}{\color{black}\mnras},
  \href{https://ui.adsabs.harvard.edu/abs/2023MNRAS.520.5016L}{520, 5016}

\bibitem[{{Louca} \& {Sellentin}(2020)}]{Louca20}
{Louca}, A.~J. \& {Sellentin}, E. 2020,
  \href{http://dx.doi.org/10.21105/astro.2007.07253}{\color{black}The Open
  Journal of Astrophysics},
  \href{https://ui.adsabs.harvard.edu/abs/2020OJAp....3E..11L}{3, 11}

\bibitem[{{LoVerde} \& {Afshordi}(2008)}]{Loverde08}
{LoVerde}, M. \& {Afshordi}, N. 2008,
  \href{http://dx.doi.org/10.1103/PhysRevD.78.123506}{\color{black}\prd},
  \href{https://ui.adsabs.harvard.edu/abs/2008PhRvD..78l3506L}{78, 123506}

\bibitem[{{Masters} {et~al.}(2015){Masters}, {Capak}, {Stern}, {Ilbert},
  {Salvato}, {Schmidt}, {Longo}, {Rhodes}, {Paltani}, {Mobasher}, {Hoekstra},
  {Hildebrandt}, {Coupon}, {Steinhardt}, {Speagle}, {Faisst}, {Kalinich},
  {Brodwin}, {Brescia}, \& {Cavuoti}}]{Masters15}
{Masters}, D., {Capak}, P., {Stern}, D., {et~al.} 2015,
  \href{http://dx.doi.org/10.1088/0004-637X/813/1/53}{\color{black}\apj},
  \href{https://ui.adsabs.harvard.edu/abs/2015ApJ...813...53M}{813, 53}

\bibitem[{{McCullough} {et~al.}(2024){McCullough}, {Amon}, {Legnani}, {Gruen},
  {Roodman}, {Friedrich}, {MacCrann}, {Becker}, {Myles}, {Dodelson},
  {Samuroff}, {Blazek}, {Prat}, {Honscheid}, {Pieres}, {Fert{\'e}}, {Alarcon},
  {Drlica-Wagner}, {Choi}, {Navarro-Alsina}, {Campos}, {Plazas Malag{\'o}n},
  {Porredon}, {Farahi}, {Ross}, {Carnero Rosell}, {Yin}, {Flaugher}, {Yanny},
  {S{\'a}nchez}, {Chang}, {Davis}, {To}, {Doux}, {Brooks}, {James}, {Sanchez
  Cid}, {Hollowood}, {Huterer}, {Rykoff}, {Gaztanaga}, {Huff}, {Suchyta},
  {Sheldon}, {Sanchez}, {Tarsitano}, {Andrade-Oliveira}, {Castander},
  {Bernstein}, {Gutierrez}, {Giannini}, {Tarle}, {Diehl}, {Huang}, {Harrison},
  {Sevilla-Noarbe}, {Tutusaus}, {Ferrero}, {Elvin-Poole}, {Marshall}, {Muir},
  {Weller}, {Zuntz}, {Carretero}, {DeRose}, {Frieman}, {Cordero}, {De Vicente},
  {Garc{\'\i}a-Bellido}, {Mena-Fern{\'a}ndez}, {Eckert}, {Romer}, {Bechtol},
  {Herner}, {Kuehn}, {Secco}, {da Costa}, {Paterno}, {Soares-Santos}, {Gatti},
  {Raveri}, {Yamamoto}, {Smith}, {Carrasco Kind}, {Troxel}, {Aguena}, {Jarvis},
  {Swanson}, {Weaverdyck}, {Lahav}, {Doel}, {Wiseman}, {Miquel}, {Gruendl},
  {Cawthon}, {Allam}, {Hinton}, {Bridle}, {Bocquet}, {Desai}, {Pandey},
  {Everett}, {Lee}, {Shin}, {Palmese}, {Conselice}, {Burke}, {Buckley-Geer},
  {Lima}, {Vincenzi}, {Pereira}, {Crocce}, {Schubnell}, {Jeffrey}, {Alves},
  {Vikram}, {Zhang}, \& {DES Collaboration}}]{McCullough24}
{McCullough}, J., {Amon}, A., {Legnani}, E., {et~al.} 2024,
  \href{https://ui.adsabs.harvard.edu/abs/2024arXiv241022272M}{\href{http://dx.doi.org/10.48550/arXiv.2410.22272}{\color{black}arXiv
  e-prints}, arXiv:2410.22272}

\bibitem[{{Miller} {et~al.}(2013){Miller}, {Heymans}, {Kitching}, {van
  Waerbeke}, {Erben}, {Hildebrandt}, {Hoekstra}, {Mellier}, {Rowe}, {Coupon},
  {Dietrich}, {Fu}, {Harnois-D{\'e}raps}, {Hudson}, {Kilbinger}, {Kuijken},
  {Schrabback}, {Semboloni}, {Vafaei}, \& {Velander}}]{Miller13}
{Miller}, L., {Heymans}, C., {Kitching}, T.~D., {et~al.} 2013,
  \href{http://dx.doi.org/10.1093/mnras/sts454}{\color{black}\mnras},
  \href{https://ui.adsabs.harvard.edu/abs/2013MNRAS.429.2858M}{429, 2858}

\bibitem[{{Neveux} {et~al.}(2020){Neveux}, {Burtin}, {de Mattia}, {Smith},
  {Ross}, {Hou}, {Bautista}, {Brinkmann}, {Chuang}, {Dawson}, {Gil-Mar{\'\i}n},
  {Lyke}, {de la Macorra}, {du Mas des Bourboux}, {Mohammad}, {M{\"u}ller},
  {Myers}, {Newman}, {Percival}, {Rossi}, {Schneider}, {Vivek}, {Zarrouk},
  {Zhao}, \& {Zhao}}]{Neveux20}
{Neveux}, R., {Burtin}, E., {de Mattia}, A., {et~al.} 2020,
  \href{http://dx.doi.org/10.1093/mnras/staa2780}{\color{black}\mnras},
  \href{https://ui.adsabs.harvard.edu/abs/2020MNRAS.499..210N}{499, 210}

\bibitem[{{Oehl} \& {Tr{\"o}ster}(2024)}]{Oehl24}
{Oehl}, V. \& {Tr{\"o}ster}, T. 2024,
  \href{https://ui.adsabs.harvard.edu/abs/2024arXiv240708718O}{\href{http://dx.doi.org/10.48550/arXiv.2407.08718}{\color{black}arXiv
  e-prints}, arXiv:2407.08718}

\bibitem[{{Planck Collaboration} {et~al.}(2020){Planck Collaboration},
  {Aghanim}, {Akrami}, {Ashdown}, {Aumont}, {Baccigalupi}, {Ballardini},
  {Banday}, {Barreiro}, {Bartolo}, {Basak}, {Battye}, {Benabed}, {Bernard},
  {Bersanelli}, {Bielewicz}, {Bock}, {Bond}, {Borrill}, {Bouchet}, {Boulanger},
  {Bucher}, {Burigana}, {Butler}, {Calabrese}, {Cardoso}, {Carron},
  {Challinor}, {Chiang}, {Chluba}, {Colombo}, {Combet}, {Contreras}, {Crill},
  {Cuttaia}, {de Bernardis}, {de Zotti}, {Delabrouille}, {Delouis}, {Di
  Valentino}, {Diego}, {Dor{\'e}}, {Douspis}, {Ducout}, {Dupac}, {Dusini},
  {Efstathiou}, {Elsner}, {En{\ss}lin}, {Eriksen}, {Fantaye}, {Farhang},
  {Fergusson}, {Fernandez-Cobos}, {Finelli}, {Forastieri}, {Frailis},
  {Fraisse}, {Franceschi}, {Frolov}, {Galeotta}, {Galli}, {Ganga},
  {G{\'e}nova-Santos}, {Gerbino}, {Ghosh}, {Gonz{\'a}lez-Nuevo}, {G{\'o}rski},
  {Gratton}, {Gruppuso}, {Gudmundsson}, {Hamann}, {Handley}, {Hansen},
  {Herranz}, {Hildebrandt}, {Hivon}, {Huang}, {Jaffe}, {Jones}, {Karakci},
  {Keih{\"a}nen}, {Keskitalo}, {Kiiveri}, {Kim}, {Kisner}, {Knox},
  {Krachmalnicoff}, {Kunz}, {Kurki-Suonio}, {Lagache}, {Lamarre}, {Lasenby},
  {Lattanzi}, {Lawrence}, {Le Jeune}, {Lemos}, {Lesgourgues}, {Levrier},
  {Lewis}, {Liguori}, {Lilje}, {Lilley}, {Lindholm}, {L{\'o}pez-Caniego},
  {Lubin}, {Ma}, {Mac{\'\i}as-P{\'e}rez}, {Maggio}, {Maino}, {Mandolesi},
  {Mangilli}, {Marcos-Caballero}, {Maris}, {Martin}, {Martinelli},
  {Mart{\'\i}nez-Gonz{\'a}lez}, {Matarrese}, {Mauri}, {McEwen}, {Meinhold},
  {Melchiorri}, {Mennella}, {Migliaccio}, {Millea}, {Mitra},
  {Miville-Desch{\^e}nes}, {Molinari}, {Montier}, {Morgante}, {Moss}, {Natoli},
  {N{\o}rgaard-Nielsen}, {Pagano}, {Paoletti}, {Partridge}, {Patanchon},
  {Peiris}, {Perrotta}, {Pettorino}, {Piacentini}, {Polastri}, {Polenta},
  {Puget}, {Rachen}, {Reinecke}, {Remazeilles}, {Renzi}, {Rocha}, {Rosset},
  {Roudier}, {Rubi{\~n}o-Mart{\'\i}n}, {Ruiz-Granados}, {Salvati}, {Sandri},
  {Savelainen}, {Scott}, {Shellard}, {Sirignano}, {Sirri}, {Spencer},
  {Sunyaev}, {Suur-Uski}, {Tauber}, {Tavagnacco}, {Tenti}, {Toffolatti},
  {Tomasi}, {Trombetti}, {Valenziano}, {Valiviita}, {Van Tent}, {Vibert},
  {Vielva}, {Villa}, {Vittorio}, {Wandelt}, {Wehus}, {White}, {White},
  {Zacchei}, \& {Zonca}}]{Planck2018}
{Planck Collaboration}, {Aghanim}, N., {Akrami}, Y., {et~al.} 2020,
  \href{http://dx.doi.org/10.1051/0004-6361/201833910}{\color{black}\aap},
  \href{https://ui.adsabs.harvard.edu/abs/2020A&A...641A...6P}{641, A6}

\bibitem[{{Prince} \& {Dunkley}(2019)}]{Prince19}
{Prince}, H. \& {Dunkley}, J. 2019,
  \href{http://dx.doi.org/10.1103/PhysRevD.100.083502}{\color{black}\prd},
  \href{https://ui.adsabs.harvard.edu/abs/2019PhRvD.100h3502P}{100, 083502}

\bibitem[{{Raichoor} {et~al.}(2021){Raichoor}, {de Mattia}, {Ross}, {Zhao},
  {Alam}, {Avila}, {Bautista}, {Brinkmann}, {Brownstein}, {Burtin}, {Chapman},
  {Chuang}, {Comparat}, {Dawson}, {Dey}, {du Mas des Bourboux}, {Elvin-Poole},
  {Gonzalez-Perez}, {Gorgoni}, {Kneib}, {Kong}, {Lang}, {Moustakas}, {Myers},
  {M{\"u}ller}, {Nadathur}, {Newman}, {Percival}, {Rezaie}, {Rossi},
  {Ruhlmann-Kleider}, {Schlegel}, {Schneider}, {Seo}, {Tamone}, {Tinker},
  {Tojeiro}, {Vivek}, {Y{\`e}che}, \& {Zhao}}]{Raichoor21}
{Raichoor}, A., {de Mattia}, A., {Ross}, A.~J., {et~al.} 2021,
  \href{http://dx.doi.org/10.1093/mnras/staa3336}{\color{black}\mnras},
  \href{https://ui.adsabs.harvard.edu/abs/2021MNRAS.500.3254R}{500, 3254}

\bibitem[{{Raichoor} {et~al.}(2023){Raichoor}, {Moustakas}, {Newman}, {Karim},
  {Ahlen}, {Alam}, {Bailey}, {Brooks}, {Dawson}, {de la Macorra}, {de Mattia},
  {Dey}, {Dey}, {Dhungana}, {Eftekharzadeh}, {Eisenstein}, {Fanning},
  {Font-Ribera}, {Garc{\'\i}a-Bellido}, {Gazta{\~n}aga}, {A Gontcho}, {Guy},
  {Honscheid}, {Ishak}, {Kehoe}, {Kisner}, {Kremin}, {Lan}, {Landriau}, {Le
  Guillou}, {Levi}, {Magneville}, {Manera}, {Martini}, {Meisner}, {Myers},
  {Nie}, {Palanque-Delabrouille}, {Percival}, {Poppett}, {Prada}, {Ross},
  {Ruhlmann-Kleider}, {Sabiu}, {Schlafly}, {Schlegel}, {Tarl{\'e}}, {Weaver},
  {Y{\`e}che}, {Zhou}, {Zhou}, \& {Zou}}]{Raichoor23}
{Raichoor}, A., {Moustakas}, J., {Newman}, J.~A., {et~al.} 2023,
  \href{http://dx.doi.org/10.3847/1538-3881/acb213}{\color{black}\aj},
  \href{https://ui.adsabs.harvard.edu/abs/2023AJ....165..126R}{165, 126}

\bibitem[{{Raveri} {et~al.}(2020){Raveri}, {Zacharegkas}, \& {Hu}}]{Raveri20}
{Raveri}, M., {Zacharegkas}, G., \& {Hu}, W. 2020,
  \href{http://dx.doi.org/10.1103/PhysRevD.101.103527}{\color{black}\prd},
  \href{https://ui.adsabs.harvard.edu/abs/2020PhRvD.101j3527R}{101, 103527}

\bibitem[{{Reischke} {et~al.}(2025){Reischke}, {Unruh}, {Asgari}, {Dvornik},
  {Hildebrandt}, {Joachimi}, {Porth}, {von Wietersheim-Kramsta}, {van den
  Busch}, {St{\"o}lzner}, {Wright}, {Yan}, {Bilicki}, {Burger}, {Chisari},
  {Harnois-D{\'e}raps}, {Georgiou}, {Heymans}, {Jalan}, {Joudaki}, {Kuijken},
  {Li}, {Linke}, {Mahony}, {Sciotti}, {Tr{\"o}ster}, \& {Yoon}}]{Reischke23}
{Reischke}, R., {Unruh}, S., {Asgari}, M., {et~al.} 2025,
  \href{http://dx.doi.org/10.1051/0004-6361/202452592}{\color{black}\aap},
  \href{https://ui.adsabs.harvard.edu/abs/2025A&A...699A.124R}{699, A124}

\bibitem[{{Ross} {et~al.}(2015){Ross}, {Samushia}, {Howlett}, {Percival},
  {Burden}, \& {Manera}}]{Ross15}
{Ross}, A.~J., {Samushia}, L., {Howlett}, C., {et~al.} 2015,
  \href{http://dx.doi.org/10.1093/mnras/stv154}{\color{black}\mnras},
  \href{https://ui.adsabs.harvard.edu/abs/2015MNRAS.449..835R}{449, 835}

\bibitem[{{Samuroff} {et~al.}(2019){Samuroff}, {Blazek}, {Troxel}, {MacCrann},
  {Krause}, {Leonard}, {Prat}, {Gruen}, {Dodelson}, {Eifler}, {Gatti},
  {Hartley}, {Hoyle}, {Larsen}, {Zuntz}, {Abbott}, {Allam}, {Annis},
  {Bernstein}, {Bertin}, {Bridle}, {Brooks}, {Carnero Rosell}, {Carrasco Kind},
  {Carretero}, {Castander}, {Cunha}, {da Costa}, {Davis}, {De Vicente},
  {DePoy}, {Desai}, {Diehl}, {Dietrich}, {Doel}, {Flaugher}, {Fosalba},
  {Frieman}, {Garc{\'\i}a-Bellido}, {Gaztanaga}, {Gerdes}, {Gruendl},
  {Gschwend}, {Gutierrez}, {Hollowood}, {Honscheid}, {James}, {Kuehn},
  {Kuropatkin}, {Lima}, {Maia}, {March}, {Marshall}, {Martini}, {Melchior},
  {Menanteau}, {Miller}, {Miquel}, {Ogando}, {Plazas}, {Sanchez}, {Scarpine},
  {Schindler}, {Schubnell}, {Serrano}, {Sevilla-Noarbe}, {Sheldon}, {Smith},
  {Sobreira}, {Suchyta}, {Tarle}, {Thomas}, {Vikram}, \& {DES
  Collaboration}}]{Samuroff19}
{Samuroff}, S., {Blazek}, J., {Troxel}, M.~A., {et~al.} 2019,
  \href{http://dx.doi.org/10.1093/mnras/stz2197}{\color{black}\mnras},
  \href{https://ui.adsabs.harvard.edu/abs/2019MNRAS.489.5453S}{489, 5453}

\bibitem[{{Samuroff} {et~al.}(2023){Samuroff}, {Mandelbaum}, {Blazek},
  {Campos}, {MacCrann}, {Zacharegkas}, {Amon}, {Prat}, {Singh}, {Elvin-Poole},
  {Ross}, {Alarcon}, {Baxter}, {Bechtol}, {Becker}, {Bernstein}, {Rosell},
  {Kind}, {Cawthon}, {Chang}, {Chen}, {Choi}, {Crocce}, {Davis}, {DeRose},
  {Dodelson}, {Doux}, {Drlica-Wagner}, {Eckert}, {Everett}, {Fert{\'e}},
  {Gatti}, {Giannini}, {Gruen}, {Gruendl}, {Harrison}, {Herner}, {Huff},
  {Jarvis}, {Kuropatkin}, {Leget}, {Lemos}, {McCullough}, {Myles},
  {Navarro-Alsina}, {Pandey}, {Porredon}, {Raveri}, {Rodriguez-Monroy},
  {Rollins}, {Roodman}, {Rossi}, {Rykoff}, {S{\'a}nchez}, {Secco},
  {Sevilla-Noarbe}, {Sheldon}, {Shin}, {Troxel}, {Tutusaus}, {Weaverdyck},
  {Yanny}, {Yin}, {Zhang}, {Zuntz}, {Aguena}, {Alves}, {Annis}, {Bacon},
  {Bertin}, {Bocquet}, {Brooks}, {Burke}, {Carretero}, {Costanzi}, {da Costa},
  {Pereira}, {De Vicente}, {Desai}, {Diehl}, {Dietrich}, {Doel}, {Ferrero},
  {Flaugher}, {Frieman}, {Garc{\'\i}a-Bellido}, {Hinton}, {Hollowood},
  {Honscheid}, {James}, {Kuehn}, {Lahav}, {Marshall}, {Melchior},
  {Mena-Fern{\'a}ndez}, {Menanteau}, {Miquel}, {Newman}, {Palmese}, {Pieres},
  {Malag{\'o}n}, {Sanchez}, {Scarpine}, {Smith}, {Suchyta}, {Swanson}, {Tarle},
  {To}, \& {DES Collaboration}}]{Samuroff24}
{Samuroff}, S., {Mandelbaum}, R., {Blazek}, J., {et~al.} 2023,
  \href{http://dx.doi.org/10.1093/mnras/stad2013}{\color{black}\mnras},
  \href{https://ui.adsabs.harvard.edu/abs/2023MNRAS.524.2195S}{524, 2195}

\bibitem[{{Schneider} {et~al.}(2010){Schneider}, {Eifler}, \&
  {Krause}}]{Schneider10}
{Schneider}, P., {Eifler}, T., \& {Krause}, E. 2010,
  \href{http://dx.doi.org/10.1051/0004-6361/201014235}{\color{black}\aap},
  \href{https://ui.adsabs.harvard.edu/abs/2010A&A...520A.116S}{520, A116}

\bibitem[{{Schneider} {et~al.}(2002){Schneider}, {van Waerbeke}, {Kilbinger},
  \& {Mellier}}]{Schneider02}
{Schneider}, P., {van Waerbeke}, L., {Kilbinger}, M., \& {Mellier}, Y. 2002,
  \href{http://dx.doi.org/10.1051/0004-6361:20021341}{\color{black}\aap},
  \href{https://ui.adsabs.harvard.edu/abs/2002A&A...396....1S}{396, 1}

\bibitem[{{Scolnic} {et~al.}(2022){Scolnic}, {Brout}, {Carr}, {Riess}, {Davis},
  {Dwomoh}, {Jones}, {Ali}, {Charvu}, {Chen}, {Peterson}, {Popovic}, {Rose},
  {Wood}, {Brown}, {Chambers}, {Coulter}, {Dettman}, {Dimitriadis},
  {Filippenko}, {Foley}, {Jha}, {Kilpatrick}, {Kirshner}, {Pan}, {Rest},
  {Rojas-Bravo}, {Siebert}, {Stahl}, \& {Zheng}}]{Scolnic22}
{Scolnic}, D., {Brout}, D., {Carr}, A., {et~al.} 2022,
  \href{http://dx.doi.org/10.3847/1538-4357/ac8b7a}{\color{black}\apj},
  \href{https://ui.adsabs.harvard.edu/abs/2022ApJ...938..113S}{938, 113}

\bibitem[{{Secco} {et~al.}(2022){Secco}, {Samuroff}, {Krause}, {Jain},
  {Blazek}, {Raveri}, {Campos}, {Amon}, {Chen}, {Doux}, {Choi}, {Gruen},
  {Bernstein}, {Chang}, {DeRose}, {Myles}, {Fert{\'e}}, {Lemos}, {Huterer},
  {Prat}, {Troxel}, {MacCrann}, {Liddle}, {Kacprzak}, {Fang}, {S{\'a}nchez},
  {Pandey}, {Dodelson}, {Chintalapati}, {Hoffmann}, {Alarcon}, {Alves},
  {Andrade-Oliveira}, {Baxter}, {Bechtol}, {Becker}, {Brandao-Souza},
  {Camacho}, {Carnero Rosell}, {Carrasco Kind}, {Cawthon}, {Cordero}, {Crocce},
  {Davis}, {Di Valentino}, {Drlica-Wagner}, {Eckert}, {Eifler}, {Elidaiana},
  {Elsner}, {Elvin-Poole}, {Everett}, {Fosalba}, {Friedrich}, {Gatti},
  {Giannini}, {Gruendl}, {Harrison}, {Hartley}, {Herner}, {Huang}, {Huff},
  {Jarvis}, {Jeffrey}, {Kuropatkin}, {Leget}, {Muir}, {Mccullough}, {Navarro
  Alsina}, {Omori}, {Park}, {Porredon}, {Rollins}, {Roodman}, {Rosenfeld},
  {Ross}, {Rykoff}, {Sanchez}, {Sevilla-Noarbe}, {Sheldon}, {Shin}, {Troja},
  {Tutusaus}, {Varga}, {Weaverdyck}, {Wechsler}, {Yanny}, {Yin}, {Zhang},
  {Zuntz}, {Abbott}, {Aguena}, {Allam}, {Annis}, {Bacon}, {Bertin}, {Bhargava},
  {Bridle}, {Brooks}, {Buckley-Geer}, {Burke}, {Carretero}, {Costanzi}, {da
  Costa}, {De Vicente}, {Diehl}, {Dietrich}, {Doel}, {Ferrero}, {Flaugher},
  {Frieman}, {Garc{\'\i}a-Bellido}, {Gaztanaga}, {Gerdes}, {Giannantonio},
  {Gschwend}, {Gutierrez}, {Hinton}, {Hollowood}, {Honscheid}, {Hoyle},
  {James}, {Jeltema}, {Kuehn}, {Lahav}, {Lima}, {Lin}, {Maia}, {Marshall},
  {Martini}, {Melchior}, {Menanteau}, {Miquel}, {Mohr}, {Morgan}, {Ogando},
  {Palmese}, {Paz-Chinch{\'o}n}, {Petravick}, {Pieres}, {Plazas Malag{\'o}n},
  {Rodriguez-Monroy}, {Romer}, {Sanchez}, {Scarpine}, {Schubnell}, {Scolnic},
  {Serrano}, {Smith}, {Soares-Santos}, {Suchyta}, {Swanson}, {Tarle}, {Thomas},
  {To}, \& {DES Collaboration}}]{Secco22}
{Secco}, L.~F., {Samuroff}, S., {Krause}, E., {et~al.} 2022,
  \href{http://dx.doi.org/10.1103/PhysRevD.105.023515}{\color{black}\prd},
  \href{https://ui.adsabs.harvard.edu/abs/2022PhRvD.105b3515S}{105, 023515}

\bibitem[{{Sellentin} {et~al.}(2018){Sellentin}, {Heymans}, \&
  {Harnois-D{\'e}raps}}]{Sellentin18}
{Sellentin}, E., {Heymans}, C., \& {Harnois-D{\'e}raps}, J. 2018,
  \href{http://dx.doi.org/10.1093/mnras/sty988}{\color{black}\mnras},
  \href{https://ui.adsabs.harvard.edu/abs/2018MNRAS.477.4879S}{477, 4879}

\bibitem[{{Sipp} {et~al.}(2021){Sipp}, {Sch{\"a}fer}, \& {Reischke}}]{Sipp21}
{Sipp}, M., {Sch{\"a}fer}, B.~M., \& {Reischke}, R. 2021,
  \href{http://dx.doi.org/10.1093/mnras/staa3710}{\color{black}\mnras},
  \href{https://ui.adsabs.harvard.edu/abs/2021MNRAS.501..683S}{501, 683}

\bibitem[{{Spergel} {et~al.}(2015){Spergel}, {Gehrels}, {Baltay}, {Bennett},
  {Breckinridge}, {Donahue}, {Dressler}, {Gaudi}, {Greene}, {Guyon}, {Hirata},
  {Kalirai}, {Kasdin}, {Macintosh}, {Moos}, {Perlmutter}, {Postman},
  {Rauscher}, {Rhodes}, {Wang}, {Weinberg}, {Benford}, {Hudson}, {Jeong},
  {Mellier}, {Traub}, {Yamada}, {Capak}, {Colbert}, {Masters}, {Penny},
  {Savransky}, {Stern}, {Zimmerman}, {Barry}, {Bartusek}, {Carpenter}, {Cheng},
  {Content}, {Dekens}, {Demers}, {Grady}, {Jackson}, {Kuan}, {Kruk}, {Melton},
  {Nemati}, {Parvin}, {Poberezhskiy}, {Peddie}, {Ruffa}, {Wallace}, {Whipple},
  {Wollack}, \& {Zhao}}]{Spergel15}
{Spergel}, D., {Gehrels}, N., {Baltay}, C., {et~al.} 2015,
  \href{https://ui.adsabs.harvard.edu/abs/2015arXiv150303757S}{arXiv e-prints,
  arXiv:1503.03757}

\bibitem[{Spiegelhalter {et~al.}(2002)Spiegelhalter, Best, Carlin, \& Van
  Der~Linde}]{Spiegelhalter02}
Spiegelhalter, D.~J., Best, N.~G., Carlin, B.~P., \& Van Der~Linde, A. 2002,
  \href{http://dx.doi.org/https://doi.org/10.1111/1467-9868.00353}{\color{black}Journal
  of the Royal Statistical Society: Series B (Statistical Methodology)}, 64,
  583

\bibitem[{{Spurio Mancini} {et~al.}(2022){Spurio Mancini}, {Piras}, {Alsing},
  {Joachimi}, \& {Hobson}}]{SpurioMancini22}
{Spurio Mancini}, A., {Piras}, D., {Alsing}, J., {Joachimi}, B., \& {Hobson},
  M.~P. 2022,
  \href{http://dx.doi.org/10.1093/mnras/stac064}{\color{black}\mnras},
  \href{https://ui.adsabs.harvard.edu/abs/2022MNRAS.511.1771S}{511, 1771}

\bibitem[{{Tamone} {et~al.}(2020){Tamone}, {Raichoor}, {Zhao}, {de Mattia},
  {Gorgoni}, {Burtin}, {Ruhlmann-Kleider}, {Ross}, {Alam}, {Percival}, {Avila},
  {Chapman}, {Chuang}, {Comparat}, {Dawson}, {de la Torre}, {du Mas des
  Bourboux}, {Escoffier}, {Gonzalez-Perez}, {Hou}, {Kneib}, {Mohammad},
  {Mueller}, {Paviot}, {Rossi}, {Schneider}, {Wang}, \& {Zhao}}]{Tamone20}
{Tamone}, A., {Raichoor}, A., {Zhao}, C., {et~al.} 2020,
  \href{http://dx.doi.org/10.1093/mnras/staa3050}{\color{black}\mnras},
  \href{https://ui.adsabs.harvard.edu/abs/2020MNRAS.499.5527T}{499, 5527}

\bibitem[{{The Dark Energy Survey Collaboration}(2005)}]{DES05}
{The Dark Energy Survey Collaboration}. 2005,
  \href{https://ui.adsabs.harvard.edu/abs/2005astro.ph.10346T}{\href{http://dx.doi.org/10.48550/arXiv.astro-ph/0510346}{\color{black}arXiv
  e-prints}, arXiv:0510346}

\bibitem[{{Tugendhat} {et~al.}(2020){Tugendhat}, {Reischke}, \&
  {Sch{\"a}fer}}]{Tugendhat20}
{Tugendhat}, T.~M., {Reischke}, R., \& {Sch{\"a}fer}, B.~M. 2020,
  \href{http://dx.doi.org/10.1093/mnras/staa641}{\color{black}\mnras},
  \href{https://ui.adsabs.harvard.edu/abs/2020MNRAS.494.2969T}{494, 2969}

\bibitem[{{van Uitert} {et~al.}(2018){van Uitert}, {Joachimi}, {Joudaki},
  {Amon}, {Heymans}, {K{\"o}hlinger}, {Asgari}, {Blake}, {Choi}, {Erben},
  {Farrow}, {Harnois-D{\'e}raps}, {Hildebrandt}, {Hoekstra}, {Kitching},
  {Klaes}, {Kuijken}, {Merten}, {Miller}, {Nakajima}, {Schneider}, {Valentijn},
  \& {Viola}}]{vanUitert18}
{van Uitert}, E., {Joachimi}, B., {Joudaki}, S., {et~al.} 2018,
  \href{http://dx.doi.org/10.1093/mnras/sty551}{\color{black}\mnras},
  \href{https://ui.adsabs.harvard.edu/abs/2018MNRAS.476.4662V}{476, 4662}

\bibitem[{{Van Waerbeke} {et~al.}(2000){Van Waerbeke}, {Mellier}, {Erben},
  {Cuillandre}, {Bernardeau}, {Maoli}, {Bertin}, {McCracken}, {Le F{\`e}vre},
  {Fort}, {Dantel-Fort}, {Jain}, \& {Schneider}}]{vanWaerbeke00}
{Van Waerbeke}, L., {Mellier}, Y., {Erben}, T., {et~al.} 2000, \aap,
  \href{https://ui.adsabs.harvard.edu/abs/2000A&A...358...30V}{358, 30}

\bibitem[{Virtanen {et~al.}(2020)Virtanen, Gommers, Oliphant, Haberland, Reddy,
  Cournapeau, Burovski, Peterson, Weckesser, Bright, van~der Walt, Brett,
  Wilson, Millman, Mayorov, Nelson, Jones, Kern, Larson, Carey, Polat, Feng,
  Moore, {VanderPlas}, Laxalde, Perktold, Cimrman, Henriksen, Quintero, Harris,
  Archibald, Ribeiro, Pedregosa, {van Mulbregt}, \& {SciPy 1.0
  Contributors}}]{Scipy}
Virtanen, P., Gommers, R., Oliphant, T.~E., {et~al.} 2020,
  \href{http://dx.doi.org/10.1038/s41592-019-0686-2}{\color{black}Nature
  Methods}, \href{https://rdcu.be/b08Wh}{17, 261}

\bibitem[{{W}es {M}c{K}inney(2010)}]{Pandas}
{W}es {M}c{K}inney. 2010, in {P}roceedings of the 9th {P}ython in {S}cience
  {C}onference, ed. {S}t\'efan van~der {W}alt \& {J}arrod {M}illman,
  \href{https://proceedings.scipy.org/articles/Majora-92bf1922-00a}{56 -- 61}

\bibitem[{{Wittman} {et~al.}(2000){Wittman}, {Tyson}, {Kirkman},
  {Dell'Antonio}, \& {Bernstein}}]{Wittman00}
{Wittman}, D.~M., {Tyson}, J.~A., {Kirkman}, D., {Dell'Antonio}, I., \&
  {Bernstein}, G. 2000,
  \href{http://dx.doi.org/10.1038/35012001}{\color{black}\nat},
  \href{https://ui.adsabs.harvard.edu/abs/2000Natur.405..143W}{405, 143}

\bibitem[{{Wright} {et~al.}(2019){Wright}, {Hildebrandt}, {Kuijken}, {Erben},
  {Blake}, {Buddelmeijer}, {Choi}, {Cross}, {de Jong}, {Edge},
  {Gonzalez-Fernandez}, {Gonz{\'a}lez Solares}, {Grado}, {Heymans}, {Irwin},
  {Kupcu Yoldas}, {Lewis}, {Mann}, {Napolitano}, {Radovich}, {Schneider},
  {Sif{\'o}n}, {Sutherland}, {Sutorius}, \& {Verdoes Kleijn}}]{Wright19}
{Wright}, A.~H., {Hildebrandt}, H., {Kuijken}, K., {et~al.} 2019,
  \href{http://dx.doi.org/10.1051/0004-6361/201834879}{\color{black}\aap},
  \href{https://ui.adsabs.harvard.edu/abs/2019A&A...632A..34W}{632, A34}

\bibitem[{{Wright} {et~al.}(2025{\natexlab{a}}){Wright}, {Hildebrandt}, {van
  den Busch}, {Bilicki}, {Heymans}, {Joachimi}, {Mahony}, {Reischke},
  {St\"olzner}, {Wittje}, {Asgari}, {Chisari}, {Dvornik}, {Georgiou}, {Giblin},
  {Hoekstra}, {Jalan}, {William}, {Joudaki}, {Kuijken}, {Lesci}, {Li}, {Linke},
  {Loureiro}, {Maturi}, {Moscardin}, {Porth}, {Radovich}, {Tr\"oster}, {von
  Wietersheim-Kramsta}, {Yan}, {Yoon}, \& {Zhang}}]{Wright23_redshifts}
{Wright}, A.~H., {Hildebrandt}, H., {van den Busch}, J.~L., {et~al.}
  2025{\natexlab{a}},
  \href{https://ui.adsabs.harvard.edu/abs/2025arXiv250319440W}{\aap, accepted,
  arXiv:2503.19440}

\bibitem[{{Wright} {et~al.}(2020{\natexlab{a}}){Wright}, {Hildebrandt}, {van
  den Busch}, \& {Heymans}}]{Wright20}
{Wright}, A.~H., {Hildebrandt}, H., {van den Busch}, J.~L., \& {Heymans}, C.
  2020{\natexlab{a}},
  \href{http://dx.doi.org/10.1051/0004-6361/201936782}{\color{black}\aap},
  \href{https://ui.adsabs.harvard.edu/abs/2020A&A...637A.100W}{637, A100}

\bibitem[{{Wright} {et~al.}(2020{\natexlab{b}}){Wright}, {Hildebrandt}, {van
  den Busch}, {Heymans}, {Joachimi}, {Kannawadi}, \& {Kuijken}}]{Wright20b}
{Wright}, A.~H., {Hildebrandt}, H., {van den Busch}, J.~L., {et~al.}
  2020{\natexlab{b}},
  \href{http://dx.doi.org/10.1051/0004-6361/202038389}{\color{black}\aap},
  \href{https://ui.adsabs.harvard.edu/abs/2020A&A...640L..14W}{640, L14}

\bibitem[{{Wright} {et~al.}(2024){Wright}, {Kuijken}, {Hildebrandt},
  {Radovich}, {Bilicki}, {Dvornik}, {Getman}, {Heymans}, {Hoekstra}, {Li},
  {Miller}, {Napolitano}, {Xia}, {Asgari}, {Brescia}, {Buddelmeijer}, {Burger},
  {Castignani}, {Cavuoti}, {de Jong}, {Edge}, {Giblin}, {Giocoli},
  {Harnois-D{\'e}raps}, {Jalan}, {Joachimi}, {John William}, {Joudaki},
  {Kannawadi}, {Kaur}, {La Barbera}, {Linke}, {Mahony}, {Maturi}, {Moscardini},
  {Nakoneczny}, {Paolillo}, {Porth}, {Puddu}, {Reischke}, {Schneider},
  {Sereno}, {Shan}, {Sif{\'o}n}, {St{\"o}lzner}, {Tr{\"o}ster}, {Valentijn},
  {van den Busch}, {Verdoes Kleijn}, {Wittje}, {Yan}, {Yao}, {Yoon}, \&
  {Zhang}}]{Wright23_DR5}
{Wright}, A.~H., {Kuijken}, K., {Hildebrandt}, H., {et~al.} 2024,
  \href{http://dx.doi.org/10.1051/0004-6361/202346730}{\color{black}\aap},
  \href{https://ui.adsabs.harvard.edu/abs/2024A&A...686A.170W}{686, A170}

\bibitem[{{Wright} {et~al.}(2025{\natexlab{b}}){Wright}, {St\"olzner},
  {Asgari}, {Bilicki}, {Giblin}, {Heymans}, {Hildebrandt}, {Hoekstra},
  {Joachimi}, {Kuijken}, {Li}, {Reischke}, {von Wietersheim-Kramsta}, {Yoon},
  {Burger}, {Chisari}, {de Jong}, {Dvornik}, {Georgiou}, {Harnois-D\'eraps},
  {Jalan}, {William}, {Joudaki}, {Lesci}, {Linke}, {Loureiro}, {Mahony},
  {Maturi}, {Miller}, {Moscardini}, {Napolitano}, {Porth}, {Radovich},
  {Schneider}, {Tr\"oster}, {Wittje}, {Yan}, \& {Zhang}}]{Wright25}
{Wright}, A.~H., {St\"olzner}, B., {Asgari}, M., {et~al.} 2025{\natexlab{b}},
  \href{https://ui.adsabs.harvard.edu/abs/2025arXiv250319441W}{\aap, accepted,
  arXiv:2503.19441}

\bibitem[{{Yan} {et~al.}(2025){Yan}, {Wright}, {Elisa Chisari}, {Georgiou},
  {Joudaki}, {Loureiro}, {Reischke}, {Asgari}, {Bilicki}, {Dvornik}, {Heymans},
  {Hildebrandt}, {Jalan}, {Joachimi}, {Francesco Lesci}, {Li}, {Linke},
  {Mahony}, {Moscardini}, {Napolitano}, {St{\"o}lzner}, {Von
  Wietersheim-Kramsta}, \& {Yoon}}]{Yan25}
{Yan}, Z., {Wright}, A.~H., {Elisa Chisari}, N., {et~al.} 2025,
  \href{http://dx.doi.org/10.1051/0004-6361/202452808}{\color{black}\aap},
  \href{https://ui.adsabs.harvard.edu/abs/2025A&A...694A.259Y}{694, A259}

\bibitem[{{Zhou} {et~al.}(2023){Zhou}, {Dey}, {Newman}, {Eisenstein}, {Dawson},
  {Bailey}, {Berti}, {Guy}, {Lan}, {Zou}, {Aguilar}, {Ahlen}, {Alam}, {Brooks},
  {de la Macorra}, {Dey}, {Dhungana}, {Fanning}, {Font-Ribera}, {Gontcho},
  {Honscheid}, {Ishak}, {Kisner}, {Kov{\'a}cs}, {Kremin}, {Landriau}, {Levi},
  {Magneville}, {Manera}, {Martini}, {Meisner}, {Miquel}, {Moustakas}, {Myers},
  {Nie}, {Palanque-Delabrouille}, {Percival}, {Poppett}, {Prada}, {Raichoor},
  {Ross}, {Schlafly}, {Schlegel}, {Schubnell}, {Tarl{\'e}}, {Weaver},
  {Wechsler}, {Y{\'e}che}, \& {Zhou}}]{Zhou23}
{Zhou}, R., {Dey}, B., {Newman}, J.~A., {et~al.} 2023,
  \href{http://dx.doi.org/10.3847/1538-3881/aca5fb}{\color{black}\aj},
  \href{https://ui.adsabs.harvard.edu/abs/2023AJ....165...58Z}{165, 58}

\bibitem[{{Zuntz} {et~al.}(2015){Zuntz}, {Paterno}, {Jennings}, {Rudd},
  {Manzotti}, {Dodelson}, {Bridle}, {Sehrish}, \& {Kowalkowski}}]{Zuntz15}
{Zuntz}, J., {Paterno}, M., {Jennings}, E., {et~al.} 2015,
  \href{http://dx.doi.org/10.1016/j.ascom.2015.05.005}{\color{black}Astronomy
  and Computing},
  \href{https://ui.adsabs.harvard.edu/abs/2015A&C....12...45Z}{12, 45}

\end{thebibliography}
\begin{appendix}
\onecolumn
\section{Catalogue-level splits: Data properties and B-mode analysis}
\label{ap:catlevel}
\begin{figure*}[h!]
\centering
\includegraphics[width=\textwidth]{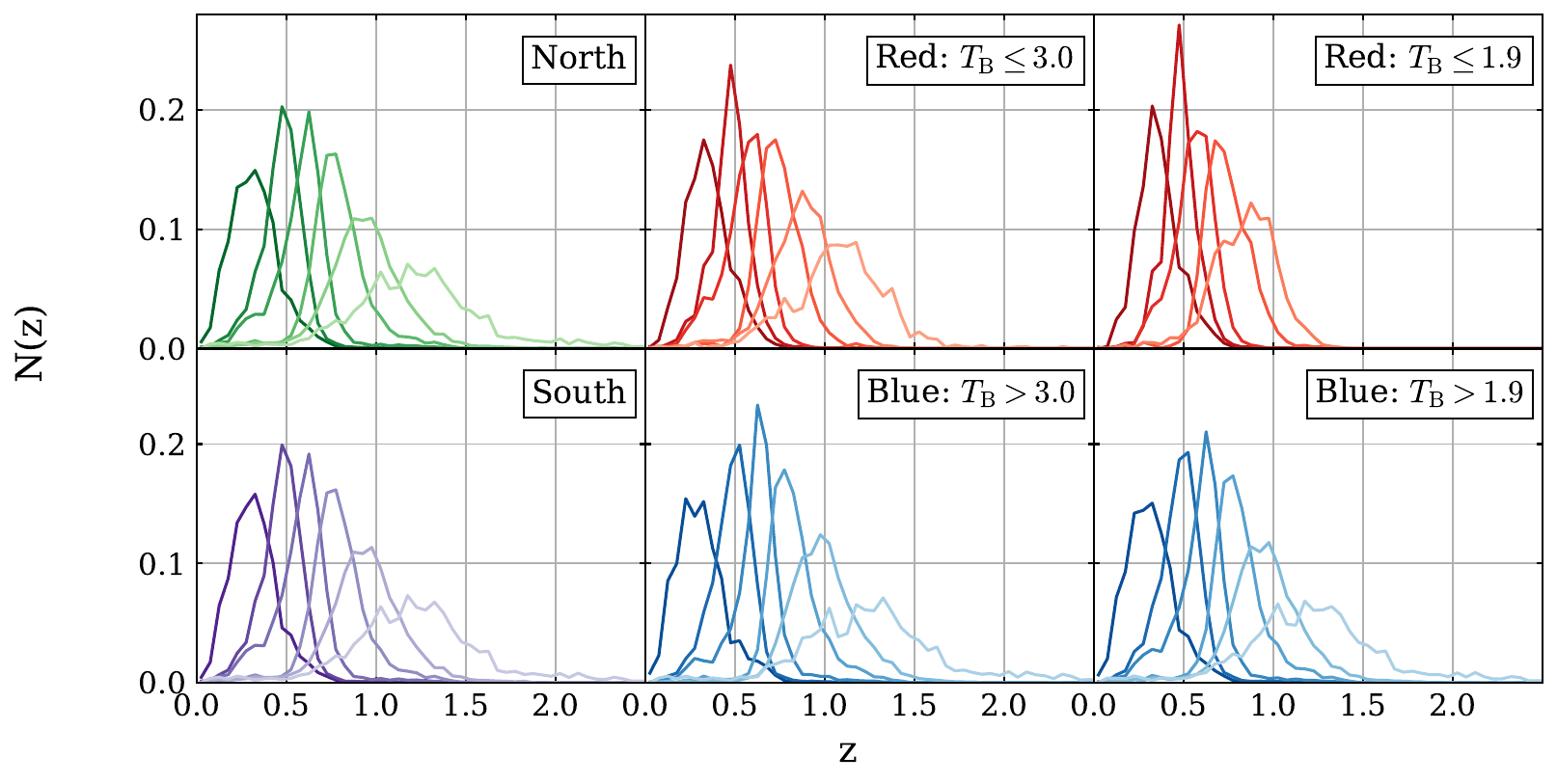}
\caption{Redshift distributions per tomographic bin for catalogue-level splits. Left panel: North-South split. Middle panel: Red-blue split defined via a cut on the spectral type of $T_{\rm B}=3.0$. Right panel: Red-blue split defined via a cut on the spectral type of $T_{\rm B}=1.9$.}
\label{fig:split_nz}
\end{figure*}

In this appendix, we summarise the data properties of the KiDS-Legacy catalogue split into mutually exclusive subsets, which are analysed in Sect. \ref{sec:results_catalogue}. Table \ref{tab:cataloguelevel_stats} lists the redshift range, the fraction of sources with respect to the total number of sources, the effective number density, the ellipticity dispersion, the shift in the mean of the redshift distribution, and the multiplicative shear bias. We provide these data for the fiducial catalogue, a catalogue split by hemisphere, and two catalogues split into samples of red and blue galaxies. The redshift distribution per tomographic bin for each catalogue-level split is illustrated in Fig. \ref{fig:split_nz}. 

For the split by hemisphere, we find the ellipticity dispersion, the shift in the mean of the redshift distribution per bin, and the multiplicative shear bias to be in agreement at the $1 \sigma$ level between both patches. Furthermore, the redshift distributions per patch in each tomographic bin show a good agreement between hemispheres.

For the colour-based split, we defined red galaxies via a selection on the spectral type $T_{\rm B}$ reported by the {\sc bpz} code. We adopted a threshold of $T_{\rm B}\leq 1.9$ from appendix B in \citetalias{Wright25}, which selects objects with contributions from an elliptical galaxy spectrum (template E1). For this split, the red galaxy sample only encompasses the first five tomographic bins since the catalogue only contains very few red galaxies with $z_{\rm B}>1.14$, making it challenging to calibrate the redshift distribution of the sixth bin. However, given the vanishingly low signal-to-noise ratio of such a sparsely populated bin, we do not expect the exclusion of this bin to make an impact on the analysis. By contrast, \citet{Li21} previously performed a consistency test between red and blue galaxies with cosmic shear data from the third KiDS data release \citep[KV450,][]{Dejong17,Wright19}, defining blue galaxies via a threshold of $T_{\rm B}\leq3.0$. This selection additionally encompasses objects with contributions from two spiral galaxy templates (Sbc and Sdc), while objects with contributions of irregular and starburst galaxy spectra (templates Im, SB2, and SB3) are labelled as blue galaxies. Originally, this threshold was chosen in order to ensure a similar constraining power per data subset, although this no longer holds in KiDS-Legacy given the addition of high redshift galaxies, which are predominantly blue.

\begin{table*}[h!]
\caption{Data properties per tomographic bin for catalogue-level splits.}
\label{tab:cataloguelevel_stats}
\centering
\begin{tabular}{cccccccc}
\hline\hline
Setup & Bin &  $z_{\rm B}$ & $f_{\rm total}$ [\%] & $n_{\rm eff}[{\rm arcmin}^{-2}]$ & $\sigma_{\rm e}$ & $\delta_{z}=z_{\rm est}-z_{\rm true}$ & $m$ \\
\hline
\multirow{6}{*}{Fiducial}
 & 1 & $0.10 < z_{\rm B} \leq 0.42$ & 18.1 & 1.77 & 0.28 & $-0.026 \pm 0.010$ & $-0.023 \pm 0.006$ \\
 & 2 & $0.42 < z_{\rm B} \leq 0.58$ & 18.0 & 1.65 & 0.27 & $\phantom{-}0.013 \pm 0.010$ & $-0.016 \pm 0.006$ \\
 & 3 & $0.58 < z_{\rm B} \leq 0.71$ & 16.6 & 1.50 & 0.29 & $-0.002 \pm 0.010$ & $-0.011 \pm 0.007$ \\
 & 4 & $0.71 < z_{\rm B} \leq 0.90$ & 16.8 & 1.46 & 0.26 & $\phantom{-}0.008 \pm 0.010$ & $\phantom{-}0.020 \pm 0.007$ \\
 & 5 & $0.90 < z_{\rm B} \leq 1.14$ & 15.8 & 1.35 & 0.28 & $-0.011 \pm 0.010$ & $\phantom{-}0.030 \pm 0.008$ \\
 & 6 & $1.14 < z_{\rm B} \leq 2.00$ & 14.6 & 1.07 & 0.30 & $-0.054 \pm 0.011$ & $\phantom{-}0.045 \pm 0.009$ \\
\hline\hline
\multirow{6}{*}{North}
 & 1 & $0.10 < z_{\rm B} \leq 0.42$ & \phantom{0}9.1 & 1.74 & 0.28 & $-0.025 \pm 0.010$ & $-0.024 \pm 0.009$ \\
 & 2 & $0.42 < z_{\rm B} \leq 0.58$ & \phantom{0}9.4 & 1.67 & 0.27 & $\phantom{-}0.013 \pm 0.010$ & $-0.021 \pm 0.009$ \\
 & 3 & $0.58 < z_{\rm B} \leq 0.71$ & \phantom{0}8.6 & 1.50 & 0.29 & $-0.001 \pm 0.010$ & $-0.008 \pm 0.011$ \\
 & 4 & $0.71 < z_{\rm B} \leq 0.90$ & \phantom{0}8.3 & 1.40 & 0.26 & $\phantom{-}0.009 \pm 0.010$ & $\phantom{-}0.016 \pm 0.010$ \\
 & 5 & $0.90 < z_{\rm B} \leq 1.14$ & \phantom{0}7.7 & 1.27 & 0.28 & $-0.011 \pm 0.010$ & $\phantom{-}0.021 \pm 0.011$ \\
 & 6 & $1.14 < z_{\rm B} \leq 2.00$ & \phantom{0}6.9 & 0.99 & 0.30 & $-0.057 \pm 0.011$ & $\phantom{-}0.040 \pm 0.013$ \\
\hline
\multirow{6}{*}{South}
 & 1 & $0.10 < z_{\rm B} \leq 0.42$ & \phantom{0}9.0 & 1.81 & 0.28 & $-0.025 \pm 0.010$ & $-0.017 \pm 0.010$ \\
 & 2 & $0.42 < z_{\rm B} \leq 0.58$ & \phantom{0}8.6 & 1.62 & 0.27 & $\phantom{-}0.013 \pm 0.010$ & $-0.010 \pm 0.010$ \\
 & 3 & $0.58 < z_{\rm B} \leq 0.71$ & \phantom{0}8.1 & 1.49 & 0.29 & $-0.001 \pm 0.010$ & $-0.015 \pm 0.012$ \\
 & 4 & $0.71 < z_{\rm B} \leq 0.90$ & \phantom{0}8.5 & 1.51 & 0.26 & $\phantom{-}0.008 \pm 0.010$ & $\phantom{-}0.018 \pm 0.011$ \\
 & 5 & $0.90 < z_{\rm B} \leq 1.14$ & \phantom{0}8.1 & 1.43 & 0.28 & $-0.012 \pm 0.010$ & $\phantom{-}0.026 \pm 0.012$ \\
 & 6 & $1.14 < z_{\rm B} \leq 2.00$ & \phantom{0}7.6 & 1.16 & 0.30 & $-0.057 \pm 0.011$ & $\phantom{-}0.047 \pm 0.013$ \\
\hline\hline
\multirow{6}{*}{Red: $T_{\rm B}\leq 3.0$} 
 & 1 & $0.10 < z_{\rm B} \leq 0.42$ & \phantom{0}6.8 & 0.73 & 0.27 & $-0.008 \pm 0.010$ & $-0.031 \pm 0.011$ \\
 & 2 & $0.42 < z_{\rm B} \leq 0.58$ & \phantom{0}6.0 & 0.60 & 0.25 & $-0.003 \pm 0.010$ & $-0.038 \pm 0.011$ \\
 & 3 & $0.58 < z_{\rm B} \leq 0.71$ & \phantom{0}6.1 & 0.62 & 0.29 & $-0.016 \pm 0.010$ & $-0.043 \pm 0.013$ \\
 & 4 & $0.71 < z_{\rm B} \leq 0.90$ & \phantom{0}6.7 & 0.62 & 0.25 & $-0.003 \pm 0.010$ & $-0.017 \pm 0.011$ \\
 & 5 & $0.90 < z_{\rm B} \leq 1.14$ & \phantom{0}6.4 & 0.58 & 0.30 & $-0.014 \pm 0.010$ & $-0.043 \pm 0.014$ \\
 & 6 & $1.14 < z_{\rm B} \leq 2.00$ & \phantom{0}2.7 & 0.21 & 0.30 & $-0.081 \pm 0.011$ & $-0.032 \pm 0.020$ \\
\hline
\multirow{6}{*}{Blue: $T_{\rm B}> 3.0$} 
 & 1 & $0.10 < z_{\rm B} \leq 0.42$ & 11.3 & 1.00 & 0.28 & $-0.025 \pm 0.010$ & $-0.016 \pm 0.007$ \\
 & 2 & $0.42 < z_{\rm B} \leq 0.58$ & 12.0 & 1.03 & 0.28 & $\phantom{-}0.030 \pm 0.010$ & $-0.005 \pm 0.007$ \\
 & 3 & $0.58 < z_{\rm B} \leq 0.71$ & 10.6 & 0.87 & 0.28 & $\phantom{-}0.019 \pm 0.010$ & $\phantom{-}0.015 \pm 0.009$ \\
 & 4 & $0.71 < z_{\rm B} \leq 0.90$ & 10.1 & 0.83 & 0.27 & $\phantom{-}0.007 \pm 0.010$ & $\phantom{-}0.047 \pm 0.009$ \\
 & 5 & $0.90 < z_{\rm B} \leq 1.14$ & \phantom{0}9.4 & 0.76 & 0.27 & $-0.023 \pm 0.010$ & $\phantom{-}0.075 \pm 0.010$ \\
 & 6 & $1.14 < z_{\rm B} \leq 2.00$ & 11.9 & 0.83 & 0.30 & $-0.060 \pm 0.011$ & $\phantom{-}0.065 \pm 0.010$ \\
\hline\hline
\multirow{5}{*}{Red: $T_{\rm B}\leq 1.9$}
 & 1 & $0.10 < z_{\rm B} \leq 0.42$ & \phantom{0}2.6 & 0.26 & 0.26 & $-0.014 \pm 0.010$ & $-0.032 \pm 0.019$ \\
 & 2 & $0.42 < z_{\rm B} \leq 0.58$ & \phantom{0}3.5 & 0.34 & 0.24 & $-0.008 \pm 0.010$ & $-0.036 \pm 0.017$ \\
 & 3 & $0.58 < z_{\rm B} \leq 0.71$ & \phantom{0}2.9 & 0.28 & 0.30 & $-0.015 \pm 0.010$ & $-0.047 \pm 0.022$ \\
 & 4 & $0.71 < z_{\rm B} \leq 0.90$ & \phantom{0}4.0 & 0.36 & 0.25 & $\phantom{-}0.001 \pm 0.010$ & $-0.016 \pm 0.019$ \\
 & 5 & $0.90 < z_{\rm B} \leq 1.14$ & \phantom{0}3.0 & 0.25 & 0.34 & $-0.005 \pm 0.010$ & $-0.086 \pm 0.028$ \\
 & 6 & $1.14 < z_{\rm B} \leq 2.00$ & \phantom{0}0.6 & --- & --- & --- & --- \\
\hline
\multirow{6}{*}{Blue: $T_{\rm B}> 1.9$}
 & 1 & $0.10 < z_{\rm B} \leq 0.42$ & 15.5 & 1.48 & 0.28 & $-0.023 \pm 0.010$ & $-0.022 \pm 0.006$ \\
 & 2 & $0.42 < z_{\rm B} \leq 0.58$ & 14.5 & 1.29 & 0.28 & $\phantom{-}0.024 \pm 0.010$ & $-0.010 \pm 0.006$ \\
 & 3 & $0.58 < z_{\rm B} \leq 0.71$ & 13.7 & 1.20 & 0.29 & $\phantom{-}0.007 \pm 0.010$ & $-0.001 \pm 0.008$ \\
 & 4 & $0.71 < z_{\rm B} \leq 0.90$ & 12.8 & 1.09 & 0.27 & $\phantom{-}0.005 \pm 0.010$ & $\phantom{-}0.030 \pm 0.008$ \\
 & 5 & $0.90 < z_{\rm B} \leq 1.14$ & 12.8 & 1.08 & 0.27 & $-0.019 \pm 0.010$ & $\phantom{-}0.044 \pm 0.008$ \\
 & 6 & $1.14 < z_{\rm B} \leq 2.00$ & 14.1 & 1.03 & 0.30 & $-0.062 \pm 0.011$ & $\phantom{-}0.048 \pm 0.009$ \\
\end{tabular}
\tablefoot{We list the index of the tomographic bin, the range in photometric redshift $z_{\rm B}$, the fraction of sources with respect to the total number of sources, the effective number density $n_{\rm eff}$, the ellipticity dispersion $\sigma_{\rm e}$, the shift in the mean of the redshift distribution $\delta_{z}$, and the multiplicative shear bias $m$.}
\end{table*}

\begin{figure*}[h!]
\centering
\includegraphics[width=\textwidth]{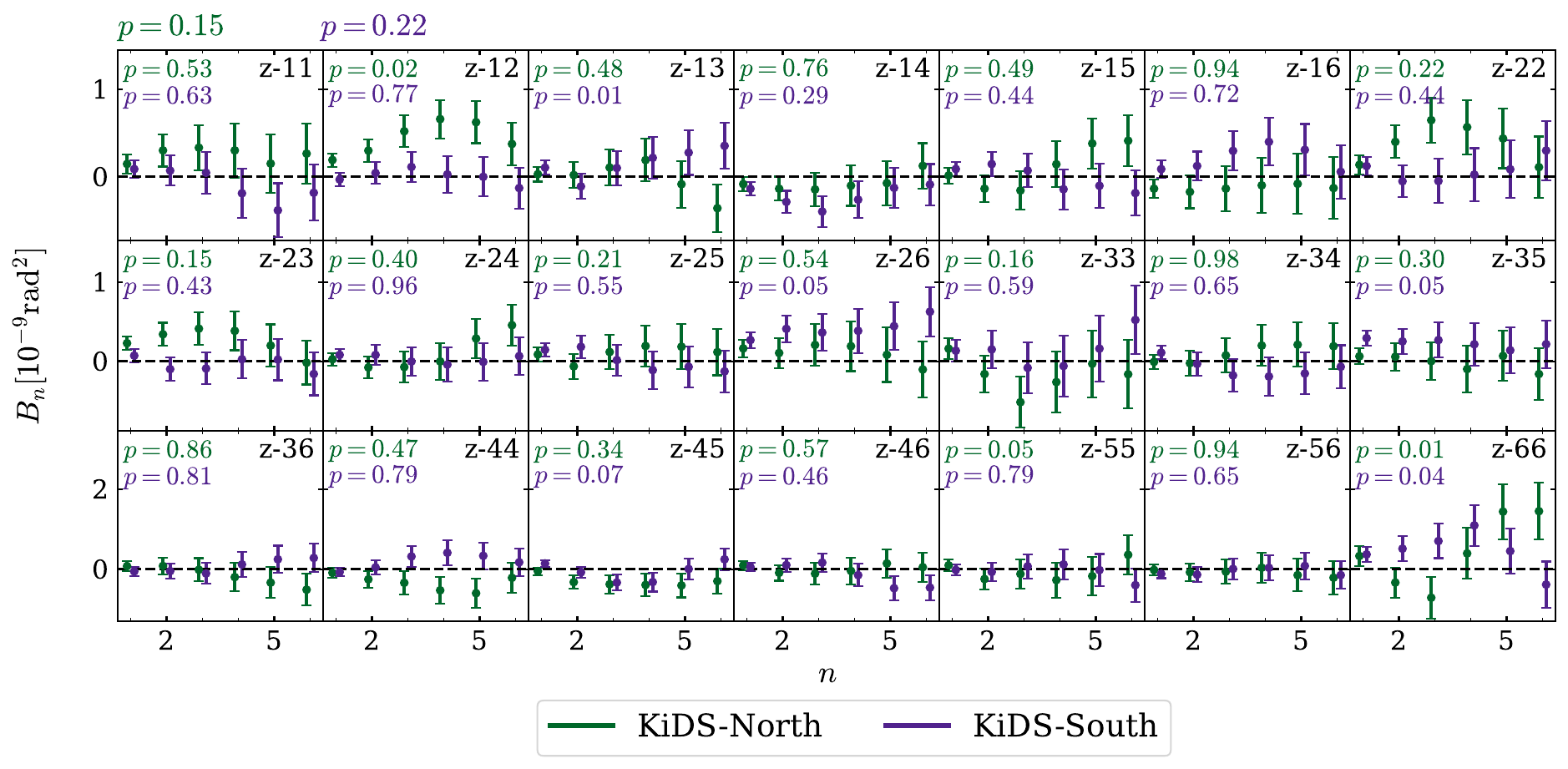}
\caption{COSEBIs B-mode measurements for the North-South split catalogue. The green and purple data points show the measurements of the KiDS-North and KiDS-South sample, respectively. Each panel represents auto- or cross-correlation between tomographic bins, as indicated by the label in the top right corner. The corresponding $p$-value is denoted in the top left corner of each panel. The $p$-value of the combined data vector is given in the top left corner of the figure. For visualisation purposes, we display the discrete $n$ modes with an offset on the x-axis. We note that the B-mode signals are highly correlated within a tomographic bin and advise against a so-called \enquote*{$\chi$-by-eye}.}
\label{fig:bmode_northsouth}
\end{figure*}
\begin{figure*}[h!]
\centering
\includegraphics[width=\textwidth]{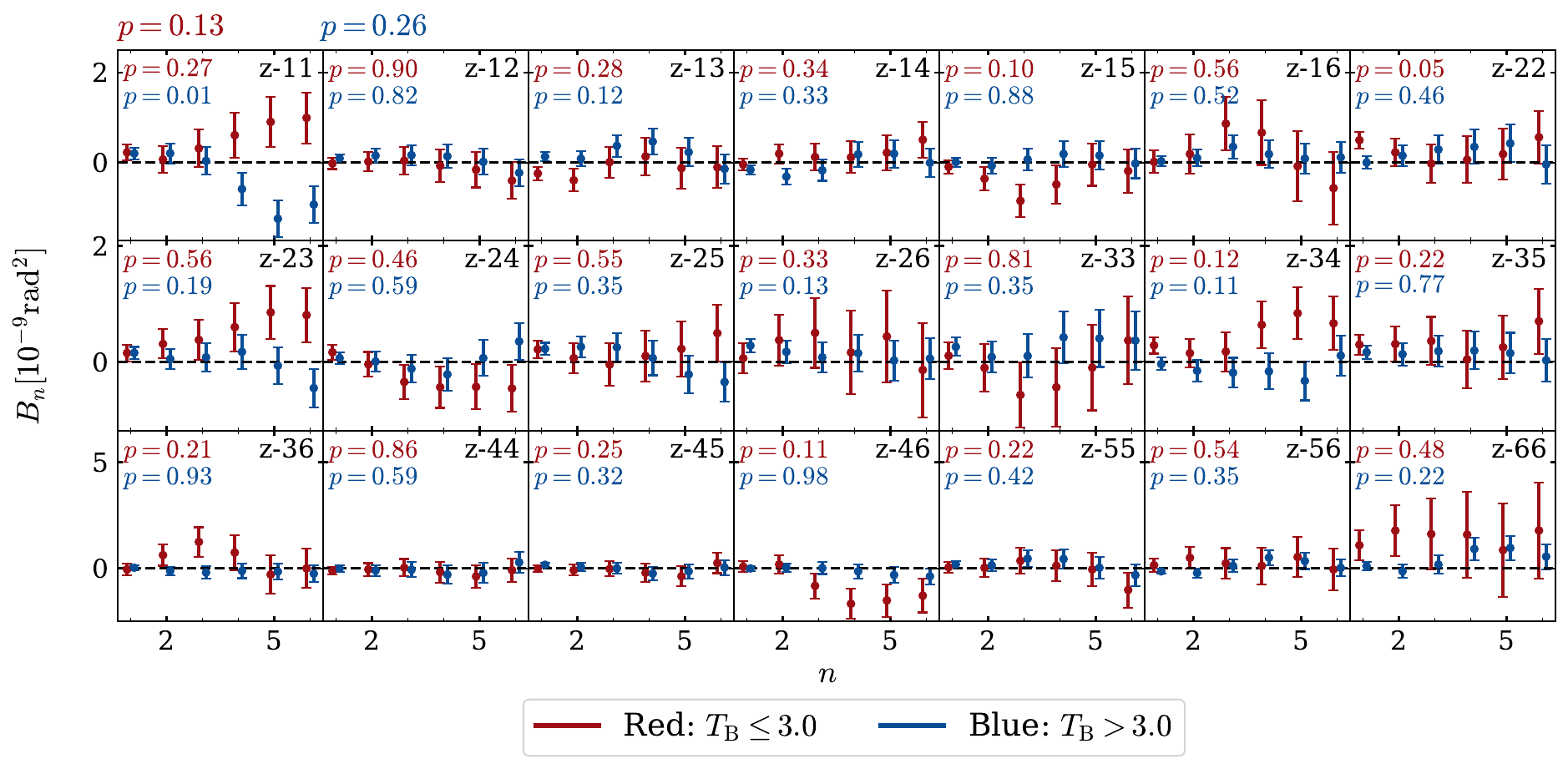}
\caption{Same as Fig. \ref{fig:bmode_northsouth} but for the red-blue split catalogue defined via a cut on the spectral type $T_{\rm B}=3.0$.}
\label{fig:bmode_redblue}
\end{figure*}
\begin{figure*}[h!]
\centering
\includegraphics[width=\textwidth]{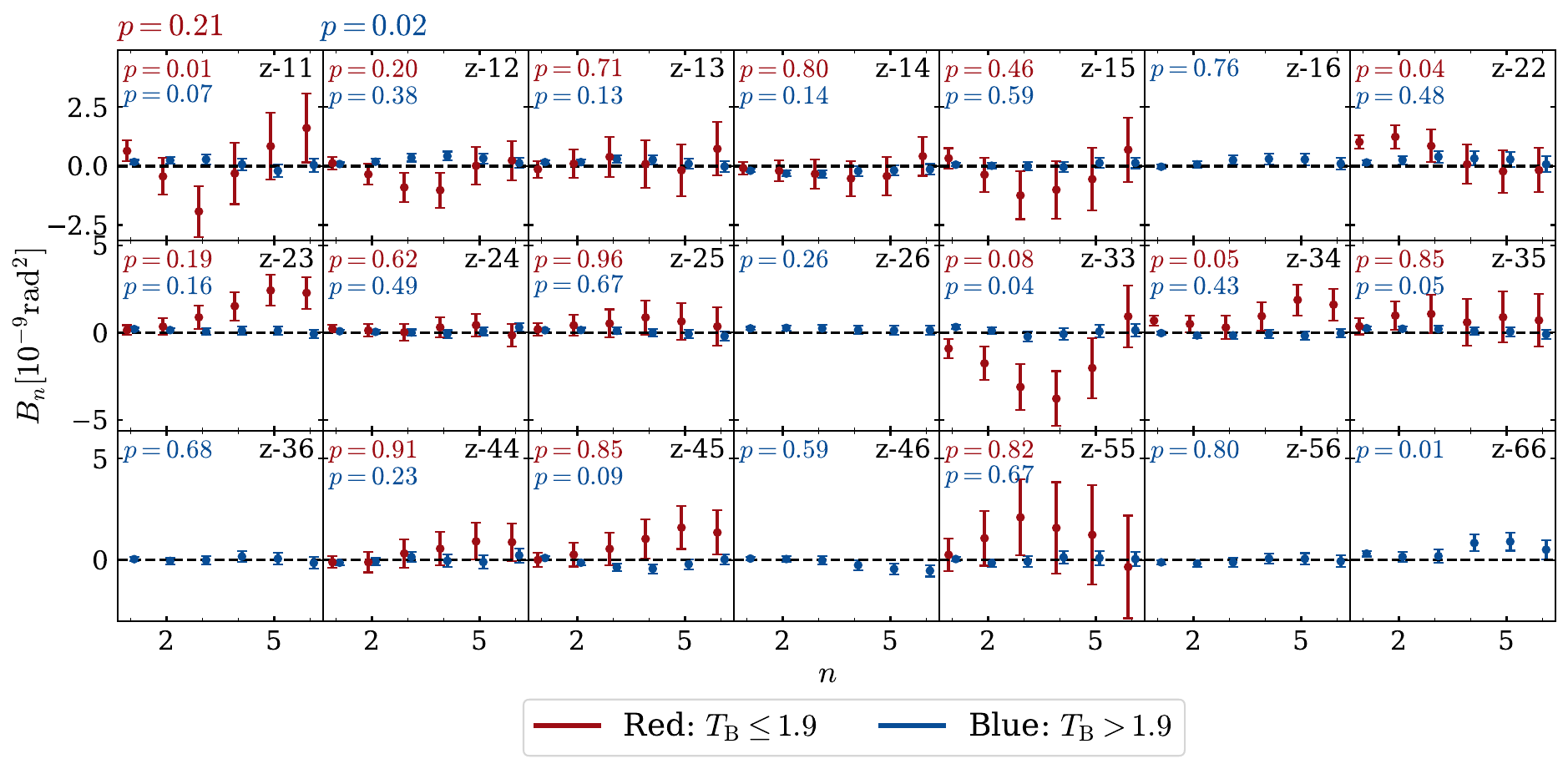}
\caption{Same as Fig. \ref{fig:bmode_northsouth} but for the red-blue split catalogue defined via a cut on the spectral type $T_{\rm B}=1.9$. For this threshold, the catalogue only contains very few red galaxies with $z_{\rm B}>1.14$. Therefore, the red galaxy sample only encompasses the first five tomographic bins.}
\label{fig:bmode_redblue_1p9}
\end{figure*}

To the first order, we expect cosmic shear to only produce E-mode signals, making B-mode signals negligible for current stage-III surveys. Therefore, B modes are a useful quantity for null-tests for residual systematics in the cosmic shear measurement. Here, we quantified the significance of the B modes for the split catalogues, referring to appendix E in \citetalias{Wright25} for a discussion of B modes in the fiducial catalogue. We employed COSEBIs as test statistic since it allows for a clean separation of E and B modes. To quantify the significance of the B mode, we computed the $\chi^2$ values assuming the null hypothesis and compute the corresponding $p$-value. Here, the $p$-value is equal to the probability of producing a B mode that is more significant than the observed signal, assuming that the B mode is randomly drawn from a Gaussian distribution with zero mean.

For each subset of the catalogue, we computed the first six COSEBIs B modes and quantified the significance for the full data vector consisting of all 21 combinations of the six tomographic redshift bins. Additionally, we quantified the B mode significance individually for each tomographic bin combination. The measured B-mode signal for the three catalogue-level splits are presented in Figs. \ref{fig:bmode_northsouth}, \ref{fig:bmode_redblue}, and \ref{fig:bmode_redblue_1p9}. All $p$-values pass our required threshold of $p>0.01$, which is the community standard for B-mode tests \citep{KiDS+DES}, for both the individual tomographic bin combination and for the combination of all tomographic bins. 
\FloatBarrier
\section{Effective number of constrained parameters}
\label{ap:neff}
For the calculation of the tension probability with the suspiciousness statistic, we require the difference between the effective number of constrained parameters $N_{\Theta}$ in the fiducial and split cosmological model, as defined in Eq. \eqref{eq:d_eff}. We followed the methodology of \citet{Joachimi21} and infer $N_{\Theta}$ from mock realisations of the cosmic shear data vector. Assuming a \citet{Planck2018} cosmology and our fiducial data covariance matrix, we generated 1000 realisations of the data vector from a multivariate Gaussian distribution. For each mock realisation, we then maximised the posterior and obtain an estimate of the best-fit $\chi^2$. As discussed in \citet{Joachimi21}, the distribution of $\chi^2_{\rm best}$ is well described by a $\chi^2$-distribution with a number of degrees of freedom $k_{\rm dof} = N_{\rm d} - N_{\Theta,\chi^2}$, where $N_{\rm d}$ denotes the dimensionality of the data vector. We therefore fit the distribution of $\chi^2_{\rm best}$-values from mock data vectors to a $\chi^2$-distribution to determine the effective number of degrees of freedom, denoted $N_{\Theta,\chi^2}$.

The inferred estimates of $N_{\Theta}$ are summarised in Table \ref{tab:ndim} for the fiducial cosmic shear analysis setup with all summary statistics, dubbed \enquote*{1cosmo}, as well as for the split cosmological analysis setups with COSEBIs considered in Sect. \ref{sec:results}. For the split between angular scales, we employ band powers and 2PCFs, as discussed in Sect. \ref{sec:angular}. Additionally, we infer $N_{\Theta}$ for catalogue- and statistic-level splits in a single cosmological setup. For comparison, we provide the BMD, $N_{\Theta,{\rm BMD}}$, computed via Eq. \eqref{eq:BMD}. For both estimates, we list the resulting difference in $N_{\Theta}$ between the fiducial cosmological model and the split cosmological model, denoted as $d_{\chi^2}$ and $d_{\rm BMD}$. Overall, we found the number of constrained parameters to be higher in the split cosmological model. We attribute this to the doubling of the cosmological parameter space, where both instances of parameters sensitive to cosmic shear have been independently constrained.
\begin{table*}[h!]
\caption{Effective number of constrained parameters.}
\label{tab:ndim}
\centering
\begin{tabular}{clcccc}
\hline\hline
Split type & Setup & $N_{\Theta,\chi^2}$ & $N_{\Theta,{\rm BMD}}$ & $d_{\chi^2}$ & $d_{\rm BMD}$\\
\hline
&Fiducial - 1cosmo (COSEBIs)& 5.60 & 6.00 &-&-\\
&Fiducial - 1cosmo (Band powers)&5.23&4.16 &-&-\\
&Fiducial - 1cosmo (2PCFs)&7.79&7.79 &-&-\\
\hline
\multirow{9}{*}{\rotatebox[origin=c]{90}{Data vector}} 
& Bin 1 & 8.05 & 7.60 & 2.45 & 1.60\\
& Bin 2 & 8.20 & 5.52 & 2.60 & -0.48\\
& Bin 3 & 8.48 & 7.83 & 2.88 & 1.83\\
& Bin 4 & 8.69 & 9.05 & 3.09 & 3.05\\
& Bin 5 & 9.01 & 8.51 & 3.41 & 2.51\\
& Bin 6 & 8.85 & 8.06 & 3.25 & 2.06\\
& Redshift bin AC vs CC & 8.53 & 8.81 & 2.93 & 2.81\\
& $E_{n=1}$ vs $E_{n>1}$ & 7.35 & 7.22 & 1.75 & 1.22\\
& Small multipoles vs large multipoles (Band powers) & 7.70 & 5.98 & 2.47 & 1.82\\
& Small scales vs large scales (2PCFs) & 10.67 & 10.35 & 2.88 & 2.56\\
\hline
\multirow{6}{*}{\rotatebox[origin=c]{90}{Catalogue}} 
& North vs South & 9.17 & 7.45 & 3.03 & 1.81\\
& Red vs blue, $T_{\rm B}=3.0$ & 9.90 & 7.55 & 3.63 & 0.80\\
& Red vs blue, $T_{\rm B}=1.9$ & 7.03 & 8.70 & 3.40 & 3.80\\
& North vs South - 1cosmo& 6.14 & 5.64 &-&-\\
& Red vs blue - 1cosmo, $T_{\rm B}=3.0$& 6.27 & 6.75 &-&-\\
& Red vs blue - 1cosmo, $T_{\rm B}=1.9$& 3.63 & 5.54 &-&-\\
\hline
\multirow{6}{*}{\rotatebox[origin=c]{90}{Statistic}}
&COSEBIs vs Band powers & 10.22 & 10.69 & 3.26 & 4.59\\
&COSEBIs vs 2PCFs & 9.45 & 11.16 & 3.80 & 3.40\\
&Band powers vs 2PCFs & 8.57 & 10.01 & 3.59 & 3.03\\
&COSEBIs vs Band powers - 1cosmo& 6.96 & 6.10 &-&-\\
&COSEBIs vs 2PCFs - 1cosmo& 5.65 & 7.76 &-&-\\
&Band powers vs 2PCFs - 1cosmo& 4.98 & 6.98 &-&-\\
\end{tabular}
\tablefoot{For each analysis setup we provide the effective number of constrained parameters inferred via $\chi^2$ minimisation of mock data vectors \citep{Joachimi21} in the third column, and the Bayesian model dimensionality \citep{handley19}, inferred directly from the chain, in the fourth column. The first three rows show the results in an analysis with the fiducial setup, dubbed `1cosmo`, in which we do not apply any splits to the data. The remaining rows present results when splitting the data into subsets. Here, we report the effective number of constrained parameters with the split cosmological model. For the splits at the catalogue level and the splits by summary statistic we additionally report the results with a shared set of parameters. The final two columns report the difference in the number of constrained parameters between the fiducial cosmological model and the split cosmological model. We employ COSEBIs as the summary statistic except when explicitly stated.}
\end{table*}
\FloatBarrier
\section{Sensitivity tests of consistency metrics}
\label{ap:sensitivity}
\begin{figure*}[h!]
    \centering
    \sidecaption
    \includegraphics[width=0.5\textwidth]{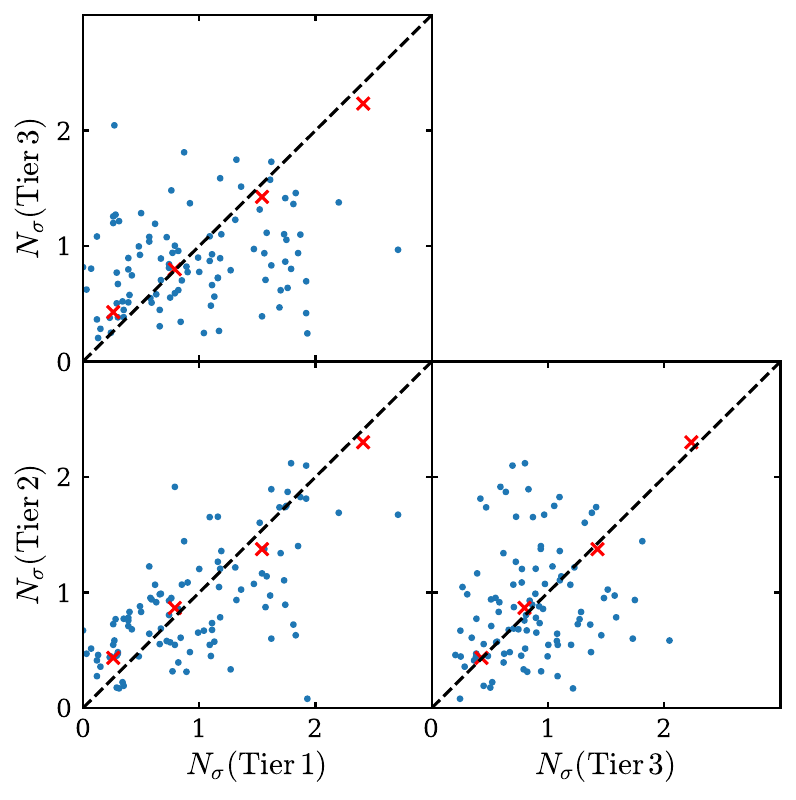}
    \caption{Distribution of the tier 1 evidence-based metric, the tier 2 multi-dimensional parameter metric, and the tier 3 PPD metric for 100 mock realisations of the fiducial data vector generated from a multivariate Gaussian distribution assuming a reference cosmology with $S_8=0.777$. The red crosses indicate the consistency metrics inferred in analyses with noise-free data vectors with systematic shifts in the input $S_8$ of the fifth bin by $\Delta S_8 = [0.01,0.02,0.03, 0.04]$.}
    \label{fig:mock_consistency}
\end{figure*}

In this appendix, we test the sensitivity of the consistency metrics. In particular, we assess the impact of noise fluctuations in a consistency analysis with internally consistent data. We generated 100 realisations of the fiducial data vector from a multivariate Gaussian distribution assuming a reference cosmology with $S_8=0.777$ and conducted consistency analyses for a split analysis of the fifth redshift bin. For this test, we computed the corresponding number of sigma from the tier 3 PPD-based metric via Eq. \eqref{eq:nsigma}. However, as discussed in Sect. \ref{sec:tier3_methods}, we emphasise that \citet{Doux21} show that the PPD metric can result in $p$-values that are biased towards low values and therefore the level of tension can potentially be overestimated. We found that the tier 1 evidence-based metric yields a tension below 1$\sigma$ (2$\sigma$) for 70\% (97\%) of the mocks, while the tier 2 multi-dimensional parameter metric and the tier 3 PPD metric yield tension below 1$\sigma$ (2$\sigma$) for 69\% (98\%), and 65\% (99\%) of the mocks, respectively. Thus, we conclude that in the absence of internal tension, the inferred consistency with each metric individually is compatible with typical noise fluctuations. The distribution of consistency metrics is shown in Fig. \ref{fig:mock_consistency}. We found the three metrics to be correlated with $\rho_{\rm Tier 1, Tier 2}=0.64$, $\rho_{\rm Tier 1, Tier 3}=0.22$, and $\rho_{\rm Tier 2, Tier 3}=0.15$. Additionally, we tested the sensitivity of the consistency metrics to internally inconsistent data by generating noise-free mock data vectors, applying shifts in the input $S_8$ of the fifth bin by $\Delta S_8 = [0.01,0.02,0.03, 0.04]$, respectively. For each mock, we conducted a consistency analysis and display the corresponding metrics as red crosses in Fig. \ref{fig:mock_consistency}. We found that all three metrics are capable of recovering the input tension in the data, with the estimate of the significance of the internal inconsistency being consistent between each metric.
\end{appendix}
\end{document}